\documentclass[showpacs,preprintnumbers,twocolumn,eqsecnum,superscriptaddress,aps,nofootinbib]{revtex4}
\usepackage{amssymb}
\usepackage[centertags]{amsmath}
\usepackage{txfonts}
\usepackage{epsfig}
\usepackage{bm}
\usepackage{color}
\usepackage{graphicx,graphics}
\usepackage{multirow}
\usepackage{float}
\usepackage[pdfstartview=FitH]{hyperref}
\hypersetup{colorlinks=true, citecolor=blue,linkcolor=blue,filecolor=black,urlcolor=blue}
\allowdisplaybreaks[2]

\usepackage{slashed}
\usepackage{ulem}

\begin{document}

\title{Leading and higher twist contributions in semi-inclusive $e^+e^-$ annihilation at high energies}

\author{Shu-yi Wei}
\affiliation{School of Physics \& Key Laboratory of Particle Physics and Particle Irradiation (MOE), Shandong University, Jinan, Shandong 250100, China}

\author{Kai-bao Chen}
\affiliation{School of Physics \& Key Laboratory of Particle Physics and Particle Irradiation (MOE), Shandong University, Jinan, Shandong 250100, China}

\author{Yu-kun Song}
\affiliation{Interdisciplinary Center for Theoretical Study and Department of Modern
Physics, University of Science and Technology of China, Anhui 230026, China}

\author{Zuo-tang Liang}
\affiliation{School of Physics \& Key Laboratory of Particle Physics and Particle Irradiation (MOE), Shandong University, Jinan, Shandong 250100, China}

\begin{abstract}
By applying the collinear expansion to the semi-inclusive hadron production process 
$e^++e^-\to h+\bar q(jet)+X$ at high energies, 
we construct a theoretical framework where leading and higher twist contributions 
at the leading perturbative QCD can be calculated systematically.
With this framework, we calculate the contributions up to twist-3 for spin-0,
spin-1/2 and spin-1 hadrons respectively.
We present the results for the hadronic tensors, the differential cross sections,
the azimuthal asymmetries, and the polarizations of the hadrons. 
\end{abstract}


\maketitle


\section{Introduction}

Since there is no hadron involved in the initial state, $e^+e^-$ annihilation is most suitable 
for the study on fragmentation functions among all different high energy reactions.
Similar to the study on parton distribution functions in deep-inelastic lepton-nucleon scattering, 
the longitudinal momentum dependence can be studied in inclusive process while  
the transverse momentum dependence can only be studied by going to semi-inclusive processes.

The study on fragmentation functions is in parallel to that on 
parton distribution and/or correlation functions in nucleon. 
It plays an important role in the description of high energy reactions and 
in studying the properties of hadronic interactions and is therefore
a standing topic in the field of high energy physics.
Many progresses have been made and summarized 
constantly in Review of Particle Properties~\cite{Beringer:1900zz} and also other recent reviews~\cite{Albino:2008gy}.
Much attention has been attracted recently in particular 
in the spin and transverse momentum dependent (TMD) sessions 
both in theory~\cite{Nachtmann:1973mr, Efremov:1983eb, Collins:1992kk, Jaffe:1993xb, Ji:1993vw, Chen:1994ar, Boer:1997mf, deFlorian:1997zj, Anselmino:2000vs, Bacchetta:2005rm, Yuan:2009dw, 
Kanazawa:2013uia, Wei:2013csa, Pitonyak:2013dsu}, 
and in experiment~\cite{Buskulic:1996vb, Ackerstaff:1997nh,Ackerstaff:1997kj, Ackerstaff:1997kd, Abreu:1997wd, ALEPH:2005ab,Abe:2005zx,Vossen:2011fk,TheBABAR:2013yha,BES3}.
This provides a new window to study the fragmentation function, 
to test the hadronization models and 
to learn the properties of Quantum Chromodynamics (QCD).

As stressed in different publications~\cite{Qiu:1991pp,Qiu:1998ia,Jaffe:1990qh}, 
to study the spin and TMD sessions of
the parton distribution or fragmentation function, 
higher twist contributions can be very significant.
It is therefore very important for such studies in high energy reactions to establish 
a suitable theoretical framework where leading and higher twist contributions can be calculated consistently.
Collinear expansion seems to be the right technique for such a purpose.

Collinear expansion was developed in 1980s 
for inclusive deep inelastic lepton-nucleon scattering~\cite{Ellis:1982wd,Qiu:1990xxa}
and has been known as the unique way to obtain a formalism where the differential cross section including
higher twist contributions is expressed in terms of the calculable hard parts and gauge invariant
parton distribution and correlation functions.
The gauge links in the parton distributions are obtained automatically in the expansion procedure
where multiple gluon scattering is taken into account.

Recently, the collinear expansion has been applied successfully to semi-inclusive deep inelastic
lepton-nucleon scattering process $l+N\to l+q {\rm (jet)}+X$~\cite{Liang:2006wp,Song:2010pf},
where $q$ denotes a quark that corresponds to a jet in experiments.
With this process, TMD parton distribution and/or correlation functions can be studied.
Calculations have been carried out where leading and higher twist contributions and also 
nuclear dependences have been obtained 
and expressed in terms of the gauge invariant parton distribution and 
correlation functions~\cite{Liang:2006wp,Song:2010pf,Liang:2008vz,Gao:2010mj,Song:2013sja,Song:2014sja}. 

To study the fragmentation function, we started with the inclusive
hadron production process in $e^+e^-$-annihilation at high energies~\cite{Wei:2013csa}. 
We have applied the collinear expansion to the process and 
obtained a theoretical framework for describing $e^++e^-\to h+X$ 
where leading and higher twist contributions can be calculated systematically.
With this process, the longitudinal momentum distribution of the fragmentation functions can be studied 
and we have made calculations up to twist three for hadrons with different spins respectively. 
Even in this simple case, we have already obtained a number of interesting features
such as the existence of transverse polarization at twist-3 for spin-$1/2$ hadrons, 
the quark polarization independence of the leading twist spin alignment  
of vector mesons and so on. 

To study the TMD session of the fragmentation functions, we need to go to semi-inclusive process
where more than a single hadron are detected. 
The simplest process in this case is $e^++e^-\to h+\bar q{\rm (jet)}+X$.
In such semi-inclusive processes, the measurable quantities sensitive to different components 
of the fragmentation function are usually the azimuthal asymmetries including 
both the spin dependent and the spin independent ones.
Higher twist effects can give significant contributions to such asymmetries, 
hence a systematic calculation to pick up the leading and higher twist contributions is important.

In this paper, we apply the collinear expansion to the semi-inclusive process $e^++e^-\to h+\bar q{\rm (jet)}+X$.
We derive the theoretical framework suitable for the description of this process,  
and carry out the calculations up to twist 3 for hadrons with different spins 
at the leading order in perturbative QCD. 
We present the results for the hadronic tensors, the differential cross sections, 
the azimuthal asymmetries and the hadron polarizations, 
and discuss the situation when confronting with experiments.
We shall note here that we can easily carry out the same calculation for process $e^++e^-\to h+ q{\rm (jet)}+X$, 
and get similar results.

The rest of this paper is organized as follows.
After this introduction, in Sec. II, we make a brief summary of the collinear expansion in 
the inclusive process $e^+ + e^-\to h+X$ and extend it to the semi-inclusive
process $e^+ + e^-\to h+\bar q+X$ to get the gauge invariant formalism
of hadronic tensor and cross section. 
In Sec. III, we present the results for the hadronic tensors up to twist 3 
for spin-0, spin-1/2 and spin-1 hadrons respectively. 
The results for the differential cross sections and the resulting azimuthal asymmetries 
and the polarizations are presented and discussed in Sec. IV.   
Finally, we make a summary and give an outlook in Sec. V.

\section{The gauge invariant formalism}

\begin{figure}[h!]
\includegraphics[width=0.4\textwidth]{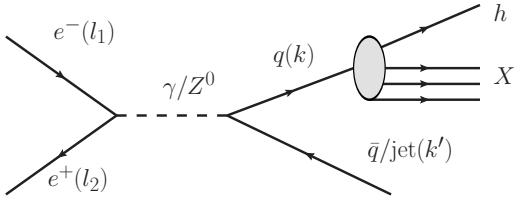}
\caption{Illustrating diagram for semi-inclusive hadron production in $e^+e^-$ annihilation.}
\label{ff1}
\end{figure}

In this section, we apply collinear expansion to the semi-inclusive $e^+e^-$
annihilation at high energies and derive the gauge invariant formalism
for the hadronic tensor including leading and higher twist contributions systematically.
To be explicit, we consider the semi-inclusive process $e^+ + e^-\to h+\bar q +X$ as illustrated in Fig.\ref{ff1}.
The differential cross section is given by the product of the leptonic tensor and the hadronic tensor,
\begin{align}
d\sigma^{(si)} = & \frac{g^4_z}{32s} L_{\mu'\nu'} (l_1,l_2) D^{\mu'\mu} (q) D^{\nu'\nu*}(q) \nonumber\\
&\times W_{\mu\nu} ^{(si)}(q,p,S,k') \frac{d^3p}{(2\pi)^2 2E_p} \frac{d^3k'}{(2\pi)^32E'}. \label{csdef}
\end{align}
The notations that we use here are the same as those in \cite{Wei:2013csa},
i.e. we use $l_1$ and $l_2$ to denote the 4-momenta of the incoming electron and positron,
and $q=l_1+l_2$ to denote the 4-momentum of the intermediate gauge boson.
The momentum of the quark is denoted by $k$ and that of the produced hadron is denoted by $p$.
Here, for the semi-inclusive process $e^+ + e^-\to h+\bar q +X$,
the superscript $si$ is introduced to denote that the quantity is for semi-inclusive
and the momentum $k'$ denotes the momentum of $\bar q$ or the corresponding jet.
Also, similar to \cite{Wei:2013csa}, we consider $e^+e^-$ annihilation into hadrons either
via electromagnetic interaction with the exchange of a virtual photon or
via weak interaction with the exchange of a $Z^0$ boson.
We do not consider the interference term and
the results apply to reactions near the $Z^0$ pole where only the
weak interaction term is considered or the energy is much lower than $Z^0$ mass
where only electromagnetic interaction is needed.

The leptonic tensor is exactly the same as that for inclusive process and is given in \cite{Wei:2013csa}.
It is defined as,
\begin{align}
  L_{\mu'\nu'}(l_1,l_2)=&\frac{1}{4} \mathrm{Tr}\left[\Gamma_{\mu'}^e \slashed l_1 \Gamma_{\nu'}^e \slashed l_2 \right],
\end{align}
where we use $\Gamma_{\mu'}^e$ instead of $\gamma_{\mu'}$ since the intermediate boson can be a photon or a $Z^0$ boson.
In the case that the intermediate boson is a $Z^0$ boson, 
i.e. for $e^++e^-\to Z^0\to h+\bar q+X$, 
we have $\Gamma_{\mu'}^e = \gamma_{\mu'}(c_V^e - c_A^e \gamma^5)$,
while $\Gamma_{\mu'}^e = \gamma_{\mu'}$ or equivalently $c_V=1$ and $c_A=0$ if it is a photon, 
i.e. for $e^++e^-\to \gamma^*\to h+\bar q+X$.
Correspondingly, the propagator is
$\hat D_{\mu'\mu} = (g_{\mu'\mu}-q_{\mu'}q_\mu /M_Z^2)/[(Q^2-M_Z^2)+i\Gamma_Z M_Z]$
and  $\hat D_{\mu'\mu} = g_{\mu'\mu}/Q^2$, respectively.
The weak coupling $g_Z=g/\cos\theta_W=e/\sin\theta_W\cos\theta_W$, where $e$ is the electron charge
and $\theta_W$ is the Weinberg angle.
For $Z^0$-exchange, $L_{\mu'\nu'}(l_1,l_2)$ is given by, 
\begin{align}
L_{\mu'\nu'}(l_1,l_2) & =c_1^e\left[l_{1\mu'} l_{2\nu'}+l_{1\nu'} l_{2\mu'}-(l_1\cdot l_2)g_{\mu'\nu'}\right] \nonumber\\
&+ic_3^e\epsilon_{\mu'\nu'\rho\sigma}l_{1}^{\rho}l_{2}^{\sigma},
\end{align}
where $c_1^e = (c_V^e)^2+(c_A^e)^2$ and $c_3^e = 2 c_V^e c_A^e$. 
We see that $L_{\mu'\nu'}(l_1,l_2)$ have both a symmetric part and an anti-symmetric part for reactions via exchange of $Z^0$.
However, for reactions via electromagnetic interaction, the results are obtained by taking $c_1=1$ and $c_3=0$ 
and we have only the symmetric part left.

The difference to the inclusive case lies in the hadronic tensor $W_{\mu\nu}^{(si)}$.
For the semi-inclusive process $e^+ + e^-\to h+\bar q +X$, it is defined as,
\begin{align}
W_{\mu\nu}^{(si)} & (q,p,S,k') = \frac{1}{2\pi}  \sum_X (2\pi)^4 \delta^4 (q-p -k'- P_X) \nonumber\\
&\times \langle 0| J_\nu (0) |k',p,S;X\rangle \langle k',p,S;X |J_\mu (0)|0\rangle.  \label{ht001}
\end{align}
It contains the fragmentation function and hard part of the hadronic interaction that will be discussed in the following.

\subsection{Multiple gluon scattering \& collinear expansion}

\begin{figure}[h!]
\includegraphics[width=0.5\textwidth]{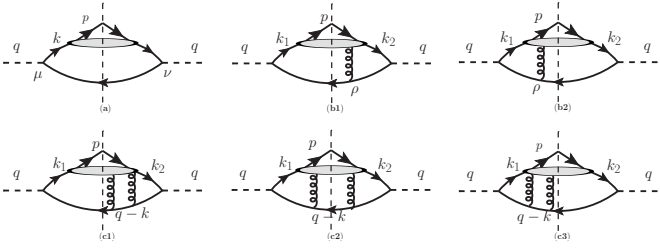}
\caption{The first few Feynman diagrams as examples of the diagram series with exchange of $j$ gluon(s).
In (a), (b) and (c), we see the case for $j=0$, $1$ and $2$ respectively.
The gluon momentum in (b) is $k_1-k_2$, while in (c), they are $k-k_1$ and $k_2-k$ respectively.} 
\label{fdmg}
\end{figure}

Similar as inclusive process, to the leading order, the hadronic tensor is shown in Fig.~\ref{fdmg}(a), and is given by,
\begin{align}
W_{\mu\nu}^{(0,si)} (q,p,S,k') = \int \frac{d^4k}{(2\pi)^4} 
\mathrm{Tr}\left[ \hat{H}_{\mu\nu}^{(0,si)} (k,k',q) \hat{\Pi}^{(0)}(k,p,S) \right], \label{W0si}
\end{align}
where the matrix element is the same as that defined in the inclusive case and is given by,
\begin{equation}
\hat{\Pi}^{(0)} (k,p,S)= \frac{1}{2\pi} \sum_X\int d^4\xi e^{-ik\xi} \langle 0 |\psi(0)|hX\rangle \langle hX| \bar\psi(\xi)|0\rangle ,
\end{equation}
while the hard part is different,
\begin{equation}
\hat{H}_{\mu\nu}^{(0,si)} (k,k',q) = \Gamma_\mu^q (\slashed q - \slashed k) \Gamma_\nu^q  (2\pi)^4 \delta^4(q-k-k'),
\label{hp01si}
\end{equation}
in contrast to that for the inclusive process,
\begin{equation}
\hat{H}_{\mu\nu}^{(0)} (k,q) = \Gamma_\mu^q (\slashed q - \slashed k) \Gamma_\nu^q  (2\pi) \delta_+ \left((q-k)^2\right). \label{hp01}
\end{equation}

It is well known that, because the two quark fields in the matrix element $\hat{\Pi}^{(0)}$ 
do not share the same space-time coordinate,
$\hat{\Pi}^{(0)}$ is not local (color) gauge invariant.
To get the gauge invariant form, we need to consider the final-state interaction in QCD,
and apply the collinear expansion technique~\cite{Ellis:1982wd,Qiu:1990xxa}.

The collinear expansion was first applied to deeply inelastic lepton-nucleon scattering (DIS) and
provides an unique way to obtain a consistent formalism that relates the gauge invariant parton
distribution and/or correlation functions to the measurable quantities such as the
differential cross section
including leading as well as higher twist contributions.
Recently, we have shown that it can be applied to semi-inclusive DIS with nucleon and nucleus targets
for jet production \cite{Liang:2006wp,Liang:2008vz,Gao:2010mj,Song:2010pf,Song:2013sja}
and corresponding expressions for the azimuthal asymmetries and nuclear dependences have been obtained.
Furthermore, we have also successfully applied it to the inclusive hadron production in $e^+e^-$ annihilation
and obtained the corresponding gauge invariant formalism for the hadronic tensor
and the differential cross section.\cite{Wei:2013csa}
This formalism provides a theoretical framework to calculate leading and higher twist contributions systematically.
It is used to obtained the relationship between the differential cross section for $e^++e^-\to h+X$
and different components of the fragmentation function.

There are two key elements in obtaining the gauge invariant formalism for the considered reaction.
The first is to consider systematically the contributions from the Feynman diagram series with multiple gluon scattering
as illustrated in Figs.~\ref{fdmg}(b) and (c) for $e^+e^-$ annihilation into hadrons
with exchange of $j=1,2,\ldots$ gluon(s) between the blob and the lower Fermion line.
In this case, the hadronic tensor is given by a sum of the contribution from each diagram,
e.g., for the semi-inclusive reaction, we have,
\begin{equation}
W_{\mu\nu}^{(si)}(q,p,S,k')=\sum_{j,c}W_{\mu\nu}^{(j,c,si)}(q,p,S,k'),
\end{equation}
where we use the superscript to denote the contribution from the Feynman diagram with
exchange of $j=0,1,2,\ldots$ gluon(s) and $c$ denotes the position of the cut line which takes
$L$ or $R$ for $j=1$, $c=L$, $M$ or $R$ for $j=2$ and corresponds to
Fig.~\ref{fdmg}(b1), (b2), (c1), (c2) and (c3), respectively.

The expression for the contribution from each diagram to the hadronic tensor
can be easily obtained and it takes exactly the same form as that for the inclusive process.
E.g., for $j=1$ or $2$, we have,
\begin{widetext}
\begin{align}
W_{\mu\nu}^{(1,c,si)}(q,p,S,k')=& \int \frac{d^4k_1}{(2\pi)^4} \frac{d^4k_2}{(2\pi)^4} \mathrm{Tr}[ \hat H^{(1,c,si)\rho}_{\mu\nu} (k_1,k_2,k',q)  \hat \Pi_{\rho}^{(1,c)}(k_1,k_2,p,S) ], \label{W1si} \\
W_{\mu\nu}^{(2,c,si)}(q,p,S,k')=& \int \frac{d^4k_1}{(2\pi)^4} \frac{d^4k_2}{(2\pi)^4} \frac{d^4k}{(2\pi)^4}
\mathrm{Tr}[ \hat H^{(2,c,si)\rho\sigma}_{\mu\nu} (k_1,k,k_2,k',q)  
\hat\Pi_{\rho\sigma}^{(2,c)}(k_1,k,k_2,p,S) ], \label{W2si}
\end{align}
where the soft matrices are the same as those defined in the inclusive case and are given by,
\begin{align}
\hat \Pi_{\rho}^{(1,L)}(k_1,k_2,p,S) =  
\frac{1}{2\pi} \sum_X & \int d^4 \xi d^4 \eta e^{-ik_1\xi} e^{-i(k_2-k_1)\eta} \langle 0 | g A_\rho(\eta) \psi(0) |hX \rangle \langle hX | \bar \psi(\xi)|0\rangle,\label{Pi1L}\\
\hat \Pi_{\rho}^{(1,R)}(k_1,k_2,p,S) =  
\frac{1}{2\pi} \sum_X & \int d^4 \xi d^4 \eta e^{-ik_1\xi} e^{-i(k_2-k_1)\eta} \langle 0 |\psi(0) |hX \rangle 
\langle hX | \bar \psi(\xi) g A_\rho(\eta) |0\rangle,\label{Pi1R}\\
%
\hat \Pi_{\rho\sigma}^{(2,L)}(k_1,k,k_2,p,S) =  \frac{1}{2\pi} \sum_X & \int d^4 \xi d^4 \eta_1 d^4 \eta_2 e^{-ik_1\xi} e^{-i(k-k_1)\eta_1} e^{-i(k_2-k)\eta_2}  
\langle 0 | g A_\rho(\eta_1) gA_\sigma (\eta_2) \psi(0) |hX \rangle \langle hX | \bar \psi(\xi)|0\rangle,\label{Pi2L}\\
\hat \Pi_{\rho\sigma}^{(2,M)}(k_1,k,k_2,p,S) =  \frac{1}{2\pi} \sum_X & \int d^4 \xi d^4 \eta_1 d^4 \eta_2 e^{-ik_1\xi} e^{-i(k-k_1)\eta_1} e^{-i(k_2-k)\eta_2} 
\langle 0 | gA_\sigma (\eta_2) \psi(0) |hX \rangle \langle hX | \bar \psi(\xi)g A_\rho(\eta_1) |0\rangle,\label{Pi2M}\\
\hat \Pi_{\rho\sigma}^{(2,R)}(k_1,k,k_2,p,S) =  \frac{1}{2\pi} \sum_X & \int d^4 \xi d^4 \eta_1 d^4 \eta_2 e^{-ik_1\xi} e^{-i(k-k_1)\eta_1} e^{-i(k_2-k)\eta_2} 
\langle 0 |\psi(0) |hX \rangle \langle hX | \bar \psi(\xi) g A_\rho(\eta_1) gA_\sigma (\eta_2) |0\rangle.\label{Pi2R}
\end{align}
The differences lie in the hard parts. For the semi-inclusive  $e^++e^-\to h+\bar q+X$, they are given by,
\begin{align}
&\hat H^{(1,L,si)\rho}_{\mu\nu}(k_1,k_2,k',q) =  \Gamma_\mu^q (\slashed q - \slashed k_1) \gamma^\rho \frac{\slashed k_2- \slashed q}{(k_2-q)^2-i\epsilon}  \Gamma^q_\nu  (2\pi)^4 \delta^4(q-k_1-k'),\\
&\hat H^{(1,R,si)\rho}_{\mu\nu}(k_1,k_2,k',q) =  \Gamma_\mu^q \frac{\slashed k_1 - \slashed q}{(k_1-q)^2+i\epsilon} \gamma^\rho (\slashed q - \slashed k_2) \Gamma^q_\nu  (2\pi)^4 \delta^4(q-k_2-k'),\\
%
 &\hat H_{\mu\nu}^{(2,L,si)\rho\sigma} (k_1,k,k_2,k', q) =  \Gamma_\mu^q (\slashed q - \slashed k_1) \gamma^\rho \frac{\slashed k- \slashed q}{(k-q)^2-i\epsilon} \gamma^\sigma  \frac{\slashed k_2- \slashed q}{(k_2-q)^2-i\epsilon}  \Gamma^q_\nu
 (2\pi)^4 \delta^4(q-k_1-k'),\\
 &\hat H_{\mu\nu}^{(2,M,si)\rho\sigma} (k_1,k,k_2, k',q) =  \Gamma_\mu^q \frac{\slashed k_1- \slashed q}{(k_1-q)^2+i\epsilon} \gamma^\rho (\slashed q - \slashed k) \gamma^\sigma  \frac{\slashed k_2- \slashed q}{(k_2-q)^2-i\epsilon}  \Gamma^q_\nu
 (2\pi)^4 \delta^4(q-k-k'),\\
  &\hat H_{\mu\nu}^{(2,R,si)\rho\sigma} (k_1,k,k_2,k', q) = \Gamma_\mu^q \frac{\slashed k_1- \slashed q}{(k_1-q)^2+i\epsilon} \gamma^\rho \frac{\slashed k- \slashed q}{(k-q)^2+i\epsilon} \gamma^\sigma  (\slashed q - \slashed k_2) \Gamma^q_\nu
 (2\pi)^4 \delta^4(q-k_2-k'),
\end{align}
in contrast to those for the inclusive process $e^++e^-\to h+X$,
\begin{align}
&\hat H^{(1,L)\rho}_{\mu\nu}(k_1,k_2,q) =  \Gamma_\mu^q (\slashed q - \slashed k_1) \gamma^\rho \frac{\slashed k_2- \slashed q}{(k_2-q)^2-i\epsilon}  \Gamma^q_\nu  (2\pi) \delta_+\left((q-k_1)^2\right),\\
&\hat H^{(1,R)\rho}_{\mu\nu}(k_1,k_2,q) =  \Gamma_\mu^q \frac{\slashed k_1 - \slashed q}{(k_1-q)^2+i\epsilon} \gamma^\rho (\slashed q - \slashed k_2) \Gamma^q_\nu  (2\pi) \delta_+ \left((q-k_2)^2\right), \\
%
 &\hat H_{\mu\nu}^{(2,L)\rho\sigma} (k_1,k,k_2, q) =  \Gamma_\mu^q (\slashed q - \slashed k_1) \gamma^\rho \frac{\slashed k- \slashed q}{(k-q)^2-i\epsilon} \gamma^\sigma  \frac{\slashed k_2- \slashed q}{(k_2-q)^2-i\epsilon}  \Gamma^q_\nu  (2\pi) \delta_+\left((q-k_1)^2\right),\\
 &\hat H_{\mu\nu}^{(2,M)\rho\sigma} (k_1,k,k_2, q) =  \Gamma_\mu^q \frac{\slashed k_1- \slashed q}{(k_1-q)^2+i\epsilon} \gamma^\rho (\slashed q - \slashed k) \gamma^\sigma  \frac{\slashed k_2- \slashed q}{(k_2-q)^2-i\epsilon}  \Gamma^q_\nu  (2\pi) \delta_+\left((q-k)^2\right),\\
  &\hat H_{\mu\nu}^{(2,R)\rho\sigma} (k_1,k,k_2, q) = \Gamma_\mu^q \frac{\slashed k_1- \slashed q}{(k_1-q)^2+i\epsilon} \gamma^\rho \frac{\slashed k- \slashed q}{(k-q)^2+i\epsilon} \gamma^\sigma  (\slashed q - \slashed k_2) \Gamma^q_\nu  (2\pi) \delta_+\left((q-k_2)^2\right).
\end{align}
\end{widetext}

We note that, compared with each other,
the inclusive hard part differs from the corresponding semi-inclusive counterpart only in the $\delta$-function.
While the $\delta$-function in the inclusive hard part is one dimensional and represents the mass shell condition,
the $\delta$-function in the hard part for the semi-inclusive process is four dimensional
and represents the energy-momentum conservation.

Since none of the soft matrices given by Eqs.~(\ref{Pi1L}-\ref{Pi2R}) is local (color) gauge invariant,
we need to perform the collinear expansion as proposed in \cite{Ellis:1982wd}
for inclusive deep inelastic lepton nucleon scattering.
Practically, collinear expansion procedure is equivalent to re-organize the contributions
into a sum of terms in the gauge invariant forms by using the Ward identities.
This is the second key element for obtaining the gauge invariant formalism.
We describe it in the following.

\subsubsection{Collinear expansion for $e^++e^-\to h+X$}

The collinear expansion is essentially a Taylor expansion of the hard parts at the collinear positions.
The procedure was summarized as four steps \cite{Liang:2006wp} and was presented in \cite{Wei:2013csa}
for inclusive hadron production $e^++e^-\to h+X$.
For the self-containence of the paper, and also for sake of comparison to the semi-inclusive case,
we briefly repeat them here by paying special attentions to those places where differences may appear when 
extending to the semi-inclusive process.

(1) We make a Taylor expansion of all the hard parts around $k_i = p/z_i $, e.g.,
\begin{align}
&\hat H_{\mu\nu}^{(0)}(k,q) =  \hat H_{\mu\nu}^{(0)} (z) +
  \frac{\partial \hat H_{\mu\nu}^{(0)}(z)}{\partial k_\rho} \omega_\rho^{\ \rho'} k_{\rho'} +\nonumber\\
&\phantom{XXXXXXX}  +\frac{1}{2} \frac{\partial^2 \hat H_{\mu\nu}^{(0)}(z)}{\partial k_\rho \partial k_\sigma} \omega_\rho^{\ \rho'} k_{\rho'} \omega_{\sigma}^{\ \sigma'} k_{\sigma'}+\cdots , \\
&\hat H_{\mu\nu}^{(1,L)\rho}(k_1,k_2,q) =  \hat H_{\mu\nu}^{(1,L)\rho} (z_1,z_2) + \nonumber\\
&\phantom{XXXXXXX} +\frac{\partial \hat H_{\mu\nu}^{(1,L)\rho}(z_1,z_2)}{\partial k_{1\sigma}} \omega_\sigma^{\ \sigma'} k_{1\sigma'} +
\cdots ,
\end{align}
where $z_i$ is defined as $z_i=p^+/k^+_i$.
The momentum of the hadron is taken as $p=p^+\bar n$, i.e., we use the light cone coordinate and
take the direction of motion of the hadron as $z$-direction,
the lepton plane as $xoz$-plane and the transverse component of
the momentum of the incident electron is taken as the $x$-direction,
the unit vectors are denoted by $\bar n$, $n$ and $n_\perp$.
The projection operator $\omega_\rho^{\ \rho'}$ is defined as
$\omega_\rho^{\ \rho'} \equiv g_\rho^{\ \rho'} -\bar n_\rho n^{\rho'}$. 

We would in particular emphasize that, 
the collinear expansion is carried out in a coordinate system 
where momentum of the hadron is taken as the $z$-direction.  
In $e^+e^-$ annihilation, this coordinate system 
is different from that we usually use in experiments 
where jet direction is usually taken as the $Z$-axis. 
We refer to the former as ``collinear frame'' of the hadron and denote it 
by $o$-$xyz$,  while the latter as ``jet frame'' and denote it by $o$-$XYZ$. 
These two frames are related to each other via a rotation. 
We now work in the collinear frame of the hadron but present the relationships  
between the quantities in the two frames at the end of Sec.~IV where results for 
measurable quantities such as the differential cross section,  azimuthal asymmetries 
and polarizations  are presented.

(2) We decompose the gluon fields into longitudinal and transverse components, i.e.,
\begin{equation}
A_\rho(y) = A^+(y) \bar n_\rho + \omega_\rho^{\ \rho'} A_{\rho'}(y).
\end{equation}

(3) We apply the Ward identities such as,
\begin{align}
& \frac{\partial \hat H^{(0)}_{\mu\nu}(z)}{\partial k_\rho}=  - \hat H^{(1,L)\rho}_{\mu\nu} (z,z) - \hat H^{(1,R)\rho}_{\mu\nu}(z,z),\label{ward1}\\
& \frac{\partial \hat H^{(1,L)\rho}_{\mu\nu}(z_1,z_2)}{\partial k_{2,\sigma}}=  - \hat H^{(2,R)\rho\sigma}_{\mu\nu} (z_1,z_2,z_2) ,\label{ward2}\\
&p_\rho \hat H^{(1,L)\rho}_{\mu\nu}(z_1,z_2) = - \frac{z_1 z_2}{z_2-z_1-i\epsilon} H^{(0)}_{\mu\nu} (z_1),\label{ward3} \\
&p_\rho \hat H^{(2,L)\rho\sigma}_{\mu\nu}(z_1,z,z_2) = - \frac{z_1 z}{z-z_1-i\epsilon} H^{(1,L)\sigma}_{\mu\nu} (z_1,z_2). \label{ward4}
\end{align}

(4) We add all the terms with the same hard part together and obtain the hadronic tensor
in the gauge invariant form $W_{\mu\nu} = \sum_{j,c}\tilde{W}^{(j,c)}_{\mu\nu}$, and,
\begin{align}
&\tilde{W}_{\mu\nu}^{(0)}(q,p,S) = \int \frac{d^4k}{(2\pi)^4} \mathrm{Tr}[ \hat H^{(0)}_{\mu\nu}(z) \hat \Xi^{(0)} (k,p,S)],\label{Wtilde0}\\
&\tilde{W}_{\mu\nu}^{(1,c)}(q,p,S)= \int \frac{d^4k_1}{(2\pi)^4 }\frac{d^4k_2}{(2\pi)^4 }  \nonumber\\
&~~~~~~~~~ \times \mathrm{Tr} [ \hat H^{(1,c)\rho}_{\mu\nu}(z_1,z_2) \omega_\rho^{\ \rho'} \hat \Xi_{\rho'}^{(1,c)} (k_1,k_2,p,S)],\label{Wtilde1}\\
&\tilde{W}_{\mu\nu}^{(2,c)} (q,p,S)= \int \frac{d^4k_1}{(2\pi)^4 }\frac{d^4k_2}{(2\pi)^4}\frac{d^4k}{(2\pi)^4}  \nonumber\\
& ~~\times\mathrm{Tr} [\hat H^{(2,c)\rho\sigma}_{\mu\nu}(z_1,z,z_2) \omega_\rho^{\ \rho'}\omega_\sigma^{\ \sigma'} \hat \Xi_{\rho'\sigma'}^{(2,c)} (k_1,k,k_2,p,S)].\label{Wtilde2}
\end{align}
Here, the new un-integrated correlators $\hat\Xi^{(j)}$'s are given by,
\begin{widetext}
\begin{align}
\hat \Xi^{(0)} (k,p,S)=
&\frac{1}{2\pi} \sum_X \int d^4 \xi e^{-ik\xi}
   \langle 0 | \mathcal{L}^\dagger(0,\infty)\psi(0) |hX\rangle \langle hX| \bar\psi(\xi) \mathcal{L}(\xi,\infty) |0\rangle, \label{Xi0} \\
\hat \Xi^{(1,L)}_{\rho} (k_1,k_2,p,S) =
&\frac{1}{2\pi} \sum_X \int d^4 \xi d^4 \eta e^{-ik_1\xi-i(k_2-k_1)\eta}
  \langle 0 | \mathcal{L}^\dagger (\eta,\infty) D_\rho(\eta) \mathcal{L}^\dagger (0,\eta) \psi(0) |hX\rangle
  \langle hX|  \bar\psi(\xi) \mathcal{L}(\xi,\infty) |0\rangle, \label{Xi1L} \\
\hat \Xi^{(1,R)}_{\rho} (k_1,k_2,p,S)  =
&\frac{1}{2\pi} \sum_X \int d^4 \xi d^4 \eta e^{-ik_1\xi-i(k_2-k_1)\eta}
   \langle 0 | \mathcal{L}^\dagger (0,\infty)\psi(0) |hX\rangle
   \langle hX| \bar\psi(\xi) \mathcal{L}(\xi,\eta) {\overleftarrow{D}}_{\rho}(\eta) \mathcal{L}(\eta,\infty) |0\rangle, \label{Xi1R}\\
\hat \Xi^{(2,L)}_{\rho\sigma} (k_1,k,k_2,p,S) =
&\frac{1}{2\pi} \sum_X \int d^4 \xi d^4 \eta_1 d^4 \eta_2  e^{-ik_1\xi-i(k-k_1)\eta_1-i(k_2-k)\eta_2} \nonumber\\
&\times \langle 0 | \mathcal{L}^\dagger (\eta_1,\infty) D_\rho(\eta_1) \mathcal{L}^\dagger (\eta_2,\eta_1) D_\sigma(\eta_2) \mathcal{L}^\dagger (0,\eta_2) \psi(0) |hX\rangle
  \langle hX| \bar\psi(\xi) \mathcal{L}(\xi,\infty) |0\rangle, \label{Xi2L}  \\
\hat \Xi^{(2,M)}_{\rho\sigma}  (k_1,k,k_2,p,S) =
&\frac{1}{2\pi} \sum_X \int d^4 \xi d^4 \eta_1 d^4 \eta_2 e^{-ik_1\xi-i(k-k_1)\eta_1-i(k_2-k)\eta_2} \nonumber\\
&\times \langle 0 | \mathcal{L}^\dagger (\eta_2,\infty) D_\sigma(\eta_2) \mathcal{L}^\dagger (0,\eta_2) \psi(0) |hX\rangle
  \langle hX| \bar\psi(\xi) \mathcal{L}(\xi,\eta_1) D_\rho(\eta_1)\mathcal{L}(\eta_1,\infty)  |0\rangle, \label{Xi2M} \\
\hat \Xi^{(2,R)}_{\rho\sigma} (k_1,k,k_2,p,S) =
& \frac{1}{2\pi} \sum_X \int d^4 \xi d^4 \eta_1 d^4 \eta_2 e^{-ik_1\xi-i(k-k_1)\eta_1-i(k_2-k)\eta_2} \nonumber\\
&\times \langle 0 | \mathcal{L}^\dagger (0,\infty) \psi(0) |hX\rangle
  \langle hX| \bar\psi(\xi) \mathcal{L}(\xi,\eta_1) D_\rho(\eta_1)\mathcal{L}(\eta_1,\eta_2) D_\sigma(\eta_2)  \mathcal{L} (\eta_2,\infty) |0\rangle, \label{Xi2R}
\end{align}
where $D_\rho (\eta)= -i\partial_\rho + gA_{\rho}(\eta)$ is the covariant derivative.
The gauge link $\mathcal{L}(\xi,\eta) = \mathcal{L}(\xi,\infty) \mathcal{L}^\dagger(\eta,\infty)$
and $\mathcal{L}(\xi,\infty)$ is given by the following path integral,
\begin{align}
\mathcal{L}(\xi,\infty) &=\mathcal{P}e^{ig\int_{\xi^-}^\infty d\eta^-A^+(\eta_-;\xi^+,\vec{\xi}_\perp)} \nonumber\\
&= 1 + ig\int_{\xi^-}^\infty d \eta^- A^+(\eta^-;\xi^+,\vec{\xi}_\perp)) + (ig)^2 \int_{\xi^-}^\infty d \eta_1^- \int_{\xi^-}^{\eta_1^-} d \eta_2^- A^+(\eta_2^-;\xi^+,\vec{\xi}_\perp))A^+(\eta_1^-;\xi^+,\vec{\xi}_\perp)) + \cdots.
\end{align}
\end{widetext}
In contrast to \cite{Wei:2013csa}, we use the un-integrated correlators here for comparison to the results in
the semi-inclusive case.

We note that these un-integrated correlators have the following properties as demanded by the hermiticity,
\begin{align}
 &\hat \Xi^{(0)\dag} (k,p,S) = \gamma^0 \hat \Xi^{(0)} (k,p,S) \gamma^0, \label{hermiticity0}\\
 &\hat \Xi^{(1,L)\dag}_\rho (k_1,k_2,p,S) = \gamma^0 \hat \Xi^{(1,R)}_\rho (k_1,k_2,p,S) \gamma^0, \label{hermiticity1}\\
 &\hat \Xi^{(2,L)\dag}_{\rho\sigma} (k_1,k_2,k,p,S) = \gamma^0 \hat \Xi^{(2,R)}_{\sigma\rho} (k_1,k_2,k,p,S) \gamma^0, \label{hermiticity2L}\\
 &\hat \Xi^{(2,M)\dag}_{\rho\sigma} (k_1,k_2,k,p,S) = \gamma^0 \hat \Xi^{(2,M)}_{\sigma\rho} (k_1,k_2,k,p,S) \gamma^0, \label{hermiticity2M}
 \end{align}
and space reflection invariance,
\begin{align}
 & \gamma^0 \hat \Xi^{(0)} (\tilde k,\tilde p,-\tilde S) \gamma^0= \hat \Xi^{(0)} (k,p,S), \label{space0}\\
 & \gamma^0\hat \Xi^{(1,L)}_\rho (\tilde k_1,\tilde k_2,\tilde p,-\tilde S)  = \hat \Xi^{(1,L)}_\rho (k_1,k_2,p,S) , \label{space1}\\
 &\gamma^0 \hat \Xi^{(2,L)\dag}_{\rho\sigma} (\tilde k_1,\tilde k_2,\tilde k,\tilde p,-\tilde S) \gamma^0  =  \hat \Xi^{(2,L)}_{\rho\sigma} (k_1,k_2,k,p,S) , \label{space2}\\
 &\gamma^0 \hat \Xi^{(2,M)\dag}_{\rho\sigma} (\tilde k_1,\tilde k_2,\tilde k,\tilde p,-\tilde S) \gamma^0  =  \hat \Xi^{(2,M)}_{\rho\sigma} (k_1,k_2,k,p,S). \label{space3}
 \end{align}
Here a tilded four vector denotes $\tilde k=(k_0,-\vec k)$ and $\tilde S=(S_0,-\vec S)$.
As mentioned in e.g. \cite{Collins:1992kk, Jaffe:1993xb, Ji:1993vw}, 
the time reversal invariance does not lead to direct constraints to the form of these correlators.

\subsubsection{Collinear expansion for the semi-inclusive process $e^++e^-\to h+\bar q+X$}

As can be seen from the four steps of the collinear expansion for the inclusive process $e^++e^-\to h+X$,
a crucial ingredient is the validity of the Ward identities for the hard parts as given by Eqs. (\ref{ward1}-\ref{ward4}).
By using these identities, we replace the derivatives by the corresponding hard parts
with more gluon(s) exchange.
Also,  the longitudinal gluon field parts are changed to the corresponding terms in the gauge link
for contribution where the hard parts contain less gluon exchange.
We also see that, for the semi-inclusive process $e^++e^-\to h+\bar q+X$,
the hadronic tensor is expressed in the same form as that for the inclusive process.
However, the hard parts in the sum of the contributions from the separate graphs
are different from those in the inclusive process.
Because of this, there is no similar Ward identities valid for the semi-inclusive hard parts.
This implies that, if one would perform the collinear expansion directly of the semi-inclusive hard parts,
one would {\it not} be able to obtain the similar results as those for the inclusive case.

This problem was solved in \cite{Liang:2006wp} by using the identity,
\begin{align}
&(2\pi)^4 \delta^4 (q-k-k') =2\pi \delta_+\left((q-k)^2\right) K(k,k'),\\
&K(k,k')=(2\pi)^3 2E'\delta^3 (\vec q-\vec k-\vec k').
\end{align}
Using this identity, we obtain that
\begin{equation}
\hat H_{\mu\nu}^{(j,c,si)}(k_i,k',q)= \hat H_{\mu\nu}^{(j.c)}(k_i,q) K(k_c,k'),
\end{equation}
where $k_i$ denotes all the parton 4-momenta involved and $k_c$ is the momentum of the cut parton.
The semi-inclusive hard part is written as a product of the corresponding inclusive hard part and a 
kinematical factor $K(k_c,k')$. 
Hence,
\begin{align}
W_{\mu\nu}^{(0,si)}=& \int \frac{d^4k}{(2\pi)^4}   K(k,k') \mathrm{Tr}[ \hat H^{(0)}_{\mu\nu} (k,q) \hat \Pi^{(0)}(k,p,S) ], \label{Wsi0rew} \\
W_{\mu\nu}^{(1,c,si)}=& \int \frac{d^4k_1}{(2\pi)^4} \frac{d^4k_2}{(2\pi)^4} K(k_c,k') \nonumber\\
&\times \mathrm{Tr}[ \hat H^{(1,c)\rho}_{\mu\nu} (k_1,k_2,q)  \hat \Pi_{\rho}^{(1,c)}(k_1,k_2,p,S) ], \label{Wsi1rew} \\
W_{\mu\nu}^{(2,c,si)}=& \int \frac{d^4k_1}{(2\pi)^4} \frac{d^4k_2}{(2\pi)^4} \frac{d^4k}{(2\pi)^4} K(k_c,k')\nonumber\\
&\times\mathrm{Tr}[ \hat H^{(2,c)\rho\sigma}_{\mu\nu} (k_1,k,k_2,q)  \hat \Pi_{\rho\sigma}^{(2,c)}(k_1,k,k_2,p,S) ]. \label{Wsi2rew}
\end{align}

We apply the same collinear expansion as that for the inclusive case,
i.e. make collinear expansion of the inclusive hard part
rather than the semi-inclusive hard part.
The kinematic factor $K(k_c,k')$ remains unchanged, so we obtain, 
\begin{align}
W_{\mu\nu}^{(si)}& (q,p,S,k')= \sum_{j, c}\tilde{W}^{(j,c)si}_{\mu\nu}(q,p,S,k'), \label{Wsitotal}\\
\tilde{W}_{\mu\nu}^{(0,si)} &= \int \frac{d^4k}{(2\pi)^4}  K(k,k') \mathrm{Tr}\Big[\hat H^{(0)}_{\mu\nu}(z) \hat\Xi^{(0)} (k,p,S)\Big],\label{Wsitilde0}\\
\tilde{W}_{\mu\nu}^{(1,c,si)} &= \int \frac{d^4k_1}{(2\pi)^4 }\frac{d^4k_2}{(2\pi)^4 } K(k_c,k') 
\mathrm{Tr} \Big[ \hat H^{(1,c)\rho}_{\mu\nu}(z_1,z_2) \nonumber\\
&\times\omega_\rho^{\ \rho'} \hat \Xi_{\rho'}^{(1,c)} (k_1,k_2,p,S) \Big],\label{Wsitilde1L}\\
\tilde{W}_{\mu\nu}^{(2,c,si)}&= \int \frac{d^4k_1}{(2\pi)^4 }\frac{d^4k_2}{(2\pi)^4}\frac{d^4k}{(2\pi)^4} K(k_c,k') 
\mathrm{Tr} \Big[ \hat H^{(2,c)\rho\sigma}_{\mu\nu}(z_1,z,z_2) \nonumber\\
&\times \omega_\rho^{\ \rho'}\omega_\sigma^{\ \sigma'} \hat \Xi_{\rho'\sigma'}^{(2,c)} (k_1,k,k_2,p,S) \Big]. \label{Wsitilde2L}
\end{align}
We see that, except for the extra kinematical factor $K(k_c,k')$, they are just the same as those for
the inclusive process $e^++e^-\to h+X$ as given by Eqs. (\ref{Wtilde0}-\ref{Wtilde2}).
These equations form a framework that can be used to calculate the leading and higher twist contributions to
the hadronic tensor systematically.
The nice feature of this framework is that the hard parts depends only on the longitudinal components of the parton momenta,
so that we can further simplify these equation in a great deal.
We show this in the following.

\subsection{The simplified expressions}

Since these collinear expanded hard parts depend on less parton momenta, and the momentum dependences
are usually $\delta$-functions or $1/(z-z_B)$ form, where $z_B\equiv 2p\cdot q /Q^2 $, 
we can carry out the integrations over the parton momenta.
We simply use $W_{\mu\nu} ^{(si)}(q,p,S,k'_\perp)$ to denote
$W_{\mu\nu} ^{(si)}(q,p,S,k')$ after integration over $dk'_z/{(2\pi)2E'}$, i.e.,
\begin{align}
W_{\mu\nu} ^{(si)}(q,p,S,k'_\perp)=\int \frac{dk'_z}{(2\pi)2E'} W_{\mu\nu} ^{(si)}(q,p,S,k'), \label{Wperp}
\end{align}
and we have,
\begin{align}
d\sigma^{(si)} &=  \frac{g^4_z}{32s} L_{\mu'\nu'} (l_1,l_2) D^{\mu'\mu} (q) \nonumber\\
&\times D^{\nu'\nu*}(q) W_{\mu\nu}^{(si)} (q,p,S,k'_\perp) \frac{d^3p}{(2\pi)^2 2E_p} \frac{d^2k'_\perp}{(2\pi)^2}.
\end{align}

By inserting Eqs. (\ref{Wsitotal}-\ref{Wsitilde2L}) into (\ref{Wperp}), 
we obtain the following simplified form for the hadronic tensor,
\begin{align}
&W_{\mu\nu}^{(si)} (q,p,S,k'_\perp)= \sum_{j,c}\tilde{W}^{(j,c)si}_{\mu\nu}(q,p,S,k'_\perp), \\
&\tilde{W}_{\mu\nu}^{(0,si)} (q,p,S,k'_\perp) = \int \frac{d^4 k}{(2\pi)^4} (2\pi)^2 \delta^2 ( \vec{k}_\perp + \vec{k}'_\perp) \nonumber\\
&\phantom{XXXX}\times \mathrm{Tr}[\hat H^{(0)}_{\mu\nu}(z) \hat\Xi^{(0)} (k,p,S)],\\
&\tilde{W}_{\mu\nu}^{(1,c,si)} (q,p,S,k'_\perp) = \int \frac{d^4k_1}{(2\pi)^4 }\frac{d^4k_2}{(2\pi)^4 } (2\pi)^2 \delta^2 ( \vec{k}_{c\perp} + \vec{k}'_\perp) \nonumber\\
&\phantom{XXX}\times \mathrm{Tr} \left[ \hat H^{(1,c)\rho}_{\mu\nu}(z_1,z_2) \omega_\rho^{\ \rho'} \hat \Xi_{\rho'}^{(1,c)} (k_1,k_2,p,S) \right],\\
&\tilde{W}_{\mu\nu}^{(2,c,si)} (q,p,S,k'_\perp) = \int \frac{d^4k_1}{(2\pi)^4 }\frac{d^4k_2}{(2\pi)^4}\frac{d^4k}{(2\pi)^4}  (2\pi)^2 \delta^2 ( \vec{k}_{c\perp} + \vec{k}'_\perp) \nonumber\\
&\phantom{}\times \mathrm{Tr} \left[ \hat H^{(2,c)\rho\sigma}_{\mu\nu}(z_1,z,z_2) \omega_\rho^{\ \rho'}\omega_\sigma^{\ \sigma'} \hat\Xi_{\rho'\sigma'}^{(2,c)} (k_1,k,k_2,p,S)\right].
\end{align}

The collinear expanded hard parts contain only the longitudinal variables and also multiplied by the
projection operator $\omega_\rho^{\ \rho'}$.
They reduce to very simple forms such as,
\begin{align}
&\hat H_{\mu\nu}^{(0)} (z) = z_B^2\pi \hat h^{(0)}_{\mu\nu} \delta (z-z_B), \\
&\hat H_{\mu\nu}^{(1,L)\rho} (z_1,z_2) \omega_\rho^{\ \rho'}
=-\frac{\pi z_B^2}{2p\cdot q} \hat h_{\mu\nu}^{(1)\rho} \delta (z_1-z_B) \omega_\rho^{\ \rho'}, \\
&\hat H_{\mu\nu}^{(2,M)\rho\sigma} (z_1,z,z_2) \omega_\rho^{\ \rho'} \omega_\sigma^{\ \sigma'}
= \frac{2\pi z_B^2}{(2 p\cdot q)^2} \hat h_{\mu\nu}^{(2)\rho\sigma} \delta (z-z_B) \omega_\rho^{\ \rho'} \omega_\sigma^{\ \sigma'} ,\\
&\hat H_{\mu\nu}^{(2,L)\rho\sigma} (z_1,z,z_2) \omega_\rho^{\ \rho'} \omega_\sigma^{\ \sigma'}
= \frac{2\pi z_B^2}{(2p \cdot q)^2}  \nonumber\\
&\phantom{XXX}
\times\Bigl(p^\sigma \hat h^{(1)\rho}_{\mu\nu} - \frac{z_2 z_B \hat N^{(2)\rho\sigma}_{\mu\nu}}{z_2-z_B - i\epsilon} \Bigr)
\delta (z_1-z_B) \omega_\rho^{\ \rho'} \omega_\sigma^{\ \sigma'} ,
\end{align}
where 
$\hat h^{(0)}_{\mu\nu} = \Gamma_\mu^q \slashed n \Gamma_\nu^q/p^+ $,
$\hat h_{\mu\nu}^{(1)\rho} = \Gamma_\mu^q \slashed n \gamma^\rho \slashed{\bar n}\Gamma_\nu^q$,
$\hat N^{(2)\rho\sigma}_{\mu\nu} = q^- \Gamma_\mu \gamma^\rho \slashed n \gamma^\sigma \Gamma_\nu$,
and $\hat h_{\mu\nu}^{(2)\rho\sigma}=p^+ \Gamma_\mu \slashed{\bar n} \gamma^\rho \slashed n \gamma^\sigma \slashed{\bar n} \Gamma_\nu/2$.

Hence these hadronic tensors can be further simplified as that we did for inclusive process.
\begin{align}
\tilde W_{\mu\nu}^{(0, si)} 
&= \frac{1}{2} {\rm Tr}\left[
	\hat h_{\mu\nu}^{(0)} \hat \Xi^{(0)} (z_B, k'_{\perp},p,S) \right], \label{Wtildesim0} \\
\tilde W^{(1,L,si)}_{\mu\nu}
&= -\frac{1}{4p\cdot q} {\rm Tr}
	\left[ \hat h^{(1)\rho}_{\mu\nu} \omega_{\rho}^{\ \rho'} \hat \Xi^{(1,L)}_{\rho'} (z_B, k'_{\perp},p,S) \right] , \label{Wtildesim1l} \\
\tilde W^{(1,R,si)}_{\mu\nu}
&= -\frac{1}{4p\cdot q} {\rm Tr}
	\left[ \hat h^{(1)\rho\dagger}_{\nu\mu} \omega_{\rho}^{\ \rho'} 
	\hat \Xi^{(1,R)\dagger}_{\rho'} (z_B, k'_{\perp},p,S) \right], \label{Wtildesim1r} \\
\tilde W^{(2,M,si)}_{\mu\nu} 
&= \frac{1}{4(p\cdot q)^2} {\rm Tr} \left[ \hat h^{(2)\rho\sigma}_{\mu\nu} 
\omega_\rho^{\ \rho'} \omega_\sigma^{\ \sigma'} \hat \Xi_{\rho'\sigma'}^{(2A)} (z_B, k'_\perp,p,S) \right], \label{Wtildesim2m}\\
\tilde W^{(2,L,si)}_{\mu\nu} 
&= \frac{1}{4(p\cdot q)^2} {\rm Tr} \Bigl[ \hat h^{(1)\rho}_{\mu\nu} \omega_\rho^{\ \rho'}
\hat \Xi_{\rho'}^{(2B)} (z_B, k'_\perp,p,S) \nonumber\\
&+ \hat N_{\mu\nu}^{(2)\rho\sigma} \omega_\rho^{\ \rho'} \omega_\sigma^{\ \sigma'}
\hat \Xi_{\rho'\sigma'}^{(2C)} (z_B, k'_\perp,p,S) \Bigr], \label{Wtildesim2l}\\
\tilde W^{(2,R,si)}_{\mu\nu} 
& = \frac{1}{4(p\cdot q)^2} {\rm Tr} \Bigl[ \hat h^{(1)\rho\dagger}_{\nu\mu} \omega_\rho^{\ \rho'}
\hat \Xi_{\rho'}^{(2B)\dagger} (z_B, k'_\perp,p,S) \nonumber\\
&+ \hat N_{\nu\mu}^{(2)\rho\sigma\dagger} \omega_\rho^{\ \rho'} \omega_\sigma^{\ \sigma'}
\hat \Xi_{\rho'\sigma'}^{(2C)\dagger} (z_B, k'_\perp,p,S) \Bigr], \label{Wtildesim2r}
\end{align}
where just for clarity, we omit the arguments $(q,p,S,k'_{\perp})$ for the $\tilde W$'s.
The new correlators are defined as,
\begin{align}
\hat \Xi^{(0)} & (z, k'_{\perp},p,S) =  \int \frac{d^4k}{(2\pi)^4}  \delta^2(\vec k_\perp+\vec k'_\perp) \nonumber\\
 &\times  2\pi  z^2_B \delta(z-z_B) \hat\Xi^{(0)} (k,p,S),\label{Xi0perp} \\
\hat \Xi^{(1)}_{\rho} &(z,k'_{\perp},p,S) = \int \frac{d^4k_1}{(2\pi)^4}\frac{d^4k_2}{(2\pi)^4}  \delta^2(\vec k_{1\perp}+\vec k'_\perp) \nonumber\\
 &\times  2\pi  z_B^2 \delta(z_1-z_B)  \hat\Xi^{(1,L)}_{\rho}(k_1,k_2,p,S),  \label{Xi1perp} \\
\hat \Xi^{(2A)}_{\rho\sigma} & (z,k'_\perp,p,S)  = \int  \frac{d^4k_1}{(2\pi)^4}\frac{d^4k_2}{(2\pi)^4}\frac{d^4k}{(2\pi)^4} \delta^2(\vec k_\perp+\vec k'_\perp) \nonumber\\
 &\times  2\pi  z_B^2 \delta(z-z_B)  \hat\Xi^{(2,M)}_{\rho\sigma}  (k_1,k_2,k,p,S),  \label{Xi2perp} \\
\hat \Xi^{(2B)}_{\rho} & (z,k'_{\perp},p,S) =\int \frac{d^4k_1}{(2\pi)^4}\frac{d^4k_2}{(2\pi)^4}\frac{d^4k}{(2\pi)^4} \delta^2(\vec k_{1\perp}+\vec k'_\perp) \nonumber\\
 &\times  2\pi  z_B^2 \delta(z_1-z_B)  p^\sigma\hat\Xi^{(2,L)}_{\rho\sigma}  (k_1,k_2,k,p,S), \\
\hat \Xi^{(2C)}_{\rho\sigma} & (z,k'_{\perp},p,S) = \int \frac{d^4k_1}{(2\pi)^4}\frac{d^4k_2}{(2\pi)^4}\frac{d^4k}{(2\pi)^4}\delta^2(\vec k_{1\perp}+\vec k'_\perp) \nonumber\\
 &\times  2\pi  z_B^2 \delta(z_1-z_B) \frac{zz_2}{z_2-z-i\epsilon}\hat\Xi^{(2,L)}_{\rho\sigma}  (k_1,k_2,k,p,S). \label{t4sp}
\end{align}
After we perform the integration, we obtain their field operator expressions as given by,
\begin{widetext}
\begin{align}
&\hat \Xi^{(0)} (z, k_{\perp},p,S) = \sum_X \int \frac{p^+ d\xi^- d^2 \xi_\perp}{2\pi}
	e^{-ip^+\xi^-/z+ik_{\perp} \cdot \xi_\perp}
    \langle 0| \mathcal{L}^\dagger (0,\infty) \psi(0)|hX\rangle \langle hX| \bar\psi(\xi) \mathcal{L} (\xi,\infty) |0 \rangle,\label{Xi0ope} \\
&\hat \Xi^{(1)}_{\rho} (z,k_{\perp},p,S)  = \sum_X \int \frac{p^+ d\xi^- d^2\xi_\perp}{2\pi}
	e^{-ip^+\xi^-/z+ik_{\perp} \cdot \xi_\perp}
\langle 0| \mathcal{L}^\dagger (0,\infty) D_\rho (0) \psi(0)|hX\rangle\langle hX| \bar\psi(\xi) \mathcal{L} (\xi,\infty) |0 \rangle,  \label{Xi1lope} \\
& \hat \Xi^{(2A)}_{\rho\sigma} (z,k_\perp,p,S)=
\sum_X \int  \frac{p^+d\xi^- d^2 \xi_\perp}{2\pi} e^{-ip^+\xi^-/z+ik_{\perp} \cdot \xi_\perp}
\langle 0 | \mathcal{L}^\dagger (0,\infty) D_\rho(0) \psi(0) |hX\rangle
\langle hX| \bar\psi(\xi) \overleftarrow{D}_\sigma (\xi) \mathcal{L}(\xi,\infty) |0\rangle, \\
 & \hat \Xi^{(2B)}_{\rho} (z,k_\perp,p,S)=
\sum_X \int  \frac{p^+d\xi^- d^2 \xi_\perp}{2\pi} e^{-ip^+\xi^-/z+ik_{\perp} \cdot \xi_\perp}
p^\sigma\langle 0 | \mathcal{L}^\dagger (0,\infty) D_\rho(0)  D_\sigma (0) \psi(0) |hX\rangle
\langle hX|\bar\psi(\xi) \mathcal{L}(\xi,\infty) |0\rangle,\\
 & \hat \Xi^{(2C)}_{\rho\sigma} (z,k_\perp,p,S)=
\sum_X\int  \frac{p^+d\xi^- d^2\xi_\perp p^+d\eta^-}{2\pi} \frac{dz_2}{2\pi} \frac{1}{z_2^2} \frac{z_2 z}{z_2-z - i\epsilon}
e^{-ip^+(\xi^--\eta^-)/z-ip^+\eta^-/z_2+ik_{\perp} \cdot \xi_\perp} \nonumber\\
&\phantom{XXXXXXXXXXXX}\times \langle 0 | \mathcal{L}^\dagger (\eta^-,\infty) D_\rho(\eta^-)
 D_\sigma (\eta^-) \mathcal{L}^\dagger (0^-,\eta^-)  \psi(0) |hX\rangle
 \langle hX| \bar\psi(\xi) \mathcal{L}(\xi,\infty) |0\rangle, \label{t4spope}
\end{align}
\end{widetext}
where the coordinate $\xi$ of a field operator or in the gauge link is understood as $\xi=(0,\xi^-,\vec \xi_\perp)$,
while a $\eta^-$ in such a place means $\eta=(0,\eta^-,\vec 0_\perp)$.

We emphasize here, just similar to  the inclusive and semi-inclusive deep inelastic lepton-nucleon scattering 
and inclusive $e^+e^-$ annihilation processes~\cite{Liang:2006wp,Song:2013sja, Wei:2013csa}, 
that the results presented above including the expressions for the gauge links and 
the different correlators are all derived by using the collinear expansion. 
We compare these results with those obtained directly from the Feynman diagrams,
we see the distinct differences not only in the hard parts but also in the correlators. 
We see in particular that all these correlators depend only on one parton momentum. 
Furthermore, in contrast to e.g. $\hat \Pi^{(1,L)}_{\rho} (k_1,k_2,p,S)$ or $\hat \Pi^{(1,R)}_{\rho} (k_1,k_2,p,S)$,  
in the expression of $\hat \Xi^{(1)}_{\rho} (z,k_{\perp},p,S)$, we have the covariant derivative instead of the gluon field. 

These correlators are all $4\times 4$ matrices,
and can be decomposed in terms of gamma matrices such as,
\begin{align}
&\hat \Xi^{(0)}  = \Xi^{(0)}_\alpha  \gamma^\alpha + \tilde\Xi^{(0)}_\alpha  \gamma_5 \gamma^\alpha + \cdots, \label{Xi0exp} \\
&\hat \Xi^{(1)}_\rho = \Xi^{(1)}_{\rho\alpha} \gamma^\alpha + \tilde\Xi^{(1)}_{\rho\alpha} \gamma_5 \gamma^\alpha + \cdots .\label{Xi1exp}
\end{align}
Because $\hat h^{(0)}_{\mu\nu}$, $\hat h_{\mu\nu}^{(1)\rho}$ etc. all have odd
number of $\gamma-$matrices, only $\gamma^\alpha$ and $\gamma_5 \gamma^\alpha$ terms
contribute to the hadronic tensors.
They are given by,
\begin{align}
\tilde W^{(0,si)}_{\mu\nu} 
=&\frac{1}{2} \left[ h_{\mu\nu}^{(0)\alpha} \Xi^{(0)}_\alpha (z,k'_{\perp},p,S)
+\tilde h_{\mu\nu}^{(0)\alpha} \tilde\Xi^{(0)}_\alpha (z,k'_{\perp},p,S)\right], \label{W0siexp}\\
\tilde W^{(1,L,si)}_{\mu\nu} 
=& -\frac{1}{4p\cdot q}\Big[ h^{(1)\rho\alpha}_{\mu\nu}\omega_{\rho}^{\ \rho'} \Xi^{(1)}_{\rho'\alpha} (z,k'_{\perp},p,S)  + \nonumber\\
&+\tilde h^{(1)\rho\alpha}_{\mu\nu}\omega_{\rho}^{\ \rho'}\tilde\Xi^{(1)}_{\rho'\alpha} (z,k'_{\perp},p,S) \Big],\label{W1Lsiexp}\\
\tilde W^{(2,M,si)}_{\mu\nu} 
=& \frac{1}{4(p\cdot q)^2}\Big[ h^{(2)\rho\sigma\alpha}_{\mu\nu}\omega_{\rho}^{\ \rho'} \omega_{\sigma}^{\ \sigma'}
\Xi^{(2A)}_{\rho'\sigma'\alpha} (z,k'_{\perp},p,S) + \nonumber\\
& + \tilde h^{(2)\rho\sigma\alpha}_{\mu\nu}\omega_{\rho}^{\ \rho'} \omega_{\sigma}^{\ \sigma'}
\tilde\Xi^{(2A)}_{\rho'\sigma'\alpha} (z,k'_{\perp},p,S) \Big],\label{W2Msiexp}\\
\tilde W^{(2,L,si)}_{\mu\nu} 
= & \frac{1}{4(p\cdot q)^2}\Big[ h^{(1)\rho\alpha}_{\mu\nu}\omega_{\rho}^{\ \rho'} \Xi^{(2B)}_{\rho'\alpha} (z,k'_{\perp},p,S)  + \nonumber\\
&+\tilde h^{(1)\rho\alpha}_{\mu\nu}\omega_{\rho}^{\ \rho'}\tilde\Xi^{(2B)}_{\rho'\alpha} (z,k'_{\perp},p,S)+\nonumber\\
& + N^{(2)\rho\sigma\alpha}_{\mu\nu}\omega_{\rho}^{\ \rho'} \omega_{\sigma}^{\ \sigma'}
\Xi^{(2C)}_{\rho'\sigma'\alpha} (z,k'_{\perp},p,S) + \nonumber\\
& + \tilde N^{(2)\rho\sigma\alpha}_{\mu\nu}\omega_{\rho}^{\ \rho'} \omega_{\sigma}^{\ \sigma'}
\tilde\Xi^{(2C)}_{\rho'\sigma'\alpha} (z,k'_{\perp},p,S)   \Big],\label{W2Lsiexp}
\end{align}
where we use the short handed notations
$h^{(j)\alpha}_{\mu\nu}=\mathrm{Tr}[\gamma^\alpha\hat h^{(j)}_{\mu\nu}]$,
$\tilde h^{(j)\alpha}_{\mu\nu}=\mathrm{Tr}[\gamma_5\gamma^\alpha\hat h^{(j)}_{\mu\nu}]$ and so on.
The other contributions are given by,
\begin{align}
&\tilde W^{(1,R,si)}_{\mu\nu}(q,p,S,k'_\perp)=\tilde W^{(1,L,si)*}_{\nu\mu}(q,p,S,k'_\perp),\\
&\tilde W^{(2,R,si)}_{\mu\nu}(q,p,S,k'_\perp)=\tilde W^{(2,L,si)*}_{\nu\mu}(q,p,S,k'_\perp).
\end{align}

By carrying out the traces to obtain $h^{(j)\alpha}_{\mu\nu}$ and $\tilde h^{(j)\alpha}_{\mu\nu}$,
making the Lorentz decompositions of the correlators and making the Lorentz contractions,
we obtain the contributions to the hadronic tensors.
To the leading twist (twist-2), we only need to consider $\tilde W^{(0,si)}$.
Up to twist-3, we need to consider both $\tilde W^{(0,si)}$ and $\tilde W^{(1,c,si)}$.
In this way, we can calculate the leading and higher twist contributions systematically.

The traces mentioned above can easily be carried out, we have e.g.,
\begin{align}
&p^+ h^{(0)\alpha}_{\mu\nu}= -4 c_1^q (g_{\mu\nu}n^\alpha - g_{\mu}^{\ \alpha} n_\nu - g_{\nu}^{\ \alpha} n_\mu) - 4i c_3^q \epsilon^{\alpha \beta}_{\ \ \mu\nu} n_\beta , \label{h0res}\\
&p^+ \tilde h^{(0)\alpha}_{\mu\nu}=4c_3^q (g_{\mu\nu} n^\alpha - g_{\mu}^{\ \alpha} n_\nu - g_{\nu}^{\ \alpha} n_\mu ) + 4i c_1^q \epsilon^{\alpha \beta}_{\ \ \mu\nu} n_\beta,\label{h0tres}\\
&h^{(1)\rho\alpha}_{\mu\nu} \bar n_\alpha \omega_\rho^{\ \rho'} = -8 c_1^q d_{\mu}^{\ \rho'}  \bar n_\nu - 8ic_3^q \epsilon_{\perp\mu}^{ \ \ \ \rho'} \bar n_\nu, \label{h1res}\\
&\tilde h^{(1)\rho\alpha}_{\mu\nu} \bar n_\alpha \omega_\rho^{\ \rho'} = 8c_3^q d_{\mu}^{\ \rho'} \bar n_\nu + 8 i c_1^q \epsilon_{\perp\mu}^{ \ \ \ \rho'} \bar n_\nu , \label{h1tres}
\end{align}
where $d_{\mu}^{\ \rho'}=g_\mu^{\ \rho'}-\bar n_\mu n^{\rho'} - n_\mu \bar n^{\rho'}$.

We should note that,
the correlators such as $\Xi^{(0)}_\alpha$, $\tilde\Xi^{(0)}_\alpha$, $\Xi^{(1)}_{\rho\alpha}$, $\tilde\Xi^{(1)}_{\rho\alpha}$
are Lorentz vectors and tensors respectively.
Also, from the hermiticity and the parity invariance of strong interaction as given by Eqs. (\ref{hermiticity0}-\ref{space2}),
we obtain further constraints on the forms of these correlators. 
From the hermiticity, we obtain that both
$\Xi^{(0)}_\alpha$ and $\tilde\Xi^{(0)}_\alpha$ are real, i.e.,
$\Xi^{(0)*}_\alpha = \Xi^{(0)}_\alpha$ and $\tilde\Xi^{(0)*}_\alpha = \tilde\Xi^{(0)}_\alpha$.
But for $\Xi^{(1)}$'s, hermiticity does not lead to such simple results because the two matrix elements
are not symmetric. 
They have both real and imaginary parts.
From space reflection invariance, we obtain, 
\begin{align}
& \Xi^{(0)}_\alpha (z,k_\perp,p,S) = \Xi^{(0)\alpha} (z, \tilde k_\perp, \tilde p,- \tilde S), \\
& \tilde \Xi^{(0)}_\alpha (z,k_\perp,p,S) = -\tilde \Xi^{(0)\alpha} (z, \tilde k_\perp, \tilde p,- \tilde S), \\
& \Xi^{(1)}_{\rho\alpha} (z,k_\perp,p,S) = \Xi^{(1)}_{\rho\alpha} (z, \tilde k_\perp, \tilde p, -\tilde S), \\
& \tilde \Xi^{(1)}_{\rho\alpha} (z,k_\perp,p,S) = -\tilde \Xi^{(1)}_{\rho\alpha} (z, \tilde k_\perp, \tilde p, -\tilde S).
\end{align}
We see in particular that $\Xi^{(0)}_\alpha$ is a vector and $\tilde\Xi^{(0)}_\alpha$ is an axial vector. 
These are properties that we should use when we make the Lorentz decompositions of these correlators.
The precise forms of the Lorentz decompositions of the correlators depend strongly
on the spin of the hadron involved.
We therefore consider hadrons with spin 0, $1/2$ and $1$ separately 
and present the results in these cases up to twist-3 in next section.

\section{Hadronic tensor for $e^++e^-\to h+\bar q+X$}

In this section, we present the results for the hadronic tensors for
$e^++e^-\to h+\bar q+X$ in the cases that $h$ is a spin-0, spin-$1/2$ or spin-1 particle separately.

\subsection{Hadronic tensor for spin-0 hadrons}

For semi-inclusive process with production of spin-0 hadrons,
the correlators $\Xi^{(0)}_\alpha$, $\tilde \Xi^{(0)}_\alpha$,
$\Xi^{(1)}_{\rho \alpha}$ and $\tilde \Xi^{(1)}_{\rho\alpha}$ are functions of $z,k_\perp$ and $p$.
The Lorentz structures can be constructed from the Lorentz vectors $p$, $k_\perp$ and $n$.
The $\Xi^{(0)}_\alpha$ and $\tilde\Xi^{(0)}_\alpha$ are Lorentz vector and axial vector respectively.
The $\bar n$-components $\Xi^{(0)}_+$ and $\tilde\Xi^{(0)}_+$ contribute at the leading twist level
and the $\perp$-components $\Xi^{(0)}_\perp$ and $\tilde\Xi^{(0)}_\perp$ contribute at the twist-3 level.
For $\Xi^{(1)}_{\rho\alpha}$, $\tilde\Xi^{(1)}_{\rho\alpha}$,
the existence of the projection operator $\omega_{\rho}^{\ \rho'}$ makes the contributions from
$\Xi^{(1)}_{+\alpha}$ and $\tilde\Xi^{(1)}_{+\alpha}$ components equal to zero.
The leading contributions are from $\Xi^{(1)}_{\perp +}$ and $\tilde\Xi^{(1)}_{\perp +}$,
and they are at the twist-3 level.
Hence, up to twist-3, we need to consider,
\begin{align}
&z\Xi^{(0)}_\alpha(z,k_{\perp},p) =p_\alpha \hat D_1 (z, k_{\perp})  +k_{\perp \alpha} \hat D^\perp (z,k_{\perp}) +\cdots,\label{Xi0S0}\\
&z\tilde \Xi^{(0)}_\alpha(z,k_{\perp},p) =\epsilon_{\perp \alpha k_\perp} \Delta \hat D^\perp (z,k_{\perp}) +\cdots,\label{Xi0tS0}\\
& z \Xi^{(1)}_{\rho \alpha}(z,k_{\perp},p) = p_\alpha k_{\perp \rho} \ \xi^{(1)}_\perp (z,k_{\perp}) +\cdots, \label{Xi1S0}\\
& z \tilde \Xi^{(1)}_{\rho\alpha}(z,k_{\perp},p) = i p_\alpha\epsilon_{\perp \rho k_\perp} \tilde \xi^{(1)}_\perp (z,k _{\perp}) +\cdots, \label{Xi1tS0}
\end{align}
where $\epsilon_{\perp \rho k_\perp}=\epsilon_{\perp \rho \sigma}k_\perp^\sigma$,
and $\epsilon_{\perp \rho \sigma}=\epsilon_{\alpha\beta \rho \sigma}\bar n^\alpha n^\beta$.

We note that all the fragmentation functions defined above have a dimension of $-2$. 
They are all scalar functions of $z$ and $k_\perp^2$ and 
are Lorentz boost (along the $z$ direction) invariant.
There is only one leading twist component $\hat D_1 (z,k_\perp)$ that has the following operator expression,
\begin{align}
\hat D_1 &(z,k_\perp) =\frac{z}{4} \sum_X \int \frac{d\xi^- d^2 \xi_\perp}{2\pi} e^{-ip^+\xi^-/z+ik_\perp \cdot \xi_\perp} \nonumber\\
&\times {\rm Tr} \Bigl[ \langle 0| \gamma^+   \mathcal{L}^\dagger (0,\infty) \psi(0)|hX\rangle \langle hX| \bar\psi(\xi) \mathcal{L} (\xi,\infty) |0 \rangle\Bigr].
\end{align}
The other two components $\hat D^\perp (z,k_\perp)$ and $\Delta \hat D^\perp (z,k_\perp)$ 
defined via the expansion of $\Xi^{(0)}_\alpha(z,k_{\perp},p)$ and $\tilde\Xi^{(0)}_\alpha(z,k_{\perp},p)$ 
respectively are twist 3 and have the following operator expressions,
\begin{align}
\hat D^\perp &(z,k_\perp) = \frac{z}{4} \sum_X \int \frac{p^+ d\xi^- d^2 \xi_\perp}{2\pi k_\perp^2}
e^{-ip^+\xi^-/z+ik_\perp \cdot \xi_\perp}\nonumber\\
\times&{\rm Tr}  \Bigl[ \langle 0| \slashed k_\perp \mathcal{L}^\dagger (0,\infty) \psi(0)|hX\rangle \langle hX| \bar\psi(\xi) \mathcal{L} (\xi,\infty) |0 \rangle\Bigr],\\
\Delta \hat D^\perp &(z,k_\perp) = \frac{z}{4} \sum_X \int \frac{p^+ d \xi^- d^2 \xi_\perp}{2\pi k_\perp^2}
e^{-ip^+\xi^-/z+ik_\perp \cdot \xi_\perp} \nonumber\\
\times& {\rm Tr} \Bigl[ \langle 0| \slashed{k}_\perp^n \gamma_5 \mathcal{L}^\dagger (0,\infty) \psi(0)|hX\rangle \langle hX| \bar\psi(\xi) \mathcal{L} (\xi,\infty) |0 \rangle\Bigr],
\end{align}
where $k_\perp =(0,0,k_x,k_y)$, and $k_\perp^n$ is defined as $k_\perp^n =(0,0,k_y,-k_x)$ 
which represents the other transverse direction in the collinear frame perpendicular to $k_\perp$. 
We see that $\vec k_\perp^n$ is just in the normal direction of the production plane of the hadron 
and $\vec k_\perp$ is the other transverse direction in the production plane. 
We will denote the two transverse directions by the unit vectors 
$\vec e_n=\vec k_\perp^n/|\vec k_\perp|$ and $\vec e_t=\vec k_\perp/|\vec k_\perp|$ 
in the following of this paper.

We note that $\hat D_1^{q\to h} (z,k_\perp)$ is the well-known fragmentation function 
and has a simple probability interpretation. 
However, we should note that the variables are defined in the collinear frame so that  
$\hat D_1^{q\to h} (z,k_\perp)dzd^2k_\perp/(2\pi)^2$ is the number of hadron $h$ with light cone momentum fraction $z\sim z+dz$  
produced in the fragmentation of a quark with light cone momentum $p^+/z$ and transverse momentum 
$-\vec k_\perp\sim -(\vec k_\perp+d\vec k_\perp)$. 
In this paper, we drop the superscript $q\to h$ for simplification.
The fragmentation function satisfies the following normalization condition, 
\begin{equation}
\sum_h \int z\hat D_1^{q\to h} (z,k_\perp)dz \frac{d^2k_\perp}{(2\pi)^2}=1. \label{D1normalization}
\end{equation}
We emphasize in particular that the TMD fragmentation functions such as $\hat D_1 (z,k_\perp)$ defined above 
differ from that usually defined in phenomenological studies in terms of the differential cross sections 
or the number densities~\cite{Beringer:1900zz}. 
We therefore denote them by using $\hat D$'s to distinguish from those defined phenomenologically.
We do not have simple probability interpretations for the two twist 3 components, 
$\hat D^\perp (z,k_\perp)$ and $\Delta \hat D^\perp (z,k_\perp)$. 
They come from the vector ($\gamma_\alpha$) and axial vector  ($\gamma_5\gamma_\alpha$) components 
of $\hat\Xi^{(0)}$ respectively and are addenda to the leading twist contribution. 

The other two twist 3 components $\xi_\perp^{(1)} (z,k_\perp)$ and $\tilde\xi_\perp^{(1)} (z,k_\perp)$ 
are not independent. They are related to $\hat D^\perp (z,k_\perp)$ and $\Delta \hat D^\perp (z,k_\perp)$ 
by using the QCD equation of motion, $\slashed D (x) \psi(x)=0$.
From this equation, we obtain,
\begin{align}
& k^+ \Xi^{(0)}_{\perp\rho} = - n^\alpha  \left( {\rm Re} \Xi^{(1)}_{\rho \alpha} + 
\epsilon_{\perp\rho}^{\ \ \sigma\ } {\rm Im} \tilde \Xi^{(1)}_{\sigma\alpha} \right), \label{eq:eom1}\\
& k^+ \tilde \Xi^{(0)}_{\perp\rho} = - n^\alpha  \left( {\rm Re} \tilde \Xi^{(1)}_{\rho\alpha} + 
\epsilon_{\perp\rho}^{\ \ \sigma\ } {\rm Im} \Xi^{(1)}_{\sigma\alpha} \right). \label{eom2}
\end{align}
They lead to,
\begin{align}
&\hat D^\perp (z,k_\perp) = - z{\rm Re}\left[\xi_\perp^{(1)} (z,k_\perp)  -  \tilde \xi_\perp^{(1)} (z,k_\perp)   \right], \label{kffeom1}\\
&\Delta \hat D^\perp  (z,k_\perp) = -z {\rm Im}\left[\xi_\perp^{(1)}  (z,k_\perp) - \tilde \xi_\perp^{(1)} (z,k_\perp) \right].\label{kffeom2}
\end{align}

By inserting Eqs.~(\ref{Xi0S0}-\ref{Xi1tS0}) into Eqs.~(\ref{W0siexp}-\ref{W1Lsiexp}), 
and by using the relationships given by Eqs.~(\ref{kffeom1}-\ref{kffeom2}) to
replace $\xi^{(1)}_\perp (z,k_\perp)$ and $\tilde\xi^{(1)}_\perp (z, k_\perp)$ in the results obtained,
we finally obtain the hadronic tensor up to twist-3 as,
\begin{align}
W_{\mu\nu}^{(si)}&(q,p,k_\perp') =   - \frac{2}{z} \Big[ \omega_{\mu\nu}^{(0)} \hat D_1(z,k'_\perp)+ \nonumber\\ 
&-\omega_{\mu\nu}^{(1)}(\hat t,k'_\perp)\hat D^\perp (z,k'_\perp) 
 + \tilde\omega_{\mu\nu}^{(1)}(\hat t,k'_\perp)  \Delta \hat D^\perp (z, k'_\perp)\Big].  
\label{Wsiunpfr}
\end{align}
Here $\hat t=(zq-2p)/(zp\cdot q)=-\bar n/zq^-+n/p^+$ 
and we introduce a set of short handed notations $\omega_{\mu\nu}^{(j)}$ and $\tilde\omega_{\mu\nu}^{(j)}$
to represent the Lorentz tensors constructed from the unit vectors and other Lorentz vector(s) involved.
We refer to them as the basic Lorentz tensors.
From the unit vectors, we can defined a symmetric Lorentz tensors
$d_{\mu\nu}=g_{\mu\nu}-n_\mu\bar{n}_\nu-\bar{n}_\mu n_\nu$,
and an anti-symmetric tensor $\epsilon_{\perp\mu\nu}=\epsilon_{\mu\nu\rho\sigma}\bar{n}^{\rho}n^{\sigma}$
that depend only on the transverse components.
The $\omega_{\mu\nu}^{(0)}$ and $\tilde\omega_{\mu\nu}^{(0)}$ are just linear combinations of them, i.e.,
\begin{align}
\omega_{\mu\nu}^{(0)} &= -\frac{1}{4}p_\alpha h_{\mu\nu}^{(0)\alpha} =c_1^q d_{\mu\nu}  + i c_3^q \epsilon_{\perp \mu\nu},  \label{omega0}\\
\tilde\omega_{\mu\nu}^{(0)} &= \frac{1}{4}p_\alpha \tilde h_{\mu\nu}^{(0)\alpha}=i c_1^q \epsilon_{\perp \mu\nu} + c_3^q d_{\mu\nu}.\label{omega0t}
\end{align}
From two Lorentz vectors $A_L$ (that has only longitudinal components) 
and $B_\perp$ (that has only transverse components), we define,
\begin{align}
\omega_{\mu\nu}^{(1)}(A_L,B_\perp) & =c_1^q A_{L\{\mu} B_{\perp\nu\}} - i c_3^q A_{L[\mu} \epsilon_{\perp \nu] B_\perp},\label{omega1}\\
\tilde\omega_{\mu\nu}^{(1)}(A_L,B_\perp) & =ic_1^q A_{L[\mu}B_{\perp \nu ]}+c_3^q A_{L\{\mu}\epsilon_{\perp \nu\} B_\perp}, \label{omega1t} \\
\omega_{\mu\nu}^{(2)}(A_L,B_\perp) & =c_1^q A_{L\{\mu}\epsilon_{\perp \nu\} B_\perp}+ic_3^q A_{L[\mu}B_{\perp \nu ]},\label{omega2}\\
\tilde\omega_{\mu\nu}^{(2)}(A_L,B_\perp) & =i c_1^q A_{L[\mu} \epsilon_{\perp \nu] B_\perp}-c_3^q A_{L\{\mu} B_{\perp\nu\}},
 \label{omega2t} 
\end{align}
where 
$A_{\{\mu} B_{\nu\}} = A_\mu B_\nu + A_\mu B_\nu$,
and $A_{[\mu} B_{\nu]} = A_\mu B_\nu - A_\mu B_\nu$.

We see that for these $\omega_{\mu\nu}^{(j)}$'s and $\tilde\omega_{\mu\nu}^{(j)}$'s,
the real parts are symmetric with respect to the $\mu\leftrightarrow\nu$ exchange
and the imaginary parts are anti-symmetric, i.e.,
\begin{align}
\omega_{\mu\nu}^{(j)*}=\omega_{\nu\mu}^{(j)}, \ \ \ \ \ & \tilde\omega_{\mu\nu}^{(j)*}=\tilde\omega_{\nu\mu}^{(j)}. \label{omehermite}
\end{align}

Under space reflection $\hat P$ and time reversal $\hat T$, 
we denote $\hat P A_L=A_L^P$, $\hat T A_L=A_L^T$,  
$\hat P B_\perp=B_\perp^P$ and $\hat T B_\perp=B_\perp^T$,  and we have,
for $j=1$ and $2$, 
\begin{align}
\hat P \omega_{\mu\nu}^{(j)}(A_L, B_\perp)& = \omega^{(j)}_{\mu\nu} (A_L^P, B_\perp^P),\label{omespace1}\\
\hat P \tilde \omega_{\mu\nu}^{(j)}(A_L, B_\perp)& =\tilde\omega^{(j)}_{\mu\nu} (A_L^P, B_\perp^P), \label{omespace1t}\\
\hat T \omega_{\mu\nu}^{(j)}(A_L, B_\perp)& = \omega^{(j)*}_{\mu\nu} (A_L^T, B_\perp^T),\label{ometime1}\\
\hat T \tilde \omega_{\mu\nu}^{(j)}(A_L, B_\perp)& =\tilde\omega^{(j)*}_{\mu\nu} (A_L^T, B_\perp^T), \label{ometime1t}
\end{align}
respectively.
In the case that both $A_L$ and $B_\perp$ Lorentz vectors representing four momenta of hadron or quark, 
we have $A_L^P=A_L^T=\tilde A_L$ and $B_\perp^P=B_\perp^T=\tilde B_\perp$, hence, 
\begin{align}
\hat P \omega_{\mu\nu}^{(1)}(A_L, B_\perp)& = \omega^{(1)\nu\mu} (A_L, B_\perp),\label{omespace1a}\\
\hat P \tilde \omega_{\mu\nu}^{(1)}(A_L, B_\perp)& =-\tilde\omega^{(1)\nu\mu} (A_L, B_\perp), \label{omespace1ta}\\
\hat T \omega_{\mu\nu}^{(1)}(A_L, B_\perp)& = \omega^{(1)\mu\nu} (A_L, B_\perp),\label{ometime1a}\\
\hat T \tilde \omega_{\mu\nu}^{(1)}(A_L, B_\perp)& =-\tilde\omega^{(1)\mu\nu} (A_L, B_\perp);\label{ometime1ta}\\
\hat P \omega_{\mu\nu}^{(2)}(A_L, B_\perp)& = -\omega^{(2)\nu\mu} (A_L, B_\perp),\label{omespace2a}\\
\hat P \tilde \omega_{\mu\nu}^{(2)}(A_L, B_\perp)& =\tilde\omega^{(2)\nu\mu} (A_L, B_\perp), \label{omespace2ta}\\
\hat T \omega_{\mu\nu}^{(2)}(A_L, B_\perp)& =- \omega^{(2)\mu\nu} (A_L, B_\perp),\label{ometime2a}\\
\hat T \tilde \omega_{\mu\nu}^{(2)}(A_L, B_\perp)& =\tilde\omega^{(2)\mu\nu} (A_L, B_\perp).\label{ometime2ta}
\end{align}
However, if $A_L$ is a Lorentz vector and $B_\perp$ is a axial vector such as the polarization vector $S_\perp$, 
we have $A_L^P=A_L^T=\tilde A_L$ but $B_\perp^P= - \tilde B_\perp$ and $B_\perp^T=\tilde B_\perp$. 
In this case, we have, these $\omega$'s and $\tilde\omega$'s behave also differently.

We also point out that these $\omega_{\mu\nu}^{(j)}$'s ($j=0,1$ and 2) correspond to the Lorentz contractions 
of $p_\alpha$, $k_{\perp\alpha}$ and $\epsilon_{\perp\alpha k_\perp}$ with $h^{(0)\alpha}_{\mu\nu}$ respectively 
(plus the corresponding contributions from $h^{(1)\rho\alpha}_{\mu\nu}$ terms).  
The $\tilde\omega_{\mu\nu}^{(j)}$'s correspond to those with $\tilde h^{(0)\alpha}_{\mu\nu}$. 
We see in particular that, if we consider reactions via electromagnetic interaction, 
i.e. for the case that $c_1=1$ and $c_3=0$,  we have  
all the $\omega_{\mu\nu}^{(j)}$'s are real and symmetric in $\mu\leftrightarrow\nu$ 
while $\tilde\omega_{\mu\nu}^{(j)}$'s are imaginary and anti-symmetric in $\mu\leftrightarrow\nu$.
Since in this case $L_{\mu\nu}(l_1,l_2)$ is symmetric, this implies that 
only $\omega_{\mu\nu}^{(j)}$ terms contribute to the cross section in this case. 
In another word, only the vector component of $\hat\Xi^{(0)}$ contributes to 
$e^++e^-\to\gamma^*\to h+\bar q+X$ and the axial vector part does not. 

It is also easy to verify that these basic tensors satisfy 
$q^\mu \omega_{\mu\nu}^{(j)} (\hat t,k'_\perp) = q^\nu \omega_{\mu\nu}^{(j)}(\hat t,k'_\perp) =0$, 
hence the hadronic tensor satisfies also $q^\mu W^{(si)}_{\mu\nu} = q^\nu W^{(si)}_{\mu\nu}=0$.

From Eq.~(\ref{Wsiunpfr}), we see that, for the semi-inclusive production of spinless hadron $h$ in 
$e^++e^-\to h+\bar q+X$, there exist one leading twist and two twist-3 terms for the hadronic tensor 
corresponding to one leading twist and two twist 3 components of the fragmentation function. 
However none of the twist 3 terms survives the integration over $d^2k'_\perp$. 
Since the fragmentation functions depend on ${k'}_\perp^2$, 
after the integration over $d^2 k'_\perp$, we obtain,
\begin{align}
W_{\mu\nu}(q,p) =&\int \frac{d^2k'_\perp}{(2\pi)^2} W_{\mu\nu}^{(si)}(q,p,k_\perp') \nonumber\\ 
=  & - \frac{2}{z}  \omega_{\mu\nu}^{(0)} \int \frac{d^2k'_\perp}{(2\pi)^2} \hat D_1(z,k'_\perp),
\label{Wsiunpint}
\end{align}
where only the leading twist term is left.
By comparing Eq. (\ref{Wsiunpint}) with the corresponding results [Eq.(83)] in \cite{Wei:2013csa} for inclusive process, 
we obtain, 
\begin{align}
 \int \frac{d^2k'_\perp}{(2\pi)^2} \hat D_1(z,k'_\perp) = D_1 (z).
\end{align}
This is also the same as we can derive from the operator expressions for both of them and $D_1(z)$ is just the 
fragmentation function defined in \cite{Wei:2013csa} and in phenomenological studies.\cite{Beringer:1900zz}

\subsection{Spin-$1/2$ hadrons}

For hadrons with non-zero spins, the quantities describing the spin states are involved in the
Lorentz decompositions of the correlators such as $\Xi^{(0)}_\alpha$, $\tilde \Xi^{(0)}_\alpha$,
$\Xi^{(1)}_{\rho \alpha}$ and $\tilde \Xi^{(1)}_{\rho\alpha}$.
This makes the decomposition much more complicated and the physics more interesting.
It is clear that such decompositions can be written as a sum of the spin independent part
plus a spin dependent part.
Hence, the resulting contributions to the hadronic tensor should also be expressed as
spin independent part plus a spin dependent part, i.e.,
\begin{equation}
W^{(si)}_{\mu\nu}(q,p,S,k'_\perp)=W^{(si,unp)}_{\mu\nu}(q,p,k'_\perp)+W^{(si,pol)}_{\mu\nu}(q,p,S,k'_\perp).
\end{equation}
Obviously, the spin independent part is the same for hadrons with non-zero spins
as that for spin-zero hadrons.
We therefore only present the spin dependent part $W^{(si,pol)}_{\mu\nu}(q,p,S,k'_\perp)$ in the following.

For spin-$1/2$ particles, the spin state is described by a $2\times 2$ spin density matrix $\rho$ 
that is usually decomposed as $\rho=(1+\vec S\cdot\vec\sigma)/2$. 
The polarization vector $\vec S$ is usually replaced by its Lorentz covariant extension  
$S =(0,\vec S)$ in the rest frame of the particle. 
We therefore denote this spin-dependent part by $W^{(si,Vpol)}_{\mu\nu}$.
At high energies, we denote $\vec S=(\vec S_\perp,\lambda_h)$ and decompose $S$ 
in the light cone coordinate unit vectors as,
\begin{align}
S = \lambda_h \frac{p^+}{M}\bar{n} + S_{\perp} - \lambda_h \frac{M}{2p^+}n, \label{Smudec}
\end{align}
where $S_{\perp}=(0,\vec S_\perp,0)$ and $\lambda_h$ is the helicity.
We see that, compared with the $\bar{n}$-component, the $n_{\perp}$- and
$n$-components of $S$ are suppressed by a factor of $M/p^+$ and $(M/p^+)^2$ respectively.
They contribute at higher twists.
The polarization of the particle system is given by $\vec P$ that is the average value of $\vec\sigma$. 
Any component of $\vec P$ can be calculated by using 
$P_i={\rm Tr}(\sigma_i\rho)={\cal P}(\sigma_i=1)-{\cal P}(\sigma_i=-1)$, 
where ${\cal P}(\sigma_i=\pm 1)$ denotes the probability for the particle 
to be in the eigenstate of $\sigma_i$ with eigenvalue $1$ or $-1$ respectively. 
We also recall that under space reflection, $\lambda_h\leftrightarrow -\lambda_h$,
and $\hat P S^\mu = -\tilde S^\mu = - S_\mu$. 
Hence, up to twist-3, the relevant spin-dependent terms in the Lorentz decompositions 
of $\Xi^{(0)}_\alpha$, $\tilde\Xi^{(0)}_\alpha$, $\Xi^{(1)}_{\rho\alpha}$ and $\tilde\Xi^{(1)}_{\rho\alpha}$
are given as follows,
\begin{align}
z\Xi^{(0)}_\alpha  & (z,k_\perp ,p,S)  =
 p_\alpha \frac{\epsilon_{\perp}^{k_\perp S_\perp}}{M}  \hat D_{1T}^\perp (z, k_\perp) 
+ k_{\perp\alpha} \frac{\epsilon_{\perp}^{k_\perp S_\perp} }{M}  \hat D_{T}^\perp (z, k_\perp) \nonumber\\
& + \lambda_h \epsilon_{\perp \alpha k_\perp} \hat D_L^\perp (z, k_\perp) 
+ M \epsilon_{\perp \alpha S_\perp} \hat D_T (z, k_\perp)+\cdots, \label{Xi0Vpol} \\
z\tilde \Xi^{(0)}_\alpha  & (z,k_\perp ,p,S)  =
p_\alpha \Big[ \lambda_h \Delta \hat D_{1L} (z, k_\perp)  
 + \frac{k_\perp \cdot S_\perp}{M} \Delta \hat D_{1T}^\perp (z, k_\perp)\Big] \nonumber\\
& + \frac{\epsilon_{\perp}^{k_\perp S_\perp}}{M}  \epsilon_{\perp \alpha k_\perp} \Delta \hat D_T^\perp (z, k_\perp)
 +  \lambda_h k_{\perp\alpha}\Delta \hat D_L^\perp (z, k_\perp) \nonumber\\
&+ M S_{\perp \alpha} \Delta \hat D_T (z, k_\perp)+\cdots,\label{Xi0tVpol} \\
z\Xi^{(1)}_{\rho\alpha}  &  (z,k_\perp,p,S)   =  p_\alpha \Bigl[
 M \epsilon_{\perp \rho S_\perp} \xi^{(1)}_{T}(z, k_\perp) \nonumber\\
& + k_{\perp\rho}\frac{\epsilon_{\perp}^{k_\perp S_\perp}}{M}   \xi^{(1)\perp }_{T} (z, k_\perp) 
+  \lambda_h \epsilon_{\perp \rho k_\perp} \xi^{(1)\perp}_{L} (z, k_\perp)\Bigr]+\cdots, \label{Xi1Vpol} \\
z\tilde \Xi^{(1)}_{\rho \alpha} &  (z,k_\perp,p,S)   =
i p_\alpha \Bigl[M S_{\perp \rho} \tilde \xi^{(1)}_{T} (z, k_\perp)  \nonumber\\
& +  \frac{\epsilon_{\perp}^{k_\perp S_\perp} }{M} \epsilon_{\perp \rho k_\perp}\tilde \xi^{(1)\perp }_{T} (z, k_\perp)  
+  \lambda_h k_{\perp\rho} \tilde \xi^{(1)\perp}_{ L} (z, k_\perp)\Bigr]+\cdots .\label{Xi1tVpol}
\end{align}
Here, since $\epsilon_\perp^{k_\perp S_\perp} \epsilon_{\perp\rho k_\perp}
= k_{\perp\rho} k_\perp \cdot S_\perp -k_\perp^2 S_{\perp\rho}$,
and $k_\perp \cdot S_\perp \epsilon_{\perp\rho k_\perp}
= k_{\perp\rho} \epsilon_\perp^{k_\perp S_\perp} +k_\perp^2 \epsilon_{\perp\rho S_\perp}$,
the two terms containing $k_\perp \cdot S_\perp$,  i.e.
$k_{\perp\rho} k_\perp \cdot S_\perp$ and  $\epsilon_{\perp\rho k_\perp}k_\perp \cdot S_\perp $ 
are not independent hence do not appear in the above Eqs.~(\ref{Xi0Vpol}-\ref{Xi1tVpol}).

There are three twist-2 spin dependent components of the fragmentation function involved here, 
i.e., $\hat D_{1T}^\perp (z, k_\perp)$, $\Delta\hat D_{1L} (z, k_\perp)$ and $\Delta\hat D_{1T}^\perp (z, k_\perp)$.
The other six components, i.e. 
$\hat D_{T}^\perp  (z,k_\perp)$, $\hat D_{L}^\perp  (z,k_\perp)$, $\hat D_{T}(z,k_\perp)$,
$\Delta\hat D_{T}^\perp  (z,k_\perp)$, $\Delta\hat D_{L}^\perp  (z,k_\perp)$, and $\Delta\hat D_{T}(z,k_\perp)$,
are twist 3.
The operator expressions of these spin dependent fragmentation functions 
can be obtained by solving Eqs.~(\ref{Xi0Vpol}-\ref{Xi1tVpol}) inversely.
We denote the hadron state with polarization in the two transverse directions 
$\vec e_t$ and $\vec e_n$ by $|e_t^\uparrow\rangle$, $|e_t^\downarrow\rangle$,
$|e_n^\uparrow\rangle$, $|e_n^\downarrow\rangle$, 
and those in the longitudinal direction $z$ (helicity state) by $|+ \rangle$ and $|- \rangle$, respectively. 
We obtain the expressions for three leading twist spin-dependent fragmentation functions as,
\begin{align}
\Delta \hat D_{1L} (z,k_\perp) & = \frac{z}{8} \int \frac{d\xi^- d^2\xi_\perp}{2\pi} e^{-ip^+\xi^-/z+ik_\perp \cdot \xi_\perp} \nonumber\\
\times\Bigl\{ &  {\rm Tr} \Bigl[ \langle 0| \gamma^+ \gamma_5 \mathcal{L}^\dagger (0,\infty) \psi(0)|+ \rangle \langle +| \bar\psi(\xi) \mathcal{L} (\xi,\infty) |0 \rangle\Bigr] \nonumber\\
- &{\rm Tr} \Bigl[ \langle 0| \gamma^+ \gamma_5 \mathcal{L}^\dagger (0,\infty) \psi(0)|- \rangle \langle -| \bar\psi(\xi) \mathcal{L} (\xi,\infty) |0 \rangle\Bigr]\Bigr\}, \\
\hat D_{1T}^\perp  (z,k_\perp)& =  \frac{zM}{8|\vec{k}_\perp|} \int\frac{d\xi^- d^2\xi_\perp}{2\pi} e^{-ip^+\xi^-/z+ik_\perp \cdot \xi_\perp} \nonumber\\
\times\Bigl\{ & {\rm Tr} \Bigl[ \langle 0| \gamma^+   \mathcal{L}^\dagger (0,\infty) \psi(0)|e_n^\downarrow \rangle \langle e_n^\downarrow| \bar\psi(\xi) \mathcal{L} (\xi,\infty) |0 \rangle\Bigr] \nonumber\\
- &{\rm Tr} \Bigl[ \langle 0| \gamma^+   \mathcal{L}^\dagger (0,\infty) \psi(0)|e_n^\uparrow \rangle \langle e_n^\uparrow| \bar\psi(\xi) \mathcal{L} (\xi,\infty) |0 \rangle\Bigr]\Bigr\} , \label{wei:defd1t}\\
\Delta \hat D_{1T}^\perp (z,k_\perp) & = \frac{zM}{8|\vec{k}_\perp|} \int \frac{d\xi^- d^2\xi_\perp}{2\pi} e^{-ip^+\xi^-/z+ik_\perp \cdot \xi_\perp} \nonumber\\
\times\Bigl\{ &  {\rm Tr} \Bigl[ \langle 0| \gamma^+ \gamma_5 \mathcal{L}^\dagger (0,\infty) \psi(0)|e_t^\downarrow \rangle \langle e_t^\downarrow| \bar\psi(\xi) \mathcal{L} (\xi,\infty) |0 \rangle\Bigr] \nonumber\\
- &{\rm Tr} \Bigl[ \langle 0| \gamma^+ \gamma_5 \mathcal{L}^\dagger (0,\infty) \psi(0)|e_t^\uparrow \rangle \langle e_t^\uparrow| \bar\psi(\xi) \mathcal{L} (\xi,\infty) |0 \rangle\Bigr]\Bigr\}, \label{wei:defdeltad1t}
\end{align}
for the six twist 3 components, we have,
\begin{align}
\hat D_T  (z,k_\perp) &= \frac{z}{8M |\vec{k}_\perp|} \int \frac{p^+ d\xi^- d^2 \xi_\perp}{2\pi} e^{-ip^+\xi^-/z+ik_\perp \cdot \xi_\perp} \nonumber\\ 
 \times\Bigl\{
&{\rm Tr} \Bigl[\langle 0| \slashed {k}_\perp^n \mathcal{L}^\dagger (0,\infty) \psi(0)|e_t^\uparrow \rangle \langle e_t^\uparrow| \bar\psi(\xi) \mathcal{L} (\xi,\infty) |0 \rangle\Bigr] \nonumber\\
-&{\rm Tr} \Bigl[ \langle 0| \slashed {k}_\perp^n \mathcal{L}^\dagger (0,\infty) \psi(0)|e_t^\downarrow \rangle \langle e_t^\downarrow| \bar\psi(\xi) \mathcal{L} (\xi,\infty) |0 \rangle\Bigr] \Bigr\},\\
\Delta \hat D_{T} (z,k_\perp) & = \frac{z}{8M|\vec k_\perp|} \int \frac{p^+d\xi^- d^2\xi_\perp}{2\pi} e^{-ip^+\xi^-/z+ik_\perp \cdot \xi_\perp} \nonumber\\
\times\Bigl\{ &  {\rm Tr} \Bigl[ \langle 0| \slashed {k}_\perp \gamma_5 \mathcal{L}^\dagger (0,\infty) \psi(0)| e_t^\downarrow \rangle \langle e_t^\downarrow | \bar\psi(\xi) \mathcal{L} (\xi,\infty) |0 \rangle\Bigr] \nonumber\\
- &{\rm Tr} \Bigl[ \langle 0| \slashed {k}_\perp \gamma_5 \mathcal{L}^\dagger (0,\infty) \psi(0)|e_t^\uparrow  \rangle \langle e_t^\uparrow | \bar\psi(\xi) \mathcal{L} (\xi,\infty) |0 \rangle\Bigr]\Bigr\},  \\
\hat D_T^\perp  (z, k_\perp) & = \frac{zM}{8|\vec{k}_\perp|^3}   \int \frac{p^+ d\xi d^2 \xi_\perp}{2\pi} e^{-ip^+\xi^-/z+ik_\perp \cdot \xi_\perp} \nonumber\\
\times\Bigl\{
&{\rm Tr} \Bigl[ \langle 0| \slashed k_\perp \mathcal{L}^\dagger (0,\infty) \psi(0)|e_n^\uparrow\rangle \langle e_n^\uparrow | \bar\psi(\xi) \mathcal{L} (\xi,\infty) |0 \rangle\Bigr]\nonumber\\
-&{\rm Tr} \Bigl[ \langle 0| \slashed k_\perp \mathcal{L}^\dagger (0,\infty) \psi(0)|e_n^\downarrow \rangle \langle e_n^\downarrow | \bar\psi(\xi) \mathcal{L} (\xi,\infty) |0 \rangle\Bigr] \nonumber\\
+&{\rm Tr} \Bigl[ \langle 0| \slashed {k}_\perp^n \mathcal{L}^\dagger (0,\infty) \psi(0)|e_t^\uparrow \rangle \langle e_t^\uparrow| \bar\psi(\xi) \mathcal{L} (\xi,\infty) |0 \rangle\Bigr] \nonumber\\
-&{\rm Tr} \Bigl[ \langle 0| \slashed {k}_\perp^n \mathcal{L}^\dagger (0,\infty) \psi(0)|e_t^\downarrow \rangle \langle e_t^\downarrow| \bar\psi(\xi) \mathcal{L} (\xi,\infty) |0 \rangle\Bigr]\Bigr\},\\
\Delta \hat D_{T}^\perp (z,k_\perp) & = \frac{zM}{8|\vec k_\perp|^3} \int \frac{p^+d\xi^- d^2\xi_\perp}{2\pi} e^{-ip^+\xi^-/z+ik_\perp \cdot \xi_\perp} \nonumber\\
\times\Bigl\{ &  {\rm Tr} \Bigl[ \langle 0| \slashed {k}_\perp^n \gamma_5 \mathcal{L}^\dagger (0,\infty) \psi(0)|e_n^\downarrow \rangle \langle e_n^\downarrow | \bar\psi(\xi) \mathcal{L} (\xi,\infty) |0 \rangle\Bigr] \nonumber\\
- &{\rm Tr} \Bigl[ \langle 0| \slashed {k}_\perp^n \gamma_5 \mathcal{L}^\dagger (0,\infty) \psi(0)|e_n^\uparrow  \rangle \langle e_n^\uparrow | \bar\psi(\xi) \mathcal{L} (\xi,\infty) |0 \rangle\Bigr]\nonumber\\
-&  {\rm Tr} \Bigl[ \langle 0| \slashed {k}_\perp \gamma_5 \mathcal{L}^\dagger (0,\infty) \psi(0)| e_t^\downarrow \rangle \langle e_t^\downarrow | \bar\psi(\xi) \mathcal{L} (\xi,\infty) |0 \rangle\Bigr] \nonumber\\
+ &{\rm Tr} \Bigl[ \langle 0| \slashed {k}_\perp \gamma_5 \mathcal{L}^\dagger (0,\infty) \psi(0)|e_t^\uparrow  \rangle \langle e_t^\uparrow | \bar\psi(\xi) \mathcal{L} (\xi,\infty) |0 \rangle\Bigr]\Bigr\},  \\
\hat D_L^\perp (z,k_\perp) &= \frac{z}{8|\vec{k}_\perp|^2} \int \frac{p^+ d \xi^- d^2 \xi_\perp}{2\pi} e^{-ip^+\xi^-/z+ik_\perp \cdot \xi_\perp} \nonumber\\
\times\Bigl\{
&{\rm Tr} \Bigl[ \langle 0| \slashed {k}_\perp^n \mathcal{L}^\dagger (0,\infty) \psi(0)|+ \rangle \langle +| \bar\psi(\xi) \mathcal{L} (\xi,\infty) |0 \rangle\Bigr] \nonumber\\
-&{\rm Tr} \Bigl[ \langle 0| \slashed {k}_\perp^n \mathcal{L}^\dagger (0,\infty) \psi(0)|- \rangle \langle -| \bar\psi(\xi) \mathcal{L} (\xi,\infty) |0 \rangle\Bigr]\Bigr\},\\
\Delta \hat D_{L}^\perp (z,k_\perp) & = \frac{z}{8|\vec k_\perp|^2} \int \frac{p^+d\xi^- d^2\xi_\perp}{2\pi} e^{-ip^+\xi^-/z+ik_\perp \cdot \xi_\perp} \nonumber\\
\times\Bigl\{ &  {\rm Tr} \Bigl[ \langle 0| \slashed k_\perp \gamma_5 \mathcal{L}^\dagger (0,\infty) \psi(0)|- \rangle \langle -| \bar\psi(\xi) \mathcal{L} (\xi,\infty) |0 \rangle\Bigr] \nonumber\\
- &{\rm Tr} \Bigl[ \langle 0| \slashed k_\perp \gamma_5 \mathcal{L}^\dagger (0,\infty) \psi(0)|+ \rangle \langle +| \bar\psi(\xi) \mathcal{L} (\xi,\infty) |0 \rangle\Bigr]\Bigr\}.
\end{align}
From these operator expressions, we can already see that the three leading twist components, 
$\Delta\hat D_{1L} (z, k_\perp)$, 
$\hat D_{1T}^\perp (z, k_\perp)$ and $\Delta\hat D_{1T}^\perp (z, k_\perp)$,
are related to longitudinal quark spin transfer to hadron, 
the induced transverse polarization of $h$ in the direction of $\vec k_\perp^n$ 
(the normal direction $\vec e_n$ of the production plane) when the fragmenting quark is unpolarized 
and the induced transverse polarization in the direction of $\vec k_\perp$ 
(the transverse direction $\vec e_t$ in the production plane) 
when the quark is longitudinally polarized.
They correspond to the three leading twist parton distribution functions in nucleon, 
$g_{1L}(x,k_\perp)$,  $f_{1T}^\perp(x,k_\perp)$,  and $g_{1T}^\perp(x,k_\perp)$,  
i.e. the helicity distribution, the Sivers function and the worm-gear function or trans-helicity distribution, 
involved in the semi-inclusive deep inelastic lepton-nucleon scattering $e^-+N\to e^-+q+X$~\cite{Song:2013sja}.  
The physical significances of the twist 3 components are not so obvious. 
They lead to addenda to the leading twist contributions discussed above.  
We will come back and discuss more later on. 

Just the same as that in the spin-zero case, the twist 3 components defined via $\Xi^{(1)}_{\rho\alpha}$
or $\tilde\Xi^{(1)}_{\rho\alpha}$ are not independent. 
They are related to the six twist 3 components defined via $\Xi^{(0)}_{\alpha}$ or $\tilde\Xi^{(0)}_{\alpha}$ by 
using the QCD equation of motion. They are given by,  
\begin{align}
&\frac{1}{z} \hat D_T (z, k_\perp) = - \mathrm{Re} \left( \xi^{(1)}_T (z, k_\perp) + \tilde \xi^{(1)}_T (z, k_\perp) \right), \label{sffeom1} \\
&\frac{1}{z} \Delta \hat D_T (z, k_\perp) = \mathrm{Im} \left( \xi^{(1)}_T (z, k_\perp) + \tilde \xi^{(1)}_T (z, k_\perp) \right), \label{sffeom4} \\
&\frac{1}{z} \hat D_T^\perp (z, k_\perp) = - {\rm Re} \left( \xi^{(1)\perp }_{T} (z, k_\perp) - \tilde \xi^{(1)\perp }_{T} (z, k_\perp) \right), \label{sffeom5} \\
&\frac{1}{z} \Delta \hat D_T^{\perp} (z, k_\perp) = -{\rm Im} \left( \xi^{(1)\perp }_{T} (z, k_\perp) - \tilde \xi^{(1)\perp }_{T} (z, k_\perp) \right),\label{sffeom2} \\
&\frac{1}{z} \hat D_L^\perp (z, k_\perp) = - {\rm Re} \left( \xi^{(1)\perp }_{L} (z, k_\perp) + \tilde \xi^{(1)\perp }_{L} (z, k_\perp) \right), \label{eq:sffeom6} \\
&\frac{1}{z} \Delta \hat D_L^\perp (z, k_\perp) = {\rm Im} \left( \xi^{(1)\perp}_{ L} (z, k_\perp) + \tilde \xi^{(1)\perp }_{L} (z, k_\perp) \right).\label{sffeom3}
\end{align}

By inserting Eqs.~(\ref{Xi0Vpol}-\ref{Xi1tVpol}) into Eqs.~(\ref{W0siexp}-\ref{W1Lsiexp}) and by using the 
results given by Eqs.~(\ref{h0res}-\ref{h1tres}), and by using Eqs.~(\ref{sffeom1}-\ref{sffeom3}) to replace 
the $\xi^{(1)}$'s, 
we obtain the polarization dependent part of the hadronic tensor for spin-1/2 hadrons up to twist-3 as given by,
\begin{align}
W&_{\mu\nu}^{(si,Vpol)} (q,p,S,z,k'_\perp)= \frac{2}{z} \Big\{
-  \omega^{(0)}_{\mu\nu} \frac{1}{M} \epsilon_\perp^{k'_\perp S_\perp} \hat D_{1T}^\perp(z,k'_\perp) \nonumber \\
&+\tilde\omega^{(0)}_{\mu\nu} \Big[\lambda_h \Delta \hat D_{1L}(z,k'_\perp) + \frac{k'_\perp\cdot S_\perp}{M} \Delta \hat D_{1T}^\perp(z,k'_\perp)\Big] \nonumber \\
&+\frac{\epsilon_\perp^{k'_\perp S_\perp} }{M}\Big[\omega_{\mu\nu}^{(1)}(\hat t,k'_\perp) \hat D^\perp_T (z,k'_\perp) -
\tilde\omega_{\mu\nu}^{(1)}(\hat t,k'_\perp) \Delta \hat D^\perp_T(z,k'_\perp)\Big] \nonumber\\
&+ \lambda_h \Big[\omega_{\mu\nu}^{(2)}(\hat t,k'_\perp) \hat D^\perp_L (z,k'_\perp)+\tilde\omega_{\mu\nu}^{(2)}(\hat t,k'_\perp) \Delta \hat D_L^\perp(z,k'_\perp)\Big]  \nonumber \\ 
&+ M \Big[\omega_{\mu\nu}^{(2)}(\hat t,S_\perp) \hat D_T (z,k'_\perp)+\tilde\omega_{\mu\nu}^{(2)}(\hat t,S_\perp)\Delta \hat D_T(z,k'_\perp)\Big] \Big\}, \label{WsiVpolfr}
\end{align}
where $\omega_{\mu\nu}^{(j)}$ and $\tilde\omega_{\mu\nu}^{(j)}$ are the basic Lorentz tensors 
defined by Eqs. (\ref{omega0}-\ref{omega2t}).

From Eq.~(\ref{WsiVpolfr}), we see that, at leading twist, we have three spin dependent contributions
that describe the polarizations of the hadron along the helicity direction and two transverse directions respectively.
At twist-3, we have six terms, every two of them contribute to the polarization in one direction.
We note in particular that there exist a leading twist term and also higher twist addenda 
to it characterized by $\epsilon_\perp^{k'_\perp S_\perp}$. 
This corresponds to the Sivers function in parton distribution and leads to transverse polarization of the hadron. 
Among the six twist three components of the fragmentation function, 
three of them are from the $\gamma_\alpha$-component of $\hat\Xi^{(0)}$ and the 
other three are from the $\gamma_5\gamma_\alpha$-component of $\hat\Xi^{(0)}$. 
They correspond to the $\omega$ and $\tilde\omega$ terms in Eq.~(\ref{WsiVpolfr}).
If we consider reactions via electromagnetic interaction $e^++e^-\to\gamma^*\to h+\bar q+X$, 
only the $\omega$-terms contribute to the cross section.

We also note that, upon integration over $d^2k'_\perp$, we obtain, 
\begin{align}
W_{\mu\nu}^{(Vpol)} &(q,p,S)=\int \frac{d^2k'_\perp}{(2\pi)^2} W_{\mu\nu}^{(si,Vpol)} (q,p,S,z,k'_\perp) \nonumber\\
 = & \frac{2}{z} \Bigl\{ \tilde \omega_{\mu\nu}^{(0)} \lambda_h \int \frac{d^2k'_\perp}{(2\pi)^2} \Delta \hat D_{1L} (z,k'_\perp) \nonumber\\
&  + M \omega_{\mu\nu}^{(2)}(\hat t,S_\perp) \int \frac{d^2k'_\perp}{(2\pi)^2} \hat D_T (z,k'_\perp) \nonumber \\
&  + M \tilde\omega_{\mu\nu}^{(2)}(\hat t,S_\perp)\int \frac{d^2k'_\perp}{(2\pi)^2} \Delta \hat D_T(z,k'_\perp) \nonumber \\
&  + \frac{1}{M}\int \frac{d^2k'_\perp}{(2\pi)^2} \epsilon_\perp^{k'_\perp S_\perp} \omega_{\mu\nu}^{(1)}(\hat t,k'_\perp) \hat D^\perp_T (z,k'_\perp) \nonumber\\
& - \frac{1}{M}\int \frac{d^2k'_\perp}{(2\pi)^2} \epsilon_\perp^{k'_\perp S_\perp} \tilde\omega_{\mu\nu}^{(1)}(\hat t,k'_\perp) \Delta \hat D^\perp_T(z,k'_\perp)\Bigr\}. \label{WVpolint}
\end{align}
This is to compare with the $S$-dependent part given by Eq.~(100) in \cite{Wei:2013csa}, and it follows that, 
\begin{align}
& \Delta D_{1L}(z) = \int \frac{d^2k_\perp}{(2\pi)^2} \Delta \hat D_{1L} (z,k_\perp), \label{D1Lint}\\
& D_T (z)=\int \frac{d^2k_\perp}{(2\pi)^2} \left(\hat D_T (z,k_\perp) + \frac{k_\perp^2}{2 M^2} \hat D_T^{\perp} (z,k_\perp) \right) ,\label{DTint}\\
& \Delta D_T (z)=\int \frac{d^2k_\perp}{(2\pi)^2} \left(\Delta \hat D_T (z,k_\perp) - \frac{k_\perp^2}{2 M^2} \Delta \hat D_T^{\perp } (z,k_\perp) \right), \label{dDTint}
\end{align}
which are the corresponding components of fragmentation function discussed in 
\cite{Wei:2013csa} for the inclusive process $e^++e^-\to h+X$.

\subsection{Spin-1 hadrons}

\subsubsection{Description of the polarization of spin-1 hadrons} \label{wei:secc1}

The polarization of a system of spin-1 particles is described by a $3\times3$ spin density matrix $\rho$. 
In the particle rest frame, the matrix can be decomposed in terms of the spin operator $\Sigma^i$ and 
$\Sigma^{ij}= \frac{1}{2} (\Sigma^i\Sigma^j + \Sigma^j \Sigma^i) - \frac{2}{3} \mathbf{1} \delta^{ij}$, i.e., 
\begin{align}
\rho = \frac{1}{3} (\mathbf{1} + \frac{3}{2}S^i \Sigma^i + 3 T^{ij} \Sigma^{ij}), \label{eq:spin1rho}
\end{align}
where the spin polarization tensor $T^{ij}={\rm Tr}(\rho \Sigma^{ij})$ and is parameterized as,
\begin{align}
\mathbf{T}= \frac{1}{2}
\left(
\begin{array}{ccc}
-\frac{2}{3}S_{LL} + S_{TT}^{xx} & S_{TT}^{xy} & S_{LT}^x  \\
S_{TT}^{xy}  & -\frac{2}{3} S_{LL} - S_{TT}^{xx} & S_{LT}^{y} \\
S_{LT}^x & S_{LT}^{y} & \frac{4}{3} S_{LL}
\end{array}
\right),
\label{spintensor}
\end{align}
where the five independent parameters are defined as,
\begin{align}
& S_{LL} = \frac{3}{2}\langle \Sigma_z^2 \rangle - 1, \label{wei:slldef}\\
& S_{LT}^x = \langle \Sigma_x \Sigma_z + \Sigma_z \Sigma_x \rangle,\label{wei:sltxdef}\\
& S_{LT}^y = \langle \Sigma_y \Sigma_z + \Sigma_z \Sigma_y \rangle,\label{wei:sltydef}\\
& S_{TT}^{xy} = S_{TT}^{yx} = \langle \Sigma_x \Sigma_y + \Sigma_y \Sigma_x \rangle,\label{wei:sttxydef}\\
& S_{TT}^{xx} = - S_{TT}^{yy} = \langle \Sigma_x^2 - \Sigma_y^2 \rangle. \label{wei:sttxxdef}
\end{align}

We see that, in this decomposition, the polarization of a spin-1 hadron is described 
by a vector polarization $S^\mu$ and a tensor polarization $T^{\mu\nu}$. 
The vector polarization is defined in exactly the same way as that for the spin-$1/2$ hadrons, 
i.e. $S^\mu=(0,\vec S)$ in the rest frame of the hadron. 
The tensor polarization part  $T^{\mu\nu}$ has five independent components 
and they are given by a Lorentz scalar $S_{LL}$, 
a Lorentz vector $S_{LT}^\mu=(0,S_{LT}^x,S_{LT}^y,0)$ 
and a Lorentz tensor $S_{TT}^{\mu\nu}$ that has two independent non-zero components 
$S_{TT}^{xx}$ and  $S_{TT}^{xy}$ in the rest frame of the hadron.  
We also note that, under space reflection, they behave as,
$\hat P S_{LL} = S_{LL}$, $\hat P S_{LT}^\mu = S_{LT\mu}$, and $\hat P S_{TT}^{\mu\nu}=S_{TT\mu\nu}$.
These polarization parameters can be related to the probabilities for the particles in different spin states\cite{Bacchetta:2000jk}. 
We use ${\cal P}(m;\theta_n,\phi_n)$ to denote the probability for the particle to be 
in the eigenstate $|m;\theta_n,\phi_n\rangle$ 
of $\vec\Sigma\cdot \vec n$ with eigen value $m$ and $\vec n$ is a direction specified by the polar angle $\theta_n$ 
and azimuthal angle $\phi_n$ and we have\cite{Bacchetta:2000jk},
\begin{align}
& S_{LL} = [1-3{\cal P}(0;0,0)]/2,\label{sllint}\\
& S_{LT}^x = {\cal P}(0;\pi/{4},\pi) - {\cal P}(0;{\pi}/{4},0) , \label{sltxint}\\
& S_{LT}^y = {\cal P}(0;{\pi}/{4},{3\pi}/{2}) - {\cal P}(0;{\pi}/{4},{\pi}/{2}) , \label{sltyint}\\
& S_{TT}^{xx} = - S_{TT}^{yy} = {\cal P}(0;{\pi}/{2},{\pi}/{2})-{\cal P}(0;{\pi}/{2},0), \label{sttxxint}\\
& S_{TT}^{xy} = S_{TT}^{yx} = {\cal P}(0;{\pi}/{2},-{\pi}/{4})-{\cal P}(0;{\pi}/{2},{\pi}/{4}).\label{sttxyint}
\end{align}
These relationships are used to calculate the expectations of these polarization parameters 
from the differential cross section. 
We will come back to this point later in the next section.

Under this decomposition of the spin density matrix of spin-1 hadrons, 
the spin dependent part of the hadronic tensor or the fragmentation function are divided into two parts, 
a polarization vector $S$-dependent part and a polarization tensor $T$-dependent part, 
\begin{equation}
W^{(si,pol)}_{\mu\nu}
=W^{(si,Vpol)}_{\mu\nu}(q,p,S,k'_\perp)
+W^{(si,Tpol)}_{\mu\nu}(q,p,T,k'_\perp).
\end{equation}
The $S$-dependent part $W^{(si,Vpol)}_{\mu\nu}(q,p,S,k'_\perp)$ has exactly the same form as
$W^{(si,Vpol)}_{\mu\nu}(q,p,S,k'_\perp)$ for spin $1/2$ particles as given by Eq.~(\ref{WsiVpolfr})
where only polarization vector $S$ is needed to describe the polarization state of the hadron.
The tensor polarization dependent part $W^{(si,Tpol)}_{\mu\nu}(q,p,S,k'_\perp)$ 
is further divided into $S_{LL}$, $S_{LT}$ and $S_{TT}$ dependent parts respectively.
\begin{align}
W&^{(si,Tpol)}_{\mu\nu}(q,p,T,k'_\perp)=
W^{(si,LL)}_{\mu\nu}(q,p,S_{LL},k'_\perp)+\nonumber\\
&+W^{(si,LT)}_{\mu\nu}(q,p,S_{LT},k'_\perp)+W^{(si,TT)}_{\mu\nu}(q,p,S_{TT},k'_\perp).
\end{align}
This part is new for spin-1 hadrons and we present the results for them in the following.

\subsubsection{Leading twist contributions}

In the Lorentz decompositions of the correlators for vector mesons,
the polarization vectors and tensor $S^\mu$, $S_{LL}$, $S_{LT}^{\mu}$ and $S_{TT}^{\mu\nu}$ are involved.
This makes the decomposition much more complicated than those for spin-$1/2$ hadrons where only polarization vector $S$ is needed.
We therefore first start with the leading twist contributions in the following.
To the leading twist, we need only to consider $\Xi^{(0)}_+$ and $\tilde\Xi^{(0)}_+$, i.e.,
only the $\bar n$ components of these vectors. 
We recall that $S_{LL}$ is a Lorentz scalar, $S_{LT}$ is a vector and $S_{TT}$ is a tensor, 
so we have one $S_{LL}$-dependent term from $\Xi^{(0)}_\alpha(z,k_\perp)$, i.e.,
\begin{align}
z\Xi^{(0)}_\alpha(z,k_\perp) & =p_\alpha S_{LL}  \hat D_{1LL}(z,k_\perp) +\cdots, \label{Xi0LL}
\end{align}
and the following $S_{LT}$ and $S_{TT}$ dependent terms,
\begin{align}
z\Xi^{(0)}_\alpha(z,k_\perp) & = p_\alpha \frac{S_{LT} \cdot k_\perp}{M}  \hat D_{1LT}^\perp(z,k_\perp) \nonumber\\
&+ p_\alpha \frac{k_{\perp \gamma}  k_{\perp\delta}S_{TT}^{\gamma\delta}}{M^2}  \hat D_{1TT}^\perp(z,k_\perp)+\cdots, \label{Xi0TT}\\
z\tilde\Xi^{(0)}_\alpha (z,k_\perp)& =p_\alpha  \frac{\epsilon_\perp^{k_\perp S_{LT}} }{M} \Delta \hat D_{1LT}^\perp(z,k_\perp)  \nonumber\\
&+p_\alpha  \frac{\epsilon_{\perp k_\perp \gamma}  k_{\perp\delta} S_{TT}^{\gamma\delta} }{M^2} \Delta \hat D_{1TT}^\perp(z,k_\perp)+\cdots.\label{Xi0tTT}
\end{align}
The corresponding leading twist contributions to the hadronic tensor can be obtained
by inserting Eqs.~(\ref{Xi0LL}-\ref{Xi0tTT}) into Eq.~(\ref{W0siexp}).
They are given by,
\begin{align}
W_{\mu\nu}^{(si,LL,0)}&(q,p,S_{LL},k'_\perp) = -\frac{2}{z} \omega_{\mu\nu}^{(0)}  S_{LL} \hat D_{1LL}(z,k'_\perp),
\label{W0siLL}\\
W_{\mu\nu}^{(si,LT,0)}&(q,p,S_{LT},k'_\perp) =-\frac{2}{z}\Big[ \omega_{\mu\nu}^{(0)} \frac{S_{LT}\cdot k'_\perp}{M} \hat D_{1LT}^\perp(z,k'_\perp) \nonumber\\
-&  \tilde\omega_{\mu\nu}^{(0)}
 \frac{1}{M}\epsilon_\perp^{k'_\perp S_{LT}} \Delta \hat D_{1LT}^\perp (z,k'_\perp)\Big],  \label{W0siLT}\\
 W_{\mu\nu}^{(si,TT,0)}&(q,p,S_{TT},k'_\perp) = -\frac{2}{z}\Big[ \omega_{\mu\nu}^{(0)}
  \frac{k'_{\perp\alpha} k'_{\perp\beta} S_{TT}^{\alpha\beta} }{M^2}  \hat D_{1TT}^\perp(z,k'_\perp)  \nonumber\\
-& \tilde\omega_{\mu\nu}^{(0)} \frac{\epsilon_{\perp k'_\perp \gamma} k'_{\perp\delta} S_{TT}^{\gamma \delta} }{M^2}  \Delta \hat D_{1TT}^\perp(z,k'_\perp)\Big], \label{W0siTT}
\end{align}
where we use $0$ in the superscript to denote that they are just the leading twist contributions.

We see that, even at leading twist, there are one $S_{LL}$-, two $S_{LT}$- and 
two $S_{TT}$-dependent terms in both $\mu\leftrightarrow\nu$ symmetric and anti-symmetric cases.
This shows that even in reactions with unpolarized electrons and unpolarized positrons,
we still obtain contributions depending on all these three components $S_{LL}$, $S_{LT}$ and $S_{TT}$, 
for $e^++e^-\to Z^0\to h+\bar q+X$ as well as $e^++e^-\to \gamma^*\to h+\bar q+X$.

Also, we note here that upon integration over $d^2k'_\perp$, 
$W_{\mu\nu}^{(si,LT,0)}$ and $W_{\mu\nu}^{(si,TT,0)}$ vanish.
We obtain the leading twist contributions to the corresponding components 
of the hadronic tensor for the inclusive process as, 
\begin{align}
W_{\mu\nu}^{(LL,0)}(q,p,S_{LL})&=\int \frac{d^2k'_\perp}{(2\pi)^2} W_{\mu\nu}^{(si,LL,0)}(q,p,S_{LL},k'_\perp)  \nonumber\\
 &=-\frac{2}{z} \omega_{\mu\nu}^{(0)}  S_{LL} \int \frac{d^2k'_\perp}{(2\pi)^2} \hat D_{1LL}(z,k'_\perp), \\
W_{\mu\nu}^{(LT,0)}(q,p,S_{LT})&=\int \frac{d^2k'_\perp}{(2\pi)^2} W_{\mu\nu}^{(si,LT,0)}(q,p,S_{LT},k'_\perp) = 0 ,  \\
W_{\mu\nu}^{(TT,0)}(q,p,S_{TT})&=\int \frac{d^2k'_\perp}{(2\pi)^2} W_{\mu\nu}^{(si,TT,0)}(q,p,S_{TT},k'_\perp) = 0.
\end{align}
This is to compare with Eq.~(112) in \cite{Wei:2013csa} for the inclusive process $e^++e^-\to h+X$. 
We see that the results are exactly the same and we have, 
\begin{align}
 D_{1LL}(z)&=\int \frac{d^2k_\perp}{(2\pi)^2} \hat D_{1LL}(z,k_\perp),
\end{align}
similar to other components of the fragmentation function.

\subsubsection{Twist-3 contributions}

Up to twist-3, we need to consider $\Xi^{(0)}_\perp$, $\tilde\Xi^{(0)}_\perp$, $\Xi^{(1)}_{\perp+}$ 
and $\tilde\Xi^{(1)}_{\perp+}$.
The $S_{LL}$ dependent terms are,
\begin{align}
z\Xi^{(0)}_{\alpha LL} (z, k_\perp) = & k_{\perp\alpha}  S_{LL} \hat D_{LL}^\perp(z, k_\perp)+ \cdots , \\
z\tilde \Xi^{(0)}_{\alpha LL} (z, k_\perp) = & \epsilon_{\perp\alpha k_\perp}  S_{LL} \Delta \hat D_{LL}^\perp(z, k_\perp)
+ \cdots, \\
z\Xi^{(1)}_{\rho\alpha LL} (z, k_\perp) = &p_\alpha k_{\perp\rho} S_{LL} \xi_{LL}^\perp(z, k_\perp)
+ \cdots,\\
z\tilde \Xi^{(1)}_{\rho\alpha LL} (z, k_\perp) = & 
i p^\alpha\epsilon_{\perp\rho k_\perp}  S_{LL}  \tilde \xi_{LL}^\perp(z, k_\perp)+ \cdots;
\end{align}
the $S_{LT}$ dependent terms are,
\begin{align}
z\Xi^{(0)}_{\alpha LT} (z, k_\perp) & =  M S_{LT\alpha} \hat D_{LT}(z, k_\perp) + \nonumber\\
&+ k_{\perp\alpha} \frac{k_\perp \cdot S_{LT}}{M}  \hat D_{LT}^\perp(z, k_\perp) +\cdots,\\
z\tilde \Xi^{(0)}_{\alpha LT} (z, k_\perp) & =
 M \epsilon_{\perp\alpha S_{LT}}  \Delta \hat D_{LT}(z, k_\perp)+ \nonumber\\
&+  \epsilon_{\perp\alpha k_\perp} \frac{k_\perp \cdot S_{LT}}{M} \Delta \hat D_{LT}^\perp(z, k_\perp) +\cdots, \\
z\Xi^{(1)}_{\rho\alpha LT} (z, k_\perp) & =  p_\alpha \big[
 M S_{LT\rho} \ \xi_{LT}(z, k_\perp)+ \nonumber\\
& + k_{\perp\rho}  \frac{k_\perp \cdot S_{LT}}{M} \xi_{LT}^\perp(z, k_\perp)\big] + \cdots,\\
z\tilde \Xi^{(1)}_{\rho\alpha LT} (z, k_\perp) & = i p_\alpha \big[  M \epsilon_{\perp \rho S_{LT}} \ \tilde \xi_{LT}(z, k_\perp)+ \nonumber\\
& + \epsilon_{\perp\rho k_\perp}  k_\perp \cdot S_{LT} \ \tilde \xi_{LT}^\perp(z, k_\perp)\big] + \cdots ;
\end{align}
and the $S_{TT}$ dependent terms are,
\begin{align}
z\Xi^{(0)}_{\alpha TT} (z, k_\perp) &=  S_{TT\alpha \beta} k_\perp^\beta \hat D_{TT}^{\perp A}(z, k_\perp) +  \nonumber\\
&+  k_{\perp\alpha} \frac{k_{\perp\gamma} k_{\perp\delta} S_{TT}^{\gamma\delta} }{M^2} \hat D_{TT}^{\perp C}(z, k_\perp) +\cdots,\\
z\tilde \Xi^{(0)}_{\alpha TT} (z, k_\perp) & =
\epsilon_{\perp \alpha \beta}  S_{TT}^{\beta \gamma} k_{\perp \gamma} \Delta \hat D_{TT}^{\perp A}(z, k_\perp) + \nonumber\\
&+\epsilon_{\perp\alpha k_\perp} \frac{k_{\perp\gamma} k_{\perp\delta} S_{TT}^{\gamma\delta}}{M^2}   \Delta \hat D_{TT}^{\perp C}(z, k_\perp) +\cdots, \\
z\Xi^{(1)}_{\rho\alpha TT} (z, k_\perp) &=   p_\alpha\big[  S_{TT\rho \beta} k_\perp^\beta \xi_{TT}^{\perp A}(z, k_\perp)+ \nonumber\\
&+ k_{\perp\rho} \frac{k_{\perp\gamma} k_{\perp\delta} S_{TT}^{\gamma\delta}}{M^2} \xi_{TT}^{\perp C} (z, k_\perp) \big]+\cdots,\\
z\tilde \Xi^{(1)}_{\rho\alpha TT} (z, k_\perp) &=  i p_\alpha \big[  \epsilon_{\perp\rho \beta}  S_{TT}^{\beta \gamma} k_{\perp\gamma} \tilde \xi_{TT}^{\perp A}(z, k_\perp)+ \nonumber\\
&+\epsilon_{\perp\rho k_\perp} \frac{k_{\perp\gamma} k_{\perp\delta} S_{TT}^{\gamma\delta}}{M^2}  \tilde \xi_{TT}^{\perp C}(z, k_\perp) \big] +\cdots.
\end{align}

Again, we use the equation of motion, $\slashed D \psi (x) = 0$, 
to relate these twist 3 fragmentation functions with each other.
We get,
\begin{align}
& \frac{1}{z} \hat D_{LL}^\perp (z,k_\perp) = 
- {\rm Re}  \left(\xi_{LL}^\perp (z,k_\perp) - \tilde \xi_{LL}^\perp (z,k_\perp) \right), \\
& \frac{1}{z} \Delta \hat D_{LL}^\perp (z,k_\perp) = 
- {\rm Im} \left( \xi_{LL}^\perp (z,k_\perp) - \tilde \xi_{LL}^\perp (z,k_\perp) \right),\\
& \frac{1}{z} \hat D_{LT} (z,k_\perp) = 
- {\rm Re} \left( \xi_{LT} (z,k_\perp) - \tilde \xi_{LT} (z,k_\perp) \right),\\
& \frac{1}{z} \Delta \hat D_{LT} (z,k_\perp) = 
- {\rm Im} \left( \xi_{LT} (z,k_\perp) - \tilde \xi_{LT} (z,k_\perp) \right),\\
& \frac{1}{z} \hat D_{LT}^\perp (z,k_\perp) = 
- {\rm Re} \left( \xi_{LT}^\perp (z,k_\perp) - \tilde \xi_{LT}^\perp (z,k_\perp) \right), \\
& \frac{1}{z} \Delta \hat D_{LT}^\perp (z,k_\perp) = 
- {\rm Im} \left( \xi_{LT}^\perp (z,k_\perp) - \tilde \xi_{LT}^\perp (z,k_\perp) \right),\\
& \frac{1}{z} \hat D_{TT}^{\perp A} (z,k_\perp) = 
- {\rm Re} \left( \xi_{TT}^{\perp A} (z,k_\perp) - \tilde \xi_{TT}^{\perp A} (z,k_\perp) \right), \\
& \frac{1}{z} \Delta \hat D_{TT}^{\perp A} (z,k_\perp) = 
- {\rm Im} \left( \xi_{TT}^{\perp A} (z,k_\perp) - \tilde \xi_{TT}^{\perp A} (z,k_\perp) \right),\\
& \frac{1}{z} \hat D_{TT}^{\perp C} (z,k_\perp) = 
- {\rm Re} \left( \xi_{TT}^{\perp C} (z,k_\perp) - \tilde \xi_{TT}^{\perp C} (z,k_\perp) \right), \\
& \frac{1}{z} \Delta \hat D_{TT}^{\perp C} (z,k_\perp) = 
- {\rm Im} \left( \xi_{TT}^{\perp C} (z,k_\perp) - \tilde \xi_{TT}^{\perp C} (z,k_\perp) \right).
\end{align}

The twist-3 contributions to tensor polarization dependent parts of the hadronic tensor are finally obtained as,
\begin{align}
W_{\mu\nu}^{(si,LL,1)}& (q,p,S_{LL},k'_\perp) =
\frac{2}{z}S_{LL} \Bigl[ \omega_{\mu\nu}^{(1)}(\hat t,k'_\perp) \hat D_{LL}^\perp (z,k'_\perp) \nonumber \\
 & - \tilde\omega_{\mu\nu}^{(1)}(\hat t,k'_\perp)  \Delta \hat D_{LL}^\perp(z,k'_\perp)\Bigr], \label{W1siLL} \\
W_{\mu\nu}^{(si,LT,1)}& (q,p,S_{LT},k'_\perp) =  \frac{2}{z}  \frac{k'_\perp \cdot S_{LT}}{M}
\Bigl[ \omega_{\mu\nu}^{(1)}(\hat t,k'_\perp)  \hat D_{LT}^{\perp}(z,k'_\perp) \nonumber \\
 & ~~~ -   \tilde \omega_{\mu\nu}^{(1)}(\hat t,k'_\perp) \Delta \hat D_{LT}^\perp(z,k'_\perp)\Bigr]  \nonumber\\
 &+ \frac{2}{z} M \Bigl[ \omega_{\mu\nu}^{(1)}(\hat t,S_{LT}) \hat D_{LT}(z,k'_\perp) +\nonumber\\
 & ~~~ - \tilde\omega_{\mu\nu}^{(1)}(\hat t,S_{LT}) \Delta \hat D_{LT}(z,k'_\perp)\Bigr], \label{W1siLT}\\
W_{\mu\nu}^{(si,TT,1)}&(q,p,S_{TT},k'_\perp) =  
\frac{2}{z} \frac{k'_{\perp\gamma} k'_{\perp\delta} S_{TT}^{\gamma\delta}}{M^2}  \nonumber \\
 & \times\Bigl[ \omega_{\mu\nu}^{(1)}(\hat t,k'_{\perp}) \hat D_{TT}^{\perp C}(z,k'_\perp)-
 \tilde\omega_{\mu\nu}^{(1)}(\hat t,k'_{\perp})  \Delta \hat D_{TT}^{\perp C}(z,k'_\perp)\Bigr] \nonumber\\
 &+\frac{2}{z} \Bigl[ \omega_{\mu\nu}^{(1)}(\hat t,S_{TT}\cdot k'_\perp) \hat D_{TT}^{\perp A}(z,k'_\perp) \nonumber\\
 & ~~~ - \tilde\omega_{\mu\nu}^{(1)}(\hat t,S_{TT}\cdot k'_\perp) \Delta \hat D_{TT}^{\perp A}(z,k'_\perp)\Bigr]. \label{W1siTT}
\end{align}
We see that there are two $S_{LL}$-, four $S_{LT}$- and four $S_{TT}$-dependent twist 3 terms in 
the hadronic tensor and they are all addenda to the leading twist contributions. 
Among them one $S_{LL}$-, two $S_{LT}$- and two $S_{TT}$-dependent terms 
are from the $\gamma_\alpha$ component of $\hat\Xi^{(0)}$ 
the other one $S_{LL}$-, two $S_{LT}$- and two $S_{TT}$-dependent terms 
are from the $\gamma_5\gamma_\alpha$ component of $\hat\Xi^{(0)}$. 
In $e^++e^-\to\gamma^*\to h+{\bar q}+X$, the former are all symmetric and contribute 
while the latter are anti-symmetric and do not contribute to the cross section.

We can also get the corresponding results for the inclusive process upon integration over $d^2k'_\perp$. 
We see that, after integration over $d^2k'_\perp$, 
$W_{\mu\nu}^{(si,LL,1)}$ and $W_{\mu\nu}^{(si,TT,1)}$ vanish, 
only the $S_{LT}$ dependent term survives, i.e., 
\begin{align}
W_{\mu\nu}^{(LL,1)} & (q,p,S_{LL})=\int \frac{d^2k'_\perp}{(2\pi)^2} W_{\mu\nu}^{(si,LL,1)} (q,p,S_{LL},k'_\perp) = 0, \label{W1siLLint}\\
W_{\mu\nu}^{(TT,1)} & (q,p,S_{TT})=\int \frac{d^2k'_\perp}{(2\pi)^2} W_{\mu\nu}^{(si,TT,1)}(q,p,S_{TT},k'_\perp) = 0, \label{W1siTTint}\\
W_{\mu\nu}^{(LT,1)} & (q,p,S_{LT})=\int \frac{d^2k'_\perp}{(2\pi)^2} W_{\mu\nu}^{(si,LT,1)} (q,p,S_{LT},k'_\perp) \nonumber\\
&=\int \frac{d^2k'_\perp}{(2\pi)^2} \frac{2}{z}  \frac{k'_\perp \cdot S_{LT}}{M} 
  \Bigl[ \omega_{\mu\nu}^{(1)}(\hat t,k'_\perp)  \hat D_{LT}^{\perp}(z,k'_\perp) \nonumber \\  
& \phantom{XXXXXXXXX} - \tilde\omega_{\mu\nu}^{(1)}(\hat t,k'_\perp) \Delta \hat D_{LT}^\perp(z,k'_\perp)\Bigr] \nonumber\\
& +\int \frac{d^2k'_\perp}{(2\pi)^2} \frac{2}{z}M \Bigl[\omega^{(1)}_{\mu\nu}(\hat t, S_{LT}) \hat D_{LT}(z,k'_\perp) \nonumber \\  
& \phantom{XXXXXXXXX} - \tilde\omega^{(1)}_{\mu\nu}(\hat t, S_{LT})\Delta \hat D_{LT}(z,k'_\perp)\Bigr]. \label{W1siLTint}
\end{align}
We compare Eqs.~(\ref{W1siLLint})-(\ref{W1siLTint}) with the results given by Eq.~(112) in \cite{Wei:2013csa}, 
again we obtain the following relationships,  
\begin{align}
&D_{LT}(z) = 
\int \frac{d^2k_\perp}{(2\pi)^2} \frac{2}{z} \Bigl[ \hat D_{LT}(z,k_\perp) + \frac{{k}_\perp^2}{2M^2} \hat D_{LT}^\perp (z, k_\perp) \Bigr], \label{DLTint} \\ 
& \Delta D_{LT}^\perp(z)= 
\int \frac{d^2k_\perp}{(2\pi)^2} \frac{2}{z} \Bigl[ \Delta \hat D_{LT}(z,k_\perp) + \frac{{k}_\perp^2}{2M^2}\Delta \hat D_{LT}^\perp (z, k_\perp) \Bigr]. \label{dDLTint}
\end{align}

\section{The cross sections, azimuthal asymmetries and  polarizations}

By inserting the hadronic tensors into Eq. (\ref{csdef}), making the Lorentz contractions with the leptonic tensor, 
we obtain the differential cross section for the semi-inclusive production process $e^++e^-\to h+\bar q+X$. 
From the differential cross section, we can calculate the azimuthal asymmetries and the polarizations of the hadron produced.
We present the results for hadron with different spins separately in this section. 

\subsection{The cross section}

We have seen in Sec. III that the hadronic tensor can be expressed as a sum of 
a spin independent, a vector polarization dependent and a tensor polarization dependent part. 
By inserting them into Eq.~(\ref{csdef}), we obtain 
the cross section expressed in the same way, i.e.,  
\begin{align}
\frac{d\sigma^{(si)}}{d^3pd^2 k'_\perp}=\frac{d\sigma^{(si,unp)}}{d^3pd^2 k'_\perp}
+\frac{d\sigma^{(si,Vpol)}}{d^3pd^2 k'_\perp}+\frac{d\sigma^{(si,Tpol)}}{d^3pd^2 k'_\perp}.
\end{align}
Usually, it is convenient to  introduce a Lorentz boost invariant variable $y$  
defined as the light cone momentum fraction of electron in the collinear frame, i.e., 
$y\equiv l_1\cdot n / k \cdot n =zl_1^+/p^+$ so that
$l_1 = y p^+ \bar n/z + (1-y) z Q^2 n/(2p^+) + l_\perp$,
$l_\perp=(0,0,l_{\perp x},0)$, $|\vec l_\perp|=|l_{\perp x}|=\sqrt{y(1-y)}Q$.
 $y$ can be expressed in terms of the 
angle $\theta$ between the incident electron and the produced hadron, 
i.e. between $\vec l$ and $\vec p$, as $y = (1+ \cos \theta) /2$, 
in the $e^+e^-$ center of mass frame. 
In terms of $y$, $z$ and $k'_\perp$, we obtain, in the collinear frame,  
\begin{align}
\frac{d\sigma^{(si)}}{dydzd^2 k'_\perp}=\frac{d\sigma^{(si,unp)}}{dydzd^2 k'_\perp}
+\frac{d\sigma^{(si,Vpol)}}{dydzd^2 k'_\perp}+\frac{d\sigma^{(si,Tpol)}}{dydzd^2 k'_\perp}.
\end{align}
For spinless hadron $h$, we have only the unpolarized part,  
whereas for spin-$1/2$ hadron, we have the unpolarized and vector polarization dependent parts, 
and for spin-1 hadron, we have the unpolarized, 
the vector polarization and tensor polarization dependent parts, i.e.,
\begin{align}
&\frac{d\sigma^{(si,spin0)}}{dydzd^2 k'_\perp}=\frac{d\sigma^{(si,unp)}}{dydzd^2 k'_\perp}, \\
&\frac{d\sigma^{(si,spin1/2)}}{dydzd^2 k'_\perp}=\frac{d\sigma^{(si,unp)}}{dydzd^2 k'_\perp}
+\frac{d\sigma^{(si,Vpol)}}{dydzd^2 k'_\perp},\\
&\frac{d\sigma^{(si,spin1)}}{dydzd^2 k'_\perp}=\frac{d\sigma^{(si,unp)}}{dydzd^2 k'_\perp}
+\frac{d\sigma^{(si,Vpol)}}{dydzd^2 k'_\perp}+\frac{d\sigma^{(si,Tpol)}}{dydzd^2 k'_\perp}.
\end{align}
In the following, we  calculate these three parts separately.

\subsubsection{The unpolarized part}

By inserting the unpolarized part of the hadronic tensor given by  Eq. (\ref{Wsiunpfr}) into Eq. (\ref{csdef}), 
we obtain the unpolarized part of the differential cross section as, 
\begin{align}
&E_p \frac{d\sigma^{(si,unp)}}{d^3p d^2k'_\perp}=\frac{\alpha^2\chi}{2\pi^2zQ^4}\Big\{T_0^q (y) \hat D_1 (z, k'_\perp) +
\frac{4}{zQ^2} \cdot \nonumber\\ 
& ~~~~~ \times\big[T_2^q (y) l_\perp\cdot k'_\perp \hat D^\perp (z, k'_\perp)+ 
T_3^q (y) \epsilon_\perp^{l_\perp k'_\perp} \Delta \hat D^\perp (z, k'_\perp)\big] \Big\}. \label{csunp}
\end{align}
Here we use the same notations as those defined in \cite{Wei:2013csa}, i.e., 
$\alpha=e^2/4\pi$, $\chi=Q^4/[(Q^2-M_z^2)^2 + \Gamma_z^2 M_z^2] \sin^42\theta_W$,
and the coefficient functions are defined as,
\begin{align}
& T_0^q(y) = c_1^q c_1^e A(y)-c_3^q c_3^e B(y),\\
& T_2^q(y) = -c_3^q c_3^e + c_1^q c_1^e B(y),\\
& T_3^q(y) = c_1^q c_3^e - c_3^q c_1^e B(y),
\end{align}
where $A(y)=(1-y)^2+y^2$ and $B(y)=1-2y$.
In terms of the angle $\theta$ between the incident electron and the produced quark, 
$A(y)=(1+\cos^2\theta)/2$ and $B(y)=-\cos\theta$.

We can also express the differential cross section in terms of $z$ and $y$, and we have,
\begin{align}
&\frac{d\sigma^{(si,unp)}}{dydzd^2k'_\perp}=\frac{\alpha^2\chi}{2\pi Q^2}\Big\{T_0^q (y) \hat D_1 (z, k'_\perp) 
+ \frac{4}{zQ^2} \nonumber\\ 
& ~~~ \times\big[T_2^q (y) l_\perp\cdot k'_\perp \hat D^\perp (z, k'_\perp)+ 
T_3^q (y) \epsilon_\perp^{l_\perp k'_\perp} \Delta \hat D^\perp (z, k'_\perp)\big] \Big\}. \label{csunpvsyz}
\end{align}

From Eq.~(\ref{csunp}) or (\ref{csunpvsyz}), we see that for the semi-inclusive process $e^++e^- \to Z \to h+\bar q+X$,
for spinless hadron $h$, there is one leading twist term and two twist three terms. 
We also see that one of the two twist-3 terms is space reflection even and the other is odd.  
If we consider $e^+e^-$ annihilation via electromagnetic interaction, 
i.e., $e^++e^- \to \gamma^* \to h+\bar q+X$, we have $c_3^e = c_3^q =0$, 
so that $T_0^q(y) = A(y)$, $T_2^q(y) = B(y)$ and $T_3^q(y) = 0$.
In this case, the space-reflection odd term characterized by $\epsilon_\perp^{l_\perp k'_\perp}$ 
in Eq. (\ref{csunp}) or (\ref{csunpvsyz}) vanishes,  and we obtain, 
\begin{align}
 &\frac{d\sigma^{(si,unp,em)}}{dydz d^2k'_\perp} = \frac{\alpha^2 e_q^2}{2\pi Q^2}\Big\{ A(y) \hat D_1 (z, k'_\perp)  
+ \frac{4l_\perp\cdot k'_\perp}{zQ^2} B(y)  \hat D^\perp (z, k'_\perp) \Big\}.  \label{csunpem}
\end{align}
If we integrate over $d^2k'_\perp$, both of the twist-3 terms vanish.
The result reduces to exactly the same as we obtained in \cite{Wei:2013csa} for the inclusive process $e^++e^-\to h+X$.

\subsubsection{The vector polarization dependent part}

The vector polarization dependent part can be obtained by inserting the corresponding vector polarization 
dependent hadronic tensor given by Eq.~(\ref{WsiVpolfr}) into (\ref{csdef}). 
We also change the variables to $y,z,k'_\perp$, and obtain the result as given by, 
\begin{align}
 &\frac{d\sigma^{(si,Vpol)}}{dydz d^2k'_\perp}=\frac{\alpha^2\chi }{2\pi Q^2} 
 \Bigl\{  T_0^q (y) \frac{\epsilon_\perp^{k'_\perp S_\perp}}{M}  \hat D_{1T}^\perp (z,k'_\perp)  \nonumber\\ 
&+T_1^q (y) \big[ \lambda_h \Delta \hat D_{1L}(z,k'_\perp)  + \frac{k'_\perp\cdot S_\perp}{M} \Delta \hat D_{1T}^\perp(z,k'_\perp)\big]\nonumber\\
&+\frac{4\lambda_h}{zQ^2} \big[ T_2^q (y)  \epsilon_\perp^{l_\perp k'_\perp}  \hat D_L^\perp(z,k'_\perp) + 
T_3^q (y) l_\perp \cdot k'_\perp  \Delta \hat D_L^\perp(z,k'_\perp) \big] \nonumber\\
&+\frac{4\epsilon_\perp^{k'_\perp S_\perp} }{zMQ^2} \big[ T_2^q (y) l_\perp \cdot k'_\perp \hat D_T^\perp(z,k'_\perp) + 
T_3^q (y) \epsilon_\perp^{l_\perp k'_\perp} \Delta \hat D_T^\perp(z,k'_\perp)\big] \nonumber \\
&+\frac{4M}{zQ^2} \big[ T_2^q (y)  \epsilon_\perp^{l_\perp S_\perp}  \hat D_T(z,k'_\perp) + 
T_3^q (y) l_\perp \cdot S_\perp  \Delta \hat D_T(z,k'_\perp)\big] \Bigr\}. \label{csVpolvsyz}
\end{align}
Here, as in \cite{Wei:2013csa}, the new coefficient function is defined as, 
\begin{align}
T_1^q(y) = -c_3^q c_1^e A(y) + c_1^q c_3^e B(y).
\end{align}

We recall that~\cite{Wei:2013csa} $T_0^q(y)$ represents the relative weight 
for the production of quark of flavor $q$ at the $e^+e^-$ annihilation vertex,  
and $P_q(y)=T_1^q(y)/T_0^q(y)$ is the longitudinal polarization of that quark~\cite{Augustin:1978wf}. 
We see from Eq.~(\ref{csVpolvsyz}) that, at the leading twist, 
there exists two transverse polarization dependent terms and one longitudinal polarization dependent term. 
We see in particular that $\hat D_{1T}^\perp(z, k'_\perp)$ is nothing else but a counterpart of 
Sivers function \cite{Sivers:1989cc} in fragmentation function. 
The associated term is P-even and T-odd.
The other two leading twist terms are P-odd and T-even, hence they contribute only in the case of weak interaction.
Different spin dependent terms exist at twist 3.

If we integrate over $d^2k'_\perp$, all the $k'_\perp$-odd terms vanish and we obtain the 
corresponding cross section for the inclusive process $e^++e^-\to h+X$ as, 
\begin{align}
 & \frac{d\sigma(si,Vpol)}{dydz}=\frac{2\pi\alpha^2}{Q^2}\chi 
 \Bigl\{ T_1 (y) \lambda_h \int \frac{d^2 k'_\perp}{(2\pi)^2} \Delta \hat D_{1L} (z,k'_\perp) \nonumber\\
+& T_2(y) \frac{4M}{zQ^2}  \epsilon_\perp^{l_\perp S_\perp} \int \frac{d^2 k'_\perp}{(2\pi)^2} 
\Bigl[ \hat D_T (z,k'_\perp) 
+ \frac{k'^2_\perp}{2 M^2} \hat D_T^\perp (z,k'_\perp) \Bigr] \nonumber\\
+& T_3(y) \frac{4M}{zQ^2}  l_\perp \cdot S_\perp \int \frac{d^2k'_\perp}{(2\pi)^2} 
\Bigl[\Delta \hat D_T (z,k'_\perp) 
- \frac{k'^2_\perp}{2 M^2} \Delta \hat D_T^\perp (z,k'_\perp) \Bigr] \Bigr\}. \label{csVpolint}
\end{align}
Using the relationships given by Eqs.(\ref{DTint}) and (\ref{dDTint}), 
we see that this is just the corresponding inclusive cross section obtained in \cite{Wei:2013csa}. 

If we consider the $e^+e^-$-annihilation via electromagnetic interaction, we have,
\begin{align}
&\frac{d\sigma^{(si,Vpol,em)}}{dydz d^2k'_\perp}=\frac{\alpha^2 e_q^2 }{2\pi Q^2} 
 \Bigl\{  A (y) \frac{\epsilon_\perp^{k'_\perp S_\perp}}{M}  \hat D_{1T}^\perp (z,k'_\perp)   \nonumber\\ 
&~~~~~~~+\frac{4B(y)}{zMQ^2} \big[  \lambda_h M\epsilon_\perp^{l_\perp k'_\perp}  \hat D_L^\perp(z,k'_\perp) \nonumber\\ 
&~~~~~~~+\epsilon_\perp^{k'_\perp S_\perp}  l_\perp \cdot k'_\perp \hat D_T^\perp(z,k'_\perp) 
+M^2  \epsilon_\perp^{l_\perp S_\perp}  \hat D_T(z,k'_\perp) \big] \Bigr\}. \label{csVpolem}
\end{align}
We see that all the space reflection odd terms vanish since they are parity violating.
If we integrate over $d^2k'_\perp$, we have,
\begin{align}
&\frac{d\sigma^{(si,Vpol,em)}}{dy dz}=\frac{\alpha^2 e_q^2 }{2\pi Q^2} \frac{4M}{zQ^2} 
 B(y)  \epsilon_\perp^{l_\perp S_\perp}  D_T (z), \label{csVpolemint}
\end{align}
which is also the same as that obtained in \cite{Wei:2013csa}. 

\subsubsection{The tensor polarization dependent part}

For the tensor polarization dependent part, we express it as a sum of 
the $S_{LL}$, $S_{LT}$ and $S_{TT}$ dependent parts, i.e., 
\begin{equation}
\frac{d\sigma^{(si,Tpol)}}{dydz d^2k'_\perp}=\frac{d\sigma^{(si,LL)}}{dydz d^2k'_\perp}+
\frac{d\sigma^{(si,LT)}}{dydz d^2k'_\perp}+\frac{d\sigma^{(si,TT)}}{dydz d^2k'_\perp}.\label{csTpol}
\end{equation}
Up to twist-3, each part can be obtained by inserting the corresponding 
leading twist hadronic tensors Eqs.~(\ref{W0siLL}-\ref{W0siTT}) and the twist 3 parts given by  
Eqs.~(\ref{W1siLL}-\ref{W1siTT}) into Eq. (\ref{csdef}). 
In terms of the variable $y,z,k'_\perp$, we obtain the differential cross section at twist-3 level,
\begin{widetext}
\begin{align}
\frac{d\sigma^{(si,LL)}}{dydz d^2k'_\perp}=\frac{\alpha^2\chi}{2\pi Q^2} S_{LL} &
\Big\{ T_0^q (y)\hat D_{1LL}(z,k'_\perp)+ \frac{4}{zQ^2} \big[ T_2^q (y) (l_\perp\cdot k'_\perp) \hat D_{LL}^\perp(z,k'_\perp)
+ T_3^q (y) \epsilon_{\perp }^{l_\perp {k'_\perp}} \Delta \hat D_{LL}^\perp(z,k'_\perp) \big]\Big\}, \label{csTpol1LL}\\
 \frac{d\sigma^{(si,LT)}}{dydz d^2k'_\perp}=\frac{\alpha^2\chi}{2\pi Q^2} S_{LT}^\alpha &
\Big\{  T_0^q (y)  \frac{k'_{\perp\alpha}}{M}  \hat D_{1LT}^\perp(z,k'_\perp) + 
T_1^q (y) \frac{\epsilon_{\perp k' \alpha}}{M}  \Delta \hat D_{1LT}^\perp(z,k'_\perp)\nonumber\\
&+\frac{4}{zQ^2} T_2^q (y) \big[ (l_\perp\cdot k'_\perp)\frac{k'_{\perp\alpha} }{M}  \hat D_{LT}^{\perp}(z,k'_\perp)
+ M l_{\perp\alpha} \hat D_{LT}(z,k'_\perp) \big] \nonumber\\
&+ \frac{4}{zQ^2} T_3^q (y) \big[ \epsilon_{\perp}^{ l_\perp k'_\perp} \frac{k'_{\perp\alpha}}{M}  \Delta \hat D_{LT}^\perp(z,k'_\perp) 
+M \epsilon_{\perp l \alpha} \Delta \hat D_{LT}(z,k'_\perp) \big] \Big\},\label{csTpol1LT}\\
\frac{d\sigma^{(si,TT)}}{dydz d^2k'_\perp} = \frac{\alpha^2\chi}{2\pi Q^2}  S_{TT}^{\alpha\beta} & 
\Big\{ T_0^q (y) \frac{k'_{\perp\alpha} k'_{\perp\beta}}{M^2} \hat D_{1TT}^\perp(z,k'_\perp) 
+ T_1^q (y) \frac{\epsilon_{\perp k' \alpha} k'_{\perp\beta}}{M^2} \Delta \hat D_{1TT}^\perp(z,k'_\perp)\nonumber\\
&+ \frac{4}{zQ^2} T_2^q (y) \big[ 
 (l_\perp\cdot k'_\perp) \frac{k'_{\perp\alpha} k'_{\perp\beta}}{M^2}  \hat D_{TT}^{\perp C}(z,k'_\perp)
+ l_{\perp\alpha} k'_{\perp\beta} \hat D_{TT}^{\perp A}(z,k'_\perp) \big]\nonumber\\
&+ \frac{4}{zQ^2}T_3^q (y)\big[ 
\epsilon_{\perp}^{ l_\perp {k'_\perp}} \frac{k'_{\perp\alpha} k'_{\perp\beta}}{M^2}  \Delta \hat D_{TT}^{\perp C}(z,k'_\perp)
+\epsilon_{\perp l \alpha}  k'_{\perp\beta} \Delta \hat D_{TT}^{\perp A}(z,k'_\perp) \big]  \Big\}.\label{csTpol1TT}
\end{align}
We note that we have a leading twist quark polarization independent term 
[with the same coefficient function $T_0^q(y)$] in each part. 
We also have a leading twist but quark longitudinal polarization $P_q(y)=T_1^q(y)/T_0^q(y)$ 
dependent term for $S_{LT}$ and $S_{TT}$ dependent parts. 
These two terms are both P-odd and T-odd.
We also have two twist 3 terms in the $S_{LL}$ dependent part 
and four terms for the $S_{LT}$ or $S_{TT}$ dependent part.
Here all the terms that contain the four dimensional Levi-Civita tensor are P-odd and T-odd.
The others are P-even and T-even.

By integrating over $d^2k'_\perp$ and by using the relationships given by Eqs.(\ref{DLTint}) and (\ref{dDLTint}), 
we obtain the corresponding results for the inclusive reaction $e^++e^-\to h+X$ as given by Eq.(148) in \cite{Wei:2013csa}. 

For electromagnetic process, up to twist-3, the cross section is given by, 
\begin{align}
 \frac{d\sigma^{(si,LL,em)}}{dydz d^2k'_\perp}& =\frac{\alpha^2e_q^2}{2\pi Q^2} S_{LL}
\Big\{ A(y) \hat D_{1LL}(z,k'_\perp)+ 
\frac{4 B(y) }{zQ^2}(l_\perp\cdot k'_\perp) \hat D_{LL}^\perp(z,k'_\perp) \Big\}, \label{wei:csLLem} \\
 \frac{d\sigma^{(si,LT,em)}}{dydz d^2k'_\perp}& =\frac{\alpha^2e_q^2}{2\pi Q^2}S_{LT}^\alpha
\Big\{  A(y) \frac{k'_{\perp\alpha}}{M}  \hat D_{1LT}^\perp(z,k'_\perp) 
 +\frac{4B(y)}{zQ^2} \big[ \frac{(l_\perp\cdot k'_\perp)k'_{\perp\alpha}}{M}  \hat D_{LT}^{\perp}(z,k'_\perp)
 + M l_{\perp\alpha} \hat D_{LT}(z,k'_\perp) \big] \Big\},\label{wei:csLTem}\\
\frac{d\sigma^{(si,TT,em)}}{dydz d^2k'_\perp}& = \frac{\alpha^2e_q^2}{2\pi Q^2} S_{TT}^{\alpha\beta} 
\Big\{  A(y) \frac{k'_{\perp\alpha}k'_{\perp\beta}}{M^2} \hat D_{1TT}^\perp(z,k'_\perp) +  
\frac{4B(y) }{zQ^2} \big[   \frac{(l_{\perp}\cdot k'_{\perp}) k'_{\perp\alpha} k'_{\perp\beta}}{M^2}  \hat D_{TT}^{\perp C}(z,k'_\perp)
+  l_{\perp\alpha}k'_{\perp\beta} \hat D_{TT}^{\perp A}(z,k'_\perp)  \big]  \Big\},\label{wei:csTTem}
\end{align}
\end{widetext}
all the space reflection odd terms disappear here because of parity conservation.

\subsubsection{Transforming to the jet frame}

\begin{figure}[h!]
\includegraphics[width=0.9\linewidth]{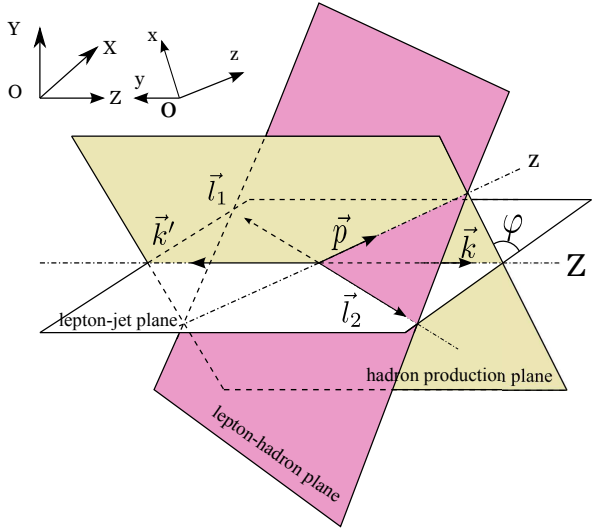}
\caption{(color online) Illustrating diagram showing the relationship between the collinear and the jet frames.} \label{fig:2frame}
\end{figure}

As we mentioned earlier, all the results presented in last sections are expressed in 
the hadron's collinear frame $o$-$xyz$ where the $z$-axis is taken along the hadron momentum direction. 
However, in experiments, one usually takes the ``jet frame'' that we denote as the $o$-$XYZ$ frame, 
where the jet direction is taken as the $Z$-direction, 
the lepton-jet plane is taken as $XZ$-plane and the $X$ direction is chosen 
so that the $X$-component of $\vec l$ is positive.   
In this reference frame, the transverse components of $k$ and $k'$ are zero 
while the  the momentum of the hadron has a transverse component $\vec p_T$. 
To distinguish them from each other, we use the notations as summarized here. 
In the collinear frame, we use, 
\begin{align}
p&=(E_h,\vec 0_\perp,p_z), \\ 
l_1&=(E,\vec l_\perp,l_z), \ \ \ \vec l_\perp=(l_x,0) = (|\vec l|\sin\theta, 0),\\
l_2&=(E,-\vec l_\perp,-l_z),\\
k&=(E,\vec k_\perp,k_z), \ \ \ \  \vec k_\perp=|\vec k_\perp|(-\cos\phi,-\sin\phi),\\
k'&=(E,-\vec k_\perp,-k_z),\\
S&=(\frac{p_z}{M} \lambda_h,\vec S_\perp,\frac{E_h}{M}\lambda_h), \ \ \ \ \vec S_\perp=|\vec S_\perp|(\cos\phi_s,\sin\phi_s),
\end{align}
while in the jet frame, we use, 
\begin{align}
p&=(E_h,\vec p_T,p_Z), \ \ \ \ \vec p_T=|\vec p_T|(\cos\varphi,\sin\varphi),\\ 
l_1&=(E,\vec l_T,l_Z), \ \ \ \vec l_T=(l_X,0),\\
l_2&=(E,-\vec l_T,-l_Z),\\
k&=(E,\vec 0_T,E), \\
k'&=(E,\vec 0_T,-E),\\
\vec{P}_h &=(\vec P_{hT}, P_{hZ}), \ \ \ \ \vec P_{hT}=|\vec P_{hT}|(\cos\varphi_s,\sin\varphi_s).
\end{align}
We show the relationship between the two frames in the illustrating diagram Fig.\ref{fig:2frame}.

A three vector expressed in these two coordinate systems are related to each other by a rotation matrix 
that we denote by ${\mathcal R}$, i.e., $\vec A_{xyz}={\mathcal R} \vec A_{XYZ}$. 
The rotation consists of two steps, $\mathcal{R} = \mathcal{R}_2 \mathcal{R}_1$. 
Here, $\mathcal{R}_1$ is a rotation of angle $\theta_h$ around the normal direction $\vec e_n$ 
of the hadron production plane, 
where $\theta_h$ is the angle between $\vec{p}$ and $\vec{k}$, i.e. $\sin\theta_h=2|\vec k_\perp|/Q$. 
This rotation turns $\vec{k}$ to be coincide with $z$ axis. 
$\mathcal{R}_2$ represents a rotation around $z$ axis that turns $\vec{l}$ into $xoz$-plane.
We should note that both $\mathcal{R}_1$ and $\mathcal{R}_2$ are small rotations 
in the sense that they differ from the unit matrix only by power suppressed factors. 
In fact, if we write $\mathcal{R}_i=1+\delta\mathcal{R}_i$ ($i$=1 and 2), 
we have, up to $1/Q$, that, 
\begin{align}
\delta\mathcal{R}_1\approx \sin \theta_h \left(
\begin{array}{ccc}
 0  & 0   & -  \cos \varphi\\
 0 & 0 & -\sin\varphi \\
 \cos\varphi & \sin\varphi & 0 \\
\end{array}
\right),
\end{align}
\begin{align}
\delta\mathcal{R}_2\approx
\sin\theta_h\left(
\begin{array}{ccc}
 0 & - \cot\theta_{lk}\sin\varphi & 0 \\
 \cot\theta_{lk}\sin\varphi & 0 \\
 0 & 0 & 0 \\
\end{array}
\right)\label{wei:rotation2}
\end{align}
where the leading terms for $\cot\theta_{lk}$ is $\cot\theta_{lk}=(2y-1)/2\sqrt{y(1-y)}+\cdots$.
If we write $\mathcal{R}=1+\delta\mathcal{R}$, we have $\delta\mathcal{R}\approx\delta\mathcal{R}_1+\delta\mathcal{R}_2$. 

We note that the second rotation $\mathcal{R}_2$ does not change the scalar products of 
the transverse components of the momenta involved here. 
It can be shown that $|\vec{k}'_\perp| = |\vec{p}_{T}/z|$.  
The transverse components of $\vec l$ and $\vec S$ in the two frames are not equal 
$\vec{l}_\perp \neq \vec{l}_T$, $\vec{S}_\perp \neq \vec{S}_T$.  
However, the differences are only at the power suppressed level. 
If we stay at the twist-3 level, i.e. we neglect terms suppressed by $1/Q^2$, 
we need to consider, 
\begin{align}
&d^2k'_\perp=d^2p_T/z^2, \label{wei:rep1}\\ 
&l_\perp \cdot k'_\perp = \frac{1}{z}  l_T \cdot p_T + \cdots , \\
&l_\perp \cdot S_\perp = l_T\cdot S_T- \lambda_h \frac{l_T\cdot p_T}{M} + \cdots,\\
&k'_\perp \cdot S_\perp =  \frac{1}{z} p_T \cdot S_T - \lambda_h \frac{p_T^2}{zM} + \cdots, \\
&\epsilon_\perp^{l_\perp k'_\perp} = \frac{1}{z} \epsilon_\perp^{l_T p_T}, \ \ \ \ \ \ 
\epsilon_\perp^{k'_\perp S_\perp} = \frac{1}{z} \epsilon_\perp^{p_T S_T}, \\
&\epsilon_\perp^{l_\perp S_\perp} = \epsilon_\perp^{l_T S_T} - \frac{\lambda_h}{M} \epsilon_\perp^{l_T p_T} + \cdots,\label{wei:rep2}
\end{align}
where the dots denote power suppressed terms.

The different components of the fragmentation function are all scalar functions 
of $z$ and $|\vec k'_\perp|^2$ with the normalization such as that given by Eq. (\ref{D1normalization}). 
Since $|\vec p_T|=z|\vec k'_\perp|$, 
if we change the transverse variable to $p_T$, 
we need to re-define the fragmentation function as, 
\begin{align}
 D (z, p_T)  = \frac{1}{z^2} \hat D (z, p_T/z) = \frac{1}{z^2} \hat D (z, k'_\perp),
\end{align}
so that we have 
\begin{align}
\sum_h \int z D^{q\to h} (z, p_T) dz \frac{d^2p_T}{(2\pi)^2} =1.
\end{align}

If we consider only the contributions up to $1/Q$, we can simply make the replacements     
given by Eqs.(\ref{wei:rep1}-\ref{wei:rep2}) in 
Eqs.(\ref{csunpvsyz}), (\ref{csVpolvsyz}), and (\ref{csTpol1LL}-{\ref{csTpol1TT}) to obtain 
the differential cross sections in the jet frame. 
We summarize the results here in the following,
\begin{widetext}
\begin{align}
\frac{d\sigma^{(si,unp)}}{dzdyd^2p_T}=\frac{\alpha^2\chi}{2\pi Q^2} &
\Big\{T_0^q (y) D_1 (z,p_T) +\frac{4}{z^2Q^2}\big[T_2^q (y) l_T\cdot p_T D^\perp (z,p_T)+
T_3^q (y) \epsilon_\perp^{l_T p_T} \Delta D^\perp (z,p_T)\big] \Big\}, \label{csunpjet} \\
\frac{d\sigma^{(si,Vpol)}}{dzdy d^2p_T}=\frac{\alpha^2\chi }{2\pi Q^2} &
 \Bigl\{ T_0^q (y) \frac{\epsilon_\perp^{p_T S_T}}{zM}  D_{1T}^\perp (z,p_T) +  
 T_1^q (y) \big[ \lambda_h \Delta D_{1L}(z,p_T)  + \frac{1}{z} ( \frac{p_T\cdot S_T}{M} - \lambda_h \frac{p_T^2}{M^2}) \Delta D_{1T}^\perp(z,p_T)\big]\nonumber\\
&+\frac{4\lambda_h}{z^2Q^2} \big[ T_2^q (y)  \epsilon_\perp^{l_T p_T}  D_L^\perp(z,p_T) + 
T_3^q (y) l_T \cdot p_T  \Delta D_L^\perp(z,p_T) \big]\nonumber\\
&+\frac{4\epsilon_\perp^{p_T S_T} }{z^3MQ^2} \big[ T_2^q (y) l_T \cdot p_T D_T^\perp(z,p_T) + 
T_3^q (y) \epsilon_\perp^{l_\perp p_T} \Delta D_T^\perp(z,p_T)\big] \nonumber \\
&+\frac{4M}{zQ^2} \big[ T_2^q (y)  (\epsilon_\perp^{l_T S_T} - \frac{\lambda_h}{M} \epsilon_\perp^{l_T p_T}) D_T(z,p_T) + 
T_3^q (y) l_T \cdot (S_T - \frac{\lambda_h}{M} p_T)  \Delta D_T(z,p_T)\big] \Bigr\}, \label{csVpoljet}\\
\frac{d\sigma^{(si,LL)}}{dydz d^2p_T}=\frac{\alpha^2\chi}{2\pi Q^2} & S_{LL}
\Big\{ T_0^q (y)D_{1LL}(z,p_T)+ \frac{4}{z^2Q^2} \big[ T_2^q (y) (l_T\cdot p_T) D_{LL}^\perp(z,p_T)
+ T_3^q (y) \epsilon_{\perp}^{ l_T {p_T}} \Delta D_{LL}^\perp(z,p_T) \big]\Big\}. \label{csLLjet}
\end{align}
\end{widetext}
Because the representation and/or physical meaning of $S_{LT}$ and $S_{TT}$ in the jet frame are rather complicated, 
the transformation of the corresponding components to jet frame does not bring much so that 
we do not present the results here. We will come back to this point when presenting the results for polarization 
in next subsection.  

If we consider the $e^+e^-$-annihilation via electromagnetic interaction, we have,
\begin{align}
\frac{d\sigma^{(si,unp,em)}}{dzdyd^2p_T}&=\frac{\alpha^2 e_q^2}{2\pi Q^2} 
\Big\{A(y) D_1 (z,p_T) +\nonumber\\ 
&+\frac{4}{z^2Q^2} B(y) l_T\cdot p_T D^\perp (z,p_T) \Big\},\label{csunpjetem} \\
\frac{d\sigma^{(si,Vpol,em)}}{dzdy d^2p_T}&= \frac{\alpha^2 e_q^2 }{2\pi Q^2} 
 \Bigl\{A(y) \frac{\epsilon_\perp^{p_T S_T}}{zM}  D_{1T}^\perp (z,p_T)  + \nonumber\\
+\frac{4}{z^2Q^2} &B(y) \Big[ \lambda_h\epsilon_\perp^{l_T p_T}  D_L^\perp(z,p_T) 
+\frac{\epsilon_\perp^{p_T S_T} }{zM}  l_T \cdot p_T D_T^\perp(z,p_T) \nonumber\\
+z (M \epsilon&_\perp^{l_T S_T} - \lambda_h \epsilon_\perp^{l_T p_T})  D_T(z,p_T) \Big] \Bigr\}, \label{csVpoljetem}\\
\frac{d\sigma^{(si,LL,em)}}{dydz d^2p_T}&=\frac{\alpha^2 e_q^2 }{2\pi Q^2}  S_{LL}
\Big\{ A (y)D_{1LL}(z,p_T)+ \nonumber\\
&+\frac{4 B(y)}{z^2Q^2} (l_T\cdot p_T) D_{LL}^\perp(z,p_T)\Big\}. \label{csLLjetem}
%
%
\end{align}

In terms of the azimuthal angle $\varphi$ and $\varphi_s$ that are defined 
as the azimuthal angle of the hadron momentum $\vec p$ 
and spin $\vec S$ respectively, we have, 
\begin{align}
&l_T\cdot p_T=-|\vec l_T||\vec p_T|\cos\varphi, \\
&\epsilon_{\perp}^{l_{T} p_{T}}=|\vec l_T||\vec p_T|\sin\varphi, \\
&l_T\cdot S_T - \lambda_h \frac{l_T\cdot p_T}{M}=-|\vec l_T||\vec P_{hT}|\cos\varphi_s, \\
&\epsilon_{\perp}^{l_{T} S_{T}} - \frac{\lambda_h}{M} \epsilon_\perp^{l_T p_T} =|\vec l_T||\vec P_{hT}|\sin\varphi_s,   \\
&p_T\cdot S_T - \lambda_h \frac{p_T^2}{M}=-|\vec p_T||\vec P_{hT}|\cos(\varphi_s-\varphi), \\
&\epsilon_{\perp}^{p_{T} S_{T}}=|\vec p_T||\vec P_{hT}|\sin(\varphi_s-\varphi),
\end{align}
where $|\vec l_T|= Q \sqrt{y(1-y)} +\cdots$. 
We express the unpolarized, the vector polarization and the $S_{LL}$ dependent parts 
of the differential cross sections in terms of these azimuthal angles. 
For $e^++e^-\to Z\to h+\bar q+X$, we have, 
\begin{widetext}
\begin{align}
\frac{d\sigma^{(si,unp)}}{dzdyd^2p_T}=\frac{\alpha^2\chi}{2\pi Q^2} &
\Big\{T_0^q (y) D_1 (z,p_T) - \frac{4|\vec p_T|}{z^2Q}\big[\tilde T_2^q (y)  D^\perp(z,p_T) \cos\varphi
 -\tilde T_3^q (y) \Delta D^\perp (z,p_T)\sin\varphi\big] \Big\},\label{csunpjetphi} \\
\frac{d\sigma^{(si,Vpol)}}{dzdy d^2p_T}=\frac{\alpha^2\chi }{2\pi Q^2} &
 \Bigl\{ T_0^q (y) \frac{|\vec p_T| }{zM} |\vec P_{hT}| D_{1T}^\perp (z,p_T) \sin (\varphi_s - \varphi) +  
 T_1^q (y) \big[ \lambda_h \Delta D_{1L}(z,p_T)  - \frac{|\vec p_T| }{zM} |\vec P_{hT}|\Delta D_{1T}^\perp(z,p_T) \cos (\varphi_s - \varphi)\big]\nonumber\\
&+\frac{4 |\vec p_T| }{z^2Q} \lambda_h\big[\tilde  T_2^q (y) D_L^\perp(z,p_T) \sin\varphi 
 - \tilde T_3^q (y) \Delta D_L^\perp(z,p_T) \cos\varphi \big] \nonumber\\
&-\frac{4 |\vec p_T|^2  }{z^3 MQ} |\vec P_{hT}|\big[ \tilde T_2^q (y)  D_T^\perp(z,p_T) \cos\varphi
-\tilde  T_3^q (y) \Delta D_T^\perp(z,p_T) \sin\varphi \big] \sin (\varphi_s - \varphi) \nonumber \\
&+\frac{4M  }{zQ} |\vec P_{hT}|\big[ \tilde T_2^q (y) D_T(z,p_T) \sin\varphi_s 
- \tilde T_3^q (y) \Delta D_T(z,p_T) \cos\varphi_s\big] \Bigr\}, \label{csVpoljetphi}\\
\frac{d\sigma^{(si,LL)}}{dydz d^2p_T}=\frac{\alpha^2\chi}{2\pi Q^2} S_{LL} &
\Big\{ T_0^q (y) D_{1LL}(z,p_T)- \frac{4 |\vec p_T|}{z^2Q}\big[\tilde T_2^q (y) D_{LL}^\perp(z,p_T) \cos\varphi
- \tilde T_3^q (y) \Delta D_{LL}^\perp(z,p_T) \sin\varphi \big]\Big\}. \label{csLLjetphi}
\end{align}
For electromagnetic process, $e^++e^-\to \gamma^* \to h+\bar q+X$, we have,
\begin{align}
\frac{d\sigma^{(si,unp,em)}}{dzdyd^2p_T}=&\frac{\alpha^2e_q^2}{2\pi Q^2} 
\Big\{A(y) D_1 (z,p_T) - \frac{4 |\vec p_T|}{z^2Q} \tilde B(y) D^\perp (z,p_T) \cos\varphi \Big\},\label{csunpjetemphi} \\
\frac{d\sigma^{(si,Vpol,em)}}{dzdy d^2p_T}=&\frac{\alpha^2 e_q^2 }{2\pi Q^2} 
 \Bigl\{A(y) \frac{|\vec p_T|}{zM}  |\vec P_{hT}|D_{1T}^\perp (z,p_T) \sin (\varphi_s - \varphi) \nonumber\\
 +&\tilde B(y) \frac{4M}{zQ}  \Bigl[ \lambda_h \frac{|\vec p_T|}{zM} D_L^\perp(z,p_T) \sin\varphi 
-\frac{|p_T|^2}{z^2M^2} |\vec P_{hT}| D_T^\perp(z,p_T) \sin (\varphi_s - \varphi) \cos \varphi 
+ |\vec P_{hT}| D_T(z,p_T) \sin\varphi_s \Big] \Bigr\}, \label{csVpoljetem}\\
\frac{d\sigma^{(si,LL,em)}}{dydz d^2p_T}=&\frac{\alpha^2 e_q^2}{2\pi Q^2}  S_{LL}
\Big\{ A (y)D_{1LL}(z,p_T) - \frac{4|\vec p_T| }{z^2Q}\tilde B(y) D_{LL}^\perp(z,p_T) \cos\varphi \Big\}, \label{csLLjetemphi}
\end{align}
\end{widetext}
where $\tilde T_i^q(y) = \sqrt{y(1-y)} T_i^q(y)$, and $\tilde B(y) =\sqrt{y(1-y)} B(y)$. 
The azimuthal symmetries with respect to these two angles are in principle measurable 
and we will discuss more in the following.

\begin{figure}[h!]
\includegraphics[width=0.96\linewidth]{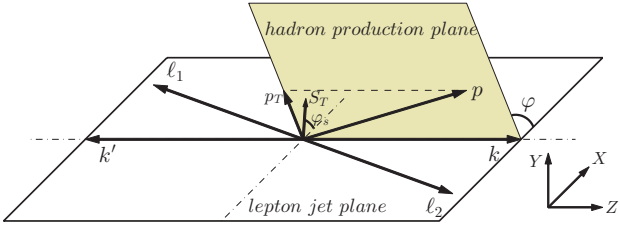}
\caption{(color online) Illustrating diagram showing the definition of the azimuthal angle $\varphi$ and $\varphi_s$ in 
the jet frame.} \label{fig:kin}
\end{figure}


\subsection{Azimuthal asymmetries}

From Eq.~(\ref{csunpjetphi}), 
we see that even for spinless hadrons or in the unpolarized case, 
there exist two independent azimuthal asymmetries at twist-3 level. 
These asymmetries can be defined as the average values of $\cos\varphi$ or $\sin\varphi$, i.e., 
\begin{align}
A^{\cos\varphi}_{unp}(z,y,p_T)\equiv  \frac{\int d\varphi  \cos\varphi\frac{d\sigma}{dzdyd^2p_T}(z,y,p_T,\varphi)}     
    {\int d\varphi\frac{d\sigma}{dzdyd^2p_T}(z,y,p_T,\varphi)}, \\
A^{\sin\varphi}_{unp}(z,y,p_T)\equiv  \frac{\int d\varphi  \sin\varphi\frac{d\sigma}{dzdyd^2p_T}(z,y,p_T,\varphi)}     
    {\int d\varphi\frac{d\sigma}{dzdyd^2p_T}(z,y,p_T,\varphi)}. 
\end{align}
They can also be measured by,  
\begin{align}
A^{\cos\varphi}_{unp}(z,y,p_T) =
\frac{\frac{d\sigma}{dzdyd^2p_T}(z,y,p_T,\varphi)-\frac{d\sigma}{dzdyd^2p_T}(z,y,p_T,\pi-\varphi)}
       {\frac{2\cos\varphi}{\pi}\int d\varphi\frac{d\sigma}{dzdyd^2p_T}(z,y,p_T,\varphi)}, \\
A^{\sin\varphi}_{unp}(z,y,p_T) = 
\frac{\frac{d\sigma}{dzdyd^2p_T}(z,y,p_T,\varphi)-\frac{d\sigma}{dzdyd^2p_T}(z,y,p_T,\pi+\varphi)}
       {\frac{2\sin\varphi}{\pi}\int d\varphi\frac{d\sigma}{dzdyd^2p_T}(z,y,p_T,\varphi)}. 
\end{align}
If we call $z$-direction as forward, the $x$-direction as left, these two asymmetries can be called 
left-right ($A^{\cos\varphi}_{unp}$) and up-down $A^{\sin\varphi}_{unp}$ asymmetries respectively. 
From Eq. (\ref{csunpjetphi}), we obtain, 
\begin{align}
A^{\cos\varphi}_{unp}(z,y,p_T)=
& -\frac{2 |\vec p_T|}{z^2 Q} \frac{\sum_q \tilde T_2^q (y) D^{\perp q\to h}(z,p_T)}{\sum_q T_0^q (y) D_1^{q\to h} (z,p_T)},\\
A^{\sin\varphi}_{unp}(z,y,p_T)=
& \frac{2 |\vec p_T|}{z^2 Q} \frac{\sum_q \tilde T_3^q (y) \Delta D^{\perp q\to h}(z,p_T)}{\sum_q T_0^q (y) D_1^{q\to h}(z,p_T)}.
\end{align}
If we consider $e^++e^- \to \gamma^* \to h+\bar q+X$, we have, 
\begin{align}
&A^{\cos\varphi}_{unp,em}(z,y,p_T)= -\frac{2 |\vec{{p}}_T|}{z^2 Q} 
 \frac{\tilde B(y) \sum_q e_q^2 D^{\perp q\to h}(z,p_T)}{A(y) \sum_q e_q^2 D_1^{q\to h}(z,p_T)}, \\
&A^{\sin\varphi}_{unp,em}(z,y,p_T)=0. 
\end{align}
We see that only $\cos\varphi$-asymmetry (the left-right asymmetry) can exist while the $\sin\varphi$-asymmetry 
(the up-down asymmetry) violates parity conservation and vanishes in the electromagnetic process. 

From Eq.~(\ref{csVpoljetphi}), we see also different azimuthal asymmetries in the polarized cases. 
At the leading twist, there are two azimuthal asymmetries,
\begin{align}
A&^{\cos(\varphi_s-\varphi)}(z,y,p_T)= - |\vec P_{hT}| \frac{|\vec p_T| }{2zM} \nonumber\\
&\times\frac{\sum_qT_0^q(y) P_q(y) \Delta D_{1T}^{\perp q\to h}(z,p_T)} 
{\sum_qT_0^q(y) \Bigl[D_{1}^{q\to h}(z,p_T)+\lambda_h P_q(y)\Delta D_{1L}^{q\to h}(z,p_T)\Bigr]},\\ 
A&^{\sin(\varphi_s-\varphi)}(z,y,p_T)= |\vec P_{hT}| \frac{|\vec p_T| }{2zM} \nonumber\\
&\times\frac{\sum_qT_0^q(y) \Delta D_{1T}^{\perp q\to h}(z,p_T)} 
{\sum_qT_0^q(y) \Bigl[D_{1}^{q\to h}(z,p_T)+\lambda_h P_q(y)\Delta D_{1L}^{q\to h}(z,p_T)\Bigr]}.
\end{align}
If we consider $e^++e^- \to \gamma^* \to h+\bar q+X$, we have, 
\begin{align}
&A_{em}^{\cos(\varphi_s-\varphi)}(z,y,p_T)= 0,\\ 
&A_{em}^{\sin(\varphi_s-\varphi)}(z,y,p_T)= |\vec P_{hT}| \frac{|\vec p_T| }{2zM} 
\frac{\sum_q e_q^2 \Delta D_{1T}^{\perp q\to h}(z,p_T)} {\sum_q e_q^2D_{1}^{q\to h}(z,p_T)}.
\end{align}

Up to twist-3, there exist also other azimuthal asymmetries in the polarized cases and 
they can easily be calculated from the differential cross section. 
These azimuthal asymmetries can be studied by measuring the corresponding hadrons with given spins. 
For example, from Eq.~(\ref{csLLjetphi}), we see that there is a $S_{LL}$ dependent term contributing to 
$\cos\varphi$ and one to $\sin\varphi$.
This means that considering the $\cos\varphi$ or $\sin\varphi$ asymmetry in 
$e^++e^-\to Z^0\to V+\bar q+X$, we have,
\begin{align}
A&^{\cos\varphi}_{V}(z,y,p_T)=-\frac{2 |\vec p_T|}{z Q} \nonumber\\
&\times \frac{\sum_q \tilde T_2^q (y) \Bigl[D^{\perp q\to V}(z,p_T)+S_{LL}D_{LL}^{\perp q\to V}(z,p_T)\Bigr]}
{\sum_q T_0^q (y) \Bigl[D_1^{q\to V} (z,p_T)+S_{LL}D_{1LL}^{\perp q\to V}(z,p_T)\Bigr]},\\
A&^{\sin\varphi}_{V}(z,y,p_T)=\frac{2 |\vec p_T|}{z Q}\nonumber\\
&\times \frac{\sum_q \tilde T_3^q (y) \Bigl[\Delta D^{\perp q\to V}(z,p_T)+S_{LL}\Delta D_{LL}^{\perp q\to V}(z,p_T)\Bigr]}{\sum_q T_0^q (y) \Bigl[D_1^{q\to V}(z,p_T)+S_{LL}D_{1LL}^{\perp q\to V}(z,p_T)\Bigr]}.
\end{align}
where $V$ denotes a vector meson such as $\rho$, $K^*$ and so on. 
If we consider $e^++e^- \to \gamma^* \to V+\bar q+X$, we have, 
\begin{align}
A^{\cos\varphi}_{V,em}&(z,y,p_T)=-\frac{2 |\vec p_T|}{z Q} \frac{\tilde B(y)}{A(y)}\nonumber\\
&\times \frac{\sum_q e_q^2 \Bigl[D^{\perp q\to V}(z,p_T)+S_{LL}D_{LL}^{\perp q\to V}(z,p_T)\Bigr]}
{\sum_q e_q^2\Bigl[D_1^{q\to V} (z,p_T)+S_{LL}D_{1LL}^{\perp q\to V}(z,p_T)\Bigr]},\\
A^{\sin\varphi}_{V,em}&(z,y,p_T)=0. 
\end{align}
Such studies are usually difficult but are helpful for deep understanding of fragmentation functions.

Here, in this subsection, in the expressions of the azimuthal asymmetries given above, 
we have written out the sum over the flavor of quarks 
and use the superscript $q\to h$ to denote the flavor dependence of the fragmentation functions explicitly. 
In the following of this paper, when presenting the expressions for polarizations, 
we will write out the sum over quark flavor explicitly to avoid the confusion that the weight factors $T_i^q(y)$ 
in the numerator and that in the denominator do not cancel with each other. 
However, we omit the superscript $q\to h$ in the fragmentation function for brevity. 
 
\subsection{Polarizations}

From Eq.~(\ref{csVpoljet}), we see that the polarizations can be measured 
in different directions and the results are different. 
We discuss them for spin-1/2 and spin-1 hadrons separately in this section.

\subsubsection{Spin-1/2 hadron} \label{wei:secpola} 

For spin-1/2 hadrons, the cross section is given by the unpolarized and the vector polarization dependent parts. 
The hadrons produced can possess a vector polarization. 
The vector polarization can be given by three components defined in different frame.  
The longitudinal component is usually defined with respect to helicity.
But there are still two directions to be chosen.
One can measure the transverse polarization, i.e., measure the polarization in the plane perpendicular to $\vec p$. 
This is the case if we study the polarization in the collinear frame. 
In this case, we describe the vector polarization of the spin-$1/2$ hadrons by 
the longitudinal component $P_{Lh}$ and two transverse components $P_{hx}$ and $P_{hy}$. 
In experiments, one may find it is easier to measure the polarization along $X$- and $Y$-direction in the jet frame. 
In this case, we describe the vector polarization of the spin-$1/2$ hadrons by 
the longitudinal component $P_{Lh}$ and other two components $P_{hX}$ and $P_{hY}$. 
We should note that these two directions $X$ and $Y$ are not perpendicular to the helicity direction. 
Nevertheless, we can use the three components to describe the polarization vector $\vec P_h$. 
We may also study the transverse polarizations in the normal direction $\vec e_n$ of the production plane 
and the transverse momentum $\vec e_t$ in the production plane. 
In this case, we describe the polarization vector $\vec P_h$ by the three components $P_{Lh}$, 
$P_{hn}=\vec e_n\cdot\vec P_h$ and $P_{ht}=\vec e_t\cdot\vec P_h$. 
From Eq.~(\ref{csVpoljet}), we see that the differential cross section 
contains terms proportional to $l_T\cdot S_T$, $\epsilon_{\perp}^{l_T S_{T}}$,
$p_T\cdot S_T$, and $\epsilon_{\perp}^{p_T S_{T}}$. 
This means that it might make sense to measure polarization with respect to $l_T$ (lepton-hadron plane) 
or $p_T$ (production plane or hadron-jet plane). 
We present the results in the three cases in the following. 

{\it In the helicity-collinear frame:} 
In this frame, we consider the longitudinal polarization $P_{Lh}$ defined with respect 
to the helicity and two transverse components $P_{hx}$ and $P_{hy}$ 
perpendicular to the helicity direction, i.e., in the $x$ and $y$ directions. 
The results can most convenient be obtained from the expression of the differential cross section 
in the collinear frame. 
From Eq. (\ref{csVpolvsyz}), we obtain, at the leading twist, 
\begin{align}
& P_{Lh}^{(0)} (y,z, k'_\perp) = 
\frac{\sum_q T_0^q (y) P_q(y)\Delta \hat D_{1L}(z,k'_\perp)}{\sum_q T_0^q(y) \hat D_1 (z,k'_\perp)},\\
& P_{hx}^{(0)} (y,z,k'_\perp) = - \frac{\sum_q T_0^q (y) \Bigl[ k'_y \hat D_{1T}^\perp(z,k'_\perp) + P_q (y) k'_x \Delta \hat D_{1T}^\perp (z, k'_\perp)\Bigr]}{M\sum_q T_0^q(y) \hat D_1 (z,k'_\perp)}, \\
& P_{hy}^{(0)} (y,z,k'_\perp) =  \frac{\sum_q T_0^q (y) \Bigl[k'_x \hat D_{1T}^\perp(z,k'_\perp) - P_q (y) k'_y \Delta \hat D_{1T}^\perp (z, k'_\perp)\Bigr]}{M\sum_q T_0^q(y) \hat D_1 (z,k'_\perp)}.
\end{align}
Here, the superscript $(0)$ is used to denote leading twist.  
We see that even at the leading twist, we have the longitudinal polarization proportional to the 
quark polarization $P_q(y)=T_1^q (y)/T_0^q (y)$ 
and transverse polarizations depending on the transverse direction of the hadron momentum. 

Up to twist 3, we have, 
\begin{align}
P_{Lh}&(y,z, k'_\perp) = P_{Lh}^{(0)} (y,z,k'_\perp)  \Bigl[ 1+ \frac{M}{Q}\hat\Delta(y,z,k'_\perp)\Bigr] \nonumber\\
&+\frac{4}{zQ}\frac{\sum_q\Bigl[\tilde T_2^q(y)k'_y \hat D_L^\perp(z,k'_\perp)-\tilde T_3^q(y)k'_x\Delta \hat D_L^\perp(z,k'_\perp)\Bigr]}
{\sum_q T_0^q(y)\hat D_1(z,k'_\perp)},\\
P_{hx} & (y,z, k'_\perp) = P_{hx}^{(0)} (y,z,k'_\perp)  \Bigl[ 1+ \frac{M}{Q}\hat\Delta(y,z,k'_\perp)\Bigr] \nonumber\\
& + \frac{4k'_y}{zQ} \frac{\sum_q \Bigl[\tilde T_2^q (y) k'_x \hat D_T^\perp (z, k'_\perp) - \tilde T_3^q (y) k'_y \Delta \hat D_T^\perp (z, k'_\perp)\Bigr]}{M\sum_q T_0^q(y) \hat D_1 (z,k'_\perp)} \nonumber\\
& - \frac{4M}{zQ} \frac{\sum_q \tilde T_3^q (y) \Delta \hat D_T (z,k'_\perp)}{\sum_q T_0^q(y) \hat D_1 (z,k'_\perp)}, \\
P_{hy} & (y,z, k'_\perp) = P_{hy}^{(0)} (y,z,k'_\perp)   \Bigl[ 1+ \frac{M}{Q}\hat\Delta(y,z,k'_\perp)\Bigr] \nonumber\\
& - \frac{4k'_x}{zQ} \frac{\sum_q \Bigl[\tilde T_2^q (y) k'_x \hat D_T^\perp (z, k'_\perp) - \tilde T_3^q (y) k'_y \Delta \hat D_T^\perp (z, k'_\perp)\Bigr]}{M\sum_q T_0^q(y) \hat D_1 (z,k'_\perp)}\nonumber\\
& + \frac{4M}{zQ} \frac{\sum_q \tilde T_2^q (y) \hat D_T (z,k'_\perp)}{\sum_q T_0^q(y) \hat D_1 (z,k'_\perp)}, 
\end{align}
where the incremental factor $\hat\Delta$ is due to the twist 3 contributions to the 
unpolarized cross section and is given by, 
\begin{align}
\hat\Delta(y,z,k'_\perp)=\frac{4\sum_q  \Bigl[\tilde T_2^q (y) k'_x \hat D^\perp (z,k'_\perp) 
- \tilde T_3^q (y) k'_y \Delta \hat D^\perp (z,k'_\perp)\Bigr]}{zM \sum_q T_0^q(y) \hat D_1(z,k'_\perp)}.
\end{align}
We see that in addition to the incrementation due to twist-3 contributions to 
the unpolarized cross section, there are twist-3 contributions to all the three components of the 
polarization due to spin-dependent twist 3 fragmentation functions. 
However, if we integrate over the transverse directions, we have, 
at the leading twist
\begin{align}
P_{Lh}^{(0)}&(y,z,|\vec k'_\perp|) = P_{Lh}^{(0)} (y,z,k'_\perp),\\
P_{hx}^{(0)} & (y,z, |\vec k'_\perp|) =  0, \\
P_{hy}^{(0)} & (y,z, |\vec k'_\perp|) = 0,
\end{align}
and up to twist 3, 
\begin{align}
P_{Lh}&(y,z,|\vec k'_\perp|) = P_{Lh}^{(0)} (y,z,k'_\perp),\\
P_{hx} & (y,z, |\vec k'_\perp|) = - \frac{4M}{zQ} 
\frac{\sum_q \tilde T_3^q (y) \Bigl[
\Delta \hat D_T (z,k'_\perp)-\frac{{k'}^2_\perp}{2M^2}\Delta \hat D_T^\perp (z, k'_\perp)\Bigr]} 
{\sum_q T_0^q(y) \hat D_1 (z,k'_\perp)}, \\
P_{hy} & (y,z, |\vec k'_\perp|) =  \frac{4M}{zQ} 
\frac{\sum_q \tilde T_2^q (y)  \Bigl[\hat D_T (z, k'_\perp)+\frac{{k'}^2_\perp}{2M^2}\hat D_T^\perp (z, k'_\perp)\Bigr]}
{\sum_q T_0^q(y) \hat D_1 (z,k'_\perp)}. 
\end{align}
We see that the transverse polarizations exist only at the twist 3 
if we integrate over the azimuthal angle of the produced hadron.

We can change the variable $k'_\perp$ to $p_T$ to obtain the polarization as function of 
the measurable quantities $z$, $y$ and $p_T$.
For this purpose, we need to find out the relationships between the two components of 
$k'_\perp$ and those of $p_T$. 
From Eq. (\ref{wei:rotation2}), we obtain that, 
\begin{align}
 zk'_x & = p_X  - \frac{(2y-1)p_Y^2}{zQ\sqrt{y(1-y)}}+\dots, \label{kperpvspTx}\\
 zk'_y & = p_Y + \frac{(2y-1)p_Xp_Y}{zQ\sqrt{y(1-y)}}+\dots,\label{kperpvspTy}
\end{align}
where the dots denote the even more power suppressed terms.
We see that, at the leading twist, we need only to replace $k'_\perp$ by $p_T/z$ and obtain,  
\begin{align}
&P_{Lh}^{(0)} (y,z, p_T) = \frac{\sum_q T_0^q (y) P_q(y) \Delta D_{1L}(z,p_T)}{\sum_q T_0^q(y) D_1 (z,p_T)},\label{Plh0}\\
&P_{hx}^{(0)} (y,z,p_T) = - \frac{\sum_q T_0^q (y) \Bigl[p_Y D_{1T}^\perp(z,p_T) + P_q (y) p_X \Delta D_{1T}^\perp (z, p_T)\Bigr]}{zM\sum_q T_0^q(y) D_1 (z,p_T)}, \label{Phx0}\\
&P_{hy}^{(0)} (y,z,p_T) =  \frac{\sum_q T_0^q (y) \Bigl[ p_X D_{1T}^\perp(z,p_T) - P_q (y) p_Y \Delta D_{1T}^\perp (z, p_T)\Bigr]}{zM\sum_q T_0^q(y) D_1 (z,p_T)}.\label{Phy0}
\end{align}
However, up to twist 3, we need to keep the next order of the power suppressed terms 
in Eqs.~(\ref{kperpvspTx}) and (\ref{kperpvspTy}) when substitute $k'_x$ and $k'_y$ in the 
leading twist contributions.
In this case, we obtain,
\begin{align}
P_{Lh}&(y,z, p_T) = P_{Lh}^{(0)} (y,z,p_T)  \Bigl[ 1+ \frac{M}{Q}\Delta(y,z,p_T)\Bigr] \nonumber\\
&+\frac{4}{z^2Q}\frac{\sum_q\Bigl[\tilde T_2^q(y)p_Y D_L^\perp(z,p_T)-\tilde T_3^q(y)p_X\Delta D_L^\perp(z,p_T)\Bigr]}
{\sum_q T_0^q(y)D_1(z,p_T)},\label{Plh}\\
P_{hx} & (y,z,p_T) = P_{hx}^{(0)} (y,z,p_T)  \Bigl[ 1+ \frac{M}{Q}\Delta(y,z,p_T)\Bigr] \nonumber\\
& - P_{hy}^{(0)} (y,z,p_T) \frac{(2y-1)p_Y}{zQ\sqrt{y(1-y)}} \nonumber \\
& + \frac{4p_Y}{z^3Q} 
\frac{\sum_q \Bigl[\tilde T_2^q (y) p_X D_T^\perp (z, p_T) - \tilde T_3^q (y) p_Y \Delta D_T^\perp (z, p_T)\Bigr]}
{M\sum_q T_0^q(y) D_1 (z,p_T)} \nonumber\\
& - \frac{4M}{zQ} \frac{\sum_q \tilde T_3^q (y) \Delta D_T (z,p_T)}{\sum_q T_0^q(y) D_1 (z,p_T)}, \label{Phx}\\
P_{hy} & (y,z,p_T) = P_{hy}^{(0)} (y,z,p_T)   \Bigl[ 1+ \frac{M}{Q}\Delta(y,z,p_T)\Bigr] \nonumber\\
& + P_{hx}^{(0)}(y,z,p_T) \frac{(2y-1)p_Y}{zQ\sqrt{y(1-y)}}\nonumber\\
& - \frac{4p_X}{z^3Q} 
\frac{\sum_q \Bigl[\tilde T_2^q (y) p_X D_T^\perp (z, p_T) - \tilde T_3^q (y) p_Y \Delta D_T^\perp (z, p_T)\Bigr]}{M\sum_q T_0^q(y) D_1 (z,p_T)}\nonumber\\
& + \frac{4M}{zQ} \frac{\sum_q \tilde T_2^q (y) D_T (z,p_T)}{\sum_q T_0^q(y) D_1 (z,p_T)},\label{Phy}
\end{align}
where $\Delta(y,z,p_T)$ is obtained by replacing $k'_\perp$ with $p_T$ in the expression of $\hat\Delta(y,z,k'_\perp)$, i.e.,  
\begin{align}
\Delta(y,z,p_T)=\frac{4\sum_q  \Bigl[\tilde T_2^q (y) p_X D^\perp (z,p_T) 
- \tilde T_3^q (y) p_Y \Delta D^\perp (z,p_T)\Bigr]}{z^2M \sum_q T_0^q(y) D_1(z,p_T)}.
\end{align}
The extra term proportional to $P_{hy}^{(0)}$ in $P_{hx}$and 
that proportional to $P_{hx}^{(0)}$ in $P_{hy}$ are due to the power suppressed terms 
in Eqs.~(\ref{kperpvspTx}) and (\ref{kperpvspTy}). 

If we consider $e^++e^-\to\gamma^*\to h+{\bar q}+X$, we have, at the leading twist, 
\begin{align}
& P_{Lh}^{(0,em)} (y,z, k'_\perp) = 0,\\
& P_{hx}^{(0,em)} (y,z,k'_\perp) = - \frac{k'_y}{M} 
\frac{\sum_q e_q^2 \hat D_{1T}^\perp(z,k'_\perp)}{\sum_q e_q^2 \hat D_1 (z,k'_\perp)}, \\
& P_{hy}^{(0,em)} (y,z,k'_\perp) = \frac{k'_x}{M} 
\frac{\sum_q e_q^2 \hat D_{1T}^\perp(z,k'_\perp)}{\sum_q e_q^2 \hat D_1 (z,k'_\perp)}.
\end{align}
and up to twist 3, 
\begin{align}
P_{Lh}^{(em)}(y,z, k'_\perp)& = \frac{4k'_y }{zQ}\frac{\tilde B(y) }{A(y)} 
\frac{\sum_qe_q^2\hat D_L^\perp(z,k'_\perp)}{\sum_q e_q^2 \hat D_1(z,k'_\perp)},\\
P_{hx}^{(em)}  (y,z, k'_\perp)& = P_{hx}^{(0,em)} (y,z,k'_\perp)  \Bigl[ 1+ \frac{M}{Q}\hat\Delta^{em}(y,z,k'_\perp)\Bigr] \nonumber\\
& + \frac{4k'_xk'_y}{zQM}\frac{\tilde B(y) }{A(y)} 
\frac{\sum_q e_q^2\hat D_T^\perp (z, k'_\perp)}{\sum_q e_q^2 \hat D_1 (z,k'_\perp)}, \\
P_{hy}^{(em)}  (y,z, k'_\perp)& = P_{hy}^{(0,em)} (y,z,k'_\perp) \Bigl[ 1+ \frac{M}{Q}\hat\Delta^{em}(y,z,k'_\perp)\Bigr] \nonumber\\
& - \frac{4{k'}^2_x}{zQM}\frac{\tilde B(y)}{A(y)} 
\frac{\sum_q e_q^2\hat D_T^\perp (z, k'_\perp)}{\sum_q e_q^2 \hat D_1 (z,k'_\perp)} \nonumber\\
& + \frac{4M}{zQ} \frac{\tilde B (y)}{A(y)} \frac{\sum_q e_q^2 \hat D_T (z,k'_\perp) }{\sum_q e_q^2 \hat D_1 (z,k'_\perp)},
\end{align}
where the incremental factor $\hat\Delta^{em}$ reduces to,
\begin{align}
\hat\Delta^{em}(y,z,k'_\perp)=\frac{4 k'_x}{z M}\frac{\tilde B(y)}{A(y)}
\frac{\sum_q  e_q^2  \hat D^\perp (z,k'_\perp)}
{ \sum_q e_q^2 \hat D_1(z,k'_\perp)}.
\end{align}
In terms of $y$, $z$ and $p_T$, we have, at the leading twist,
\begin{align}
& P_{Lh}^{(0,em)} (y,z, p_T) = 0,\label{Plh0em}\\
& P_{hx}^{(0,em)} (y,z,p_T) = - \frac{p_Y}{zM} 
\frac{\sum_q e_q^2 D_{1T}^\perp(z,p_T)}{\sum_q e_q^2 D_1 (z,p_T)}, \label{Phx0em}\\
& P_{hy}^{(0,em)} (y,z,p_T) = \frac{p_X}{zM} 
\frac{\sum_q e_q^2 D_{1T}^\perp(z,p_T)}{\sum_q e_q^2 D_1 (z,p_T)}. \label{Phy0em}
\end{align}
And up to twist 3,
\begin{align}
P_{Lh}^{(em)}(y,z, p_T)& = \frac{4p_Y }{z^2Q}\frac{\tilde B(y) }{A(y)} 
\frac{\sum_qe_q^2D_L^\perp(z,p_T)}{\sum_q e_q^2 D_1(z,p_T)},\label{Plhem}\\
P_{hx}^{(em)}  (y,z, p_T)& = P_{hx}^{(0,em)} (y,z,p_T)  \Bigl[ 1+ \frac{M}{Q}\Delta^{em}(y,z,p_T)\Bigr] \nonumber\\
& - P_{hy}^{(0,em)}(y,z,p_T) \frac{(2y-1)p_Y}{zQ\sqrt{y(1-y)}}\nonumber\\
& + \frac{4p_Xp_Y}{z^3QM} \frac{\tilde B(y) }{A(y)} 
\frac{\sum_q e_q^2D_T^\perp (z, p_T)}{\sum_q e_q^2 D_1 (z,p_T)}, \label{Phxem}\\
P_{hy}^{(em)}  (y,z, p_T)& = P_{hy}^{(0,em)} (y,z,p_T)   \Bigl[ 1+ \frac{M}{Q}\Delta^{em}(y,z,p_T)\Bigr] \nonumber\\
& + P_{hx}^{(0,em)}(y,z,p_T) \frac{(2y-1)p_Y}{zQ\sqrt{y(1-y)}}\nonumber\\
& - \frac{4p^2_X}{z^3QM}\frac{\tilde B(y) }{A(y)} 
\frac{\sum_q e_q^2D_T^\perp (z, p_T)}{\sum_q e_q^2 D_1 (z,p_T)} \nonumber\\
&+ \frac{4M}{zQ} \frac{\tilde B(y)}{A(y)} \frac{\sum_q e_q^2 D_T (z,p_T)}{\sum_q e_q^2 D_1(z,p_T)}, \label{Phyem}
\end{align}
where the incremental factor is given by,
\begin{align}
\Delta^{em}(y,z,p_T)=\frac{4 p_X}{z^2 M}\frac{\tilde B(y)}{A(y)}
\frac{\sum_q  e_q^2  D^\perp (z,p_T)}
{ \sum_q e_q^2 D_1(z,p_T)}.
\end{align}
We see that for $e^++e^-\to\gamma^*\to h+\bar q+X$, the longitudinal polarization exists only at twist 3. 
There is however leading twist transverse polarization proportional to $D_{1T}^\perp(z,p_T)$, 
the counterpart of the Sivers function in fragmentation function.

{\it In terms of ($P_{Lh}$, $P_{hn}$, $P_{ht}$)}:
From Eq.~(\ref{csVpolvsyz}), we see that the $\hat D_{1T}^\perp(z,k'_\perp)$-term 
contributes to the transverse polarization just in 
the direction of $\vec e_n=(k'_y\vec e_x-k'_x\vec e_y)/|\vec k'_\perp|$ 
while the $\Delta\hat D_{1T}^\perp(z,k'_\perp)$-term contributes 
in the transverse direction $\vec e_t=(k'_x\vec e_x+k'_y\vec e_y)/|\vec k'_\perp|$ in the production plane.
In other words, if we consider $P_{hn}=\vec e_n\cdot\vec P_h$ and $P_{ht}=\vec e_t\cdot\vec P_h$, 
we obtain, at the leading twist,
\begin{align}
& P_{hn}^{(0)}(y,z,k'_\perp) = - \frac{|\vec k'_\perp|}{M}
\frac{\sum_q T_0^q (y)  \hat D_{1T}^\perp(z,k'_\perp)}{\sum_q T_0^q(y) \hat D_1 (z,k'_\perp)}, \\
& P_{ht}^{(0)}(y,z,k'_\perp) = - \frac{|\vec k'_\perp|}{M}
\frac{\sum_q  P_q (y)T_0^q(y) \Delta \hat D_{1T}^\perp (z, k'_\perp)}{\sum_q T_0^q(y) \hat D_1 (z,k'_\perp)}.
\end{align}
We see that each of them just proportional to one leading twist spin dependent component 
of the fragmentation function defined by Eqs.~(\ref{wei:defd1t}-\ref{wei:defdeltad1t}).  
Up to twist 3, we have,
\begin{align}
P_{hn} & (y,z, k'_\perp) = P_{hn}^{(0)} (y,z,k'_\perp)  \Bigl[ 1+ \frac{M}{Q}\hat\Delta(y,z,k'_\perp)\Bigr] \nonumber\\
& + \frac{4|\vec k'_\perp|}{zQ} \frac{\sum_q \Bigl[\tilde T_2^q (y) k'_x \hat D_T^\perp (z, k'_\perp) - \tilde T_3^q (y) k'_y \Delta \hat D_T^\perp (z, k'_\perp)\Bigr]}{M\sum_q T_0^q(y) \hat D_1 (z,k'_\perp)} \nonumber\\
& - \frac{4M}{zQ} 
\frac{\sum_q \Bigl[\tilde T_2^q (y) k'_x \hat D_T (z,k'_\perp)+\tilde T_3^q (y) k'_y\Delta \hat D_T (z,k'_\perp)\Bigr]}
{|\vec k'_\perp|\sum_q T_0^q(y) \hat D_1 (z,k'_\perp)}, \\
P_{ht} & (y,z, k'_\perp) = P_{ht}^{(0)} (y,z,k'_\perp)   \Bigl[ 1+ \frac{M}{Q}\hat\Delta(y,z,k'_\perp)\Bigr] \nonumber\\
& + \frac{4M}{zQ} \frac{\sum_q \Bigl[\tilde T_2^q (y) k'_y \hat D_T (z,k'_\perp) - \tilde T_3^q (y) k'_x\Delta \hat D_T (z,k'_\perp) \Bigr]}{|\vec k'_\perp|\sum_q T_0^q(y) \hat D_1 (z,k'_\perp)}.
\end{align}
In terms of $y$, $z$ and $P_T$, we have, at the leading twist,
\begin{align}
&P_{hn}^{(0)}(y,z,p_T) = - \frac{|\vec p_T|}{zM}
\frac{\sum_q T_0^q (y) D_{1T}^\perp(z,p_T)}{\sum_q T_0^q(y) D_1 (z,p_T)}, \label{Phn0}\\
&P_{ht}^{(0)}(y,z,p_T) = - \frac{|\vec p_T|}{zM}
\frac{\sum_q  P_q (y)T_0^q(y) \Delta D_{1T}^\perp (z, p_T)}{\sum_q T_0^q(y) D_1 (z,p_T)},\label{Pht0}
\end{align}
and up to twist 3,
\begin{align}
P_{hn} & (y,z, p_T) = P_{hn}^{(0)} (y,z,p_T)  \Bigl[ 1+ \frac{M}{Q}\Delta(y,z,p_T)\Bigr] \nonumber\\
& + \frac{4|\vec p_T|}{z^3Q} \frac{\sum_q \Bigl[\tilde T_2^q (y) p_X D_T^\perp (z, p_T) - \tilde T_3^q (y) p_Y \Delta D_T^\perp (z, p_T)\Bigr]}{M\sum_q T_0^q(y) D_1 (z,p_T)} \nonumber\\
& - \frac{4M}{zQ} 
\frac{\sum_q \Bigl[\tilde T_2^q (y) p_X D_T (z,p_T) + \tilde T_3^q (y) p_Y\Delta D_T (z,p_T)\Bigr]}
{|\vec p_T|\sum_q T_0^q(y) D_1 (z,p_T)}, \label{Phn}\\
P_{ht} & (y,z, p_T) = P_{ht}^{(0)} (y,z,p_T)   \Bigl[ 1+ \frac{M}{Q}\Delta(y,z,p_T)\Bigr] \nonumber\\
& + \frac{4M}{zQ} \frac{\sum_q \Bigl[\tilde T_2^q (y) p_Y D_T (z,p_T) - \tilde T_3^q (y) p_X\Delta D_T (z,p_T) \Bigr]}{|\vec p_T|\sum_q T_0^q(y) D_1 (z,p_T)}.\label{Pht}
\end{align}

If we consider $e^++e^-\to\gamma^*\to h+{\bar q}+X$, we have, at the leading twist, 
\begin{align}
& P_{hn}^{(0,em)}(y,z,k'_\perp) = - \frac{|\vec k'_\perp|}{M}
\frac{\sum_q e_q^2 \hat D_{1T}^\perp(z,k'_\perp)}{\sum_q e_q^2 \hat D_1 (z,k'_\perp)}, \\
& P_{ht}^{(0,em)}(y,z,k'_\perp) = 0.
\end{align}
And up to twist 3, 
\begin{align}
P_{hn}^{(em)}(y,z,k'_\perp)& = P_{hn}^{(0,em)} (y,z,k'_\perp)  \Bigl[ 1+ \frac{M}{Q}\hat\Delta^{em}(y,z,k'_\perp)\Bigr] \nonumber\\
& + \frac{4|\vec{k}'_\perp| k'_x}{zQM}\frac{\tilde B(y) }{A(y)} 
\frac{\sum_q e_q^2\hat D_T^\perp (z, k'_\perp)}{\sum_q e_q^2 \hat D_1 (z,k'_\perp)} \nonumber\\
& - \frac{4M k'_x}{zQ|\vec{k}'_\perp|} \frac{\tilde B(y) }{A(y)} \frac{\sum_q e_q^2 \hat D_T (z, k'_\perp)}{\sum_q e_q^2 \hat D_1 (z,k'_\perp)},  \\
P_{ht}^{(em)}(y,z,k'_\perp)& =  \frac{4M k'_y}{zQ|\vec{k}'_\perp|} \frac{\tilde B(y) }{A(y)} \frac{\sum_q e_q^2 \hat D_T (z, k'_\perp)}{\sum_q e_q^2 \hat D_1 (z,k'_\perp)} .
\end{align}
In terms of $y$, $z$ and $p_T$, we have, at the leading twist,
\begin{align}
& P_{hn}^{(0,em)}(y,z,p_T) = - \frac{|\vec p_T|}{zM}
\frac{\sum_q e_q^2 D_{1T}^\perp(z,p_T)}{\sum_q e_q^2 D_1 (z,p_T)}, \label{Phn0em}\\
& P_{ht}^{(0,em)}(y,z,p_T) = 0.\label{Pht0em}
\end{align}
And up to twist 3,
\begin{align}
P_{hn}^{(em)}(y,z,p_T)& = P_{hn}^{(0,em)} (y,z,p_T)  \Bigl[ 1+ \frac{M}{Q}\Delta^{em}(y,z,p_T)\Bigr] \nonumber\\
& + \frac{4|\vec{p}_T|p_X}{z^3QM}\frac{\tilde B(y)}{A(y)} 
\frac{\sum_q e_q^2D_T^\perp (z, p_T)}{\sum_q e_q^2 D_1 (z,p_T)} \nonumber\\
& - \frac{4M p_X}{zQ|\vec{p}_T|} \frac{\tilde B (y)}{A (y)} \frac{\sum_q e_q^2 D_T (z,p_T)}{\sum_q e_q^2 D_1 (z,p_T)}, \label{Phnem}\\
P_{ht}^{(em)}(y,z,p_T)& = \frac{4Mp_Y}{zQ|\vec{p}_T|}\frac{\tilde B(y) }{A(y)} 
\frac{\sum_q e_q^2 D_T (z, p_T)}{\sum_q e_q^2 D_1 (z,p_T)}.\label{Phtem}
\end{align}
We see that, expressed in this way, 
the leading twist contribution has only one transverse polarization proportional to $D_{1T}^\perp(z,p_T)$ 
and is along the normal of the production plane. 

{\it In the helicity-jet frame:} 
We can also study the polarization in the jet frame where 
we calculate the longitudinal component $P_{Lh}$ with respect to helicity 
and other two components along $X$ and $Y$ directions.
The results for the longitudinal component $P_{Lh}$ are the same as those presented in last section. 
We present the results for $P_{hX}$ and $P_{hY}$ in the following. 
At the leading twist, they are given by,
\begin{align}
&P_{hX}^{(0)} (y,z,p_T) = -\frac{\sum_q T_0^q (y) \Bigl[ p_Y D_{1T}^\perp (z, p_T) + P_q (y) p_{X} \Delta D_{1T}^\perp (z, p_T)\Bigr]}{zM \sum_q T_0^q (y) D_1 (z,p_T)},\\
&P_{hY}^{(0)} (y,z,p_T) = \frac{\sum_q T_0^q (y) \Bigl[  p_{X} D_{1T}^\perp (z, p_T) - P_q (y) p_{Y} \Delta D_{1T}^\perp (z,p_T)\Bigr]}
{zM \sum_q T_0^q (y) D_1 (z,p_T)},
\end{align}
and up to twist-3, we have, 
\begin{align}
P_{hX}  &(y,z,p_T)= P_{hX}^{(0)} (y,z,p_T) \Bigl[ 1+ \frac{M}{Q}\Delta(y,z,p_T)\Bigr] + \nonumber\\
& + \frac{4p_{Y}}{z^3Q} \frac{\sum_q \Bigl[\tilde T_2^q (y) p_{X} D_T^\perp (z,p_T) - \tilde T_3^q (y) p_{Y} \Delta D_T^\perp (z,p_T)\Bigr]}{M\sum_q T_0^q (y) D_1 (z,p_T)}  \nonumber\\
&  - \frac{2}{zQ} \frac{\sum_q \Bigl[2M\tilde T_3^q (y) \Delta D_T (z,p_T) -p_X  T_1^q(y) \Delta D_{1L} (z,p_T)\Bigr] }
{\sum_q T_0^q (y) D_1 (z,p_T)},\\
P_{hY}  &(y,z,p_T)= P_{hY}^{(0)} (y,z,p_T) \Bigl[ 1+ \frac{M}{Q}\Delta(y,z,p_T)\Bigr] + \nonumber\\
& - \frac{4p_{X} }{z^3Q} \frac{\sum_q \Bigl[\tilde T_2^q (y) p_{X} D_T^\perp (z,p_T) - \tilde T_3^q (y) p_{Y} \Delta D_T^\perp (z,p_T)\Bigr]}{M\sum_q T_0^q (y) D_1 (z,p_T)}  \nonumber\\
&  + \frac{2}{zQ} \frac{\sum_q \Bigl[2M \tilde T_2^q (y)  D_T (z,p_T)+p_Y T_1^q(y) \Delta D_{1L} (z,p_T)\Bigr] }
{\sum_q T_0^q (y) D_1 (z,p_T)}.
\end{align}

If we integrate over $\varphi$, we obtain,  
\begin{align}
P_{hX}  &(y,z,|\vec p_T|)= - \frac{4M}{zQ} \nonumber\\
&\times\frac{\sum_q \tilde T_3^q (y) \Bigl[ \Delta D_T (z,p_T) - \frac{p_T^2}{2z^2M^2}\Delta D_T^\perp (z,p_T)\Bigr]} 
 {\sum_q T_0^q (y) D_1 (z,p_T)}, \label{wei:xpjett3}\\
P_{hY}  &(y,z,|\vec p_T|)=  \frac{4M}{zQ} \nonumber\\
&\times\frac{\sum_q \tilde T_2^q (y) \Bigl[ \Delta D_T (z,p_T) + \frac{p_T^2}{2z^2M^2}\Delta D_T^\perp (z,p_T)\Bigr] } 
{\sum_q T_0^q (y) D_1 (z,p_T)}. \label{wei:ypjett3}
\end{align}

If we consider the $e^+e^-$-annihilation via electromagnetic interaction, we have,
\begin{align}
P_{hX}^{(em)} & (y,z,p_T)  =  -\frac{{p}_Y}{zM} \frac{\sum_q e^2_q D_{1T}^{\perp} (z,p_T) }
{\sum_q e_q^2 D_1 (z,p_T)}  \Bigl[1+ \frac{M}{Q}\Delta^{(em)}(y,z,p_T)\Bigr]  \nonumber\\ 
+ &\frac{4 p_X p_Y}{z^3MQ}  \frac{\tilde B(y)}{A(y)}
\frac{ \sum_q e_q^2 D_T^{\perp} (z,p_T) }{\sum_q e_q^2 D_1 (z,p_T)}, \\
P_{hY}^{(em)} & (y,z,p_T) = \frac{p_X}{zM}  
\frac{\sum_q e_q^2 D_{1T}^\perp (z, p_T) }{\sum_q e_q^2 D_1 (z,p_T)} 
\Bigl[1+ \frac{M}{Q}\Delta^{(em)}(y,z,p_T)\Bigr] \nonumber\\
+ & \frac{4M}{zQ} \frac{ \tilde B(y)}{A(y)} 
\frac{ \sum_q e_q^2  \Bigl[ D_T (z,p_T) - \frac{p_X^2}{z^2M^2} D_T^\perp (z,p_T)\Bigr]}
{\sum_q e_q^2 D_1  (z,p_T)}.
\end{align}
Upon integration over the azimuthal angle $\varphi$, we obtain, 
\begin{align}
P_{hX}^{(em)} &(y,z,|\vec p_T|) =  0,\\
P_{hY}^{(em)} &(y,z,|\vec p_T|) = 
\frac{4M}{zQ}\frac{ \tilde B(y)}{A(y)}
\frac{ \sum_q e_q^2 \Bigl[ D_T (z,p_T) + \frac{p_T^2}{2z^2M^2} D_T^\perp (z,p_T)\Bigr]}
 {\sum_q e_q^2 D_1 (z,p_T)}.
\end{align}

We compare these results with those in the collinear frame and we see that
at the leading twist they are the same, i.e., 
i.e. $P_{hx}^{(0)}=P_{hX}^{(0)}$ and $P_{hy}^{(0)}=P_{hY}^{(0)}$. 
However, there exist differences at twist-3, and they are given by,
\begin{align}
P_{hX}(y,z,p_T)  & - P_{hx}(y,z,p_T) =
  \frac{2p_X}{zQ} P_{Lh}^{(0)}(y,z,p_T) \nonumber\\
&+ P_{hy}^{(0)} (y,z,p_T) \frac{(2y-1)p_Y}{zQ\sqrt{y(1-y)}}, \\
P_{hY}(y,z,p_T)  & - P_{hy}(y,z,p_T) =
 \frac{2p_Y}{zQ} P_{Lh}^{(0)}(y,z,p_T) \nonumber\\
&-P_{hx}^{(0)}(y,z,p_T) \frac{(2y-1)p_Y}{zQ\sqrt{y(1-y)}}.
\end{align}

{\it Summary and discussions:} 
As a brief summary of the results presented in this sub-section for polarization of hadron with spin-1/2, 
we note that, although the polarization can be equally studied by measuring 
($P_{Lh}$, $P_{hx}$, $P_{hy}$) or ($P_{Lh}$, $P_{hX}$, $P_{hY}$) or ($P_{Lh}$, $P_{hn}$, $P_{ht}$), 
the expressions obtained above seem to show that ($P_{Lh}$, $P_{hn}$, $P_{ht}$) are most suitable to 
use when studying the corresponding components of the fragmentation function. 
We see in particular that at the leading twist there are three spin dependent components 
of the fragmentation function, i.e.,  
$\Delta D_{1L}$, $D_{1T}^\perp$ and $\Delta D_{1T}^\perp$, are involved in $e^++e^-\to h+\bar q+X$.
From Eqs.~(\ref{Plh0}-\ref{Pht0}), we see that 
the three components, $P_{Lh}^{(0)}$, $P_{hn}^{(0)}$ and $P_{ht}^{(0)}$, 
of the polarization are just determined by each of them respectively. 
We also recall that the physical significances of these three leading twist spin dependent 
components of the fragmentation function are rather clear. 
More precisely, $\Delta D_{1L}^{q\to h}$ describes the spin transfer of the fragmenting quark $q$ 
to the produced hadron $h$ for the case that $q$ is longitudinally polarized. 
The other two components $D_{1T}^{\perp q\to h}$ and $\Delta D_{1T}^{\perp q\to h}$ describe 
the induced transverse polarization in the fragmentation in the case that the fragmenting quark $q$ 
is unpolarized or is longitudinally polarized respectively. 
We recall further that $\Delta D_{1L}^{q\to h}$ corresponds to the helicity distribution $g_{1L}(x,k_\perp)$ 
in parton distribution function, 
$D_{1T}^{\perp q\to h}$ and $\Delta D_{1T}^{\perp q\to h}$ correspond to 
Sivers function $f_{1T}^\perp(x,k_\perp)$ and worm-gear 
or tran-helicity distribution $g_{1T}^\perp(x,k_\perp)$ respectively.
Each of them has a clear counterpart in the parton distribution and clear physical meaning.
Since the quark and anti-quark are longitudinally polarized in $Z$-decay, 
$e^+e^-$ annihilation at $Z$ pole provides an ideal place to study these 
components of fragmentation function. 

We see also that, for reaction via electromagnetic interaction, 
i.e., $e^++e^-\to\gamma^*\to h+\bar q+X$, $P_q(y)=0$ and parity violating terms disappear.  
The results reduce to those given Eqs.~(\ref{Plh0em}-\ref{Pht0em}) 
where the only nonzero component of the polarization at the leading twist is $P_{hn}^{(0)}$. 
In this case, for $P_{Lh}$ and $P_{ht}$, the twist 3 contributions are the leading contributions 
so that they can be better studied via precise measurements.

\subsubsection{Spin-1 hadrons}
For spin-1 hadrons, the polarization is described by the vector polarization and the tensor polarization.
The vector polarization is exactly the same as that for spin-1/2 hadrons presented in last section. 
In this section, we present the results for tensor polarization in the following.

Similar to those for the vector polarization, different components of the tensor polarization 
can also be expressed in different coordinate system. 
Usually the longitudinal direction is just taken as the helicity direction, but the transverse 
directions have different choices. 
However, as can be seen in last subsection for vector polarized case, there will be 
mixtures of different components if we choose helicity as the longitudinal direction 
but $X$ and $Y$ as the transverse directions since the helicity direction is neither 
orthogonal to $X$ nor to $Y$-direction. 
This makes the study much more complicated and the physical meaning even unclear.
According to experiences in the case of vector polarization, 
we choose two different cases for the transverse directions, i.e. $x$ and $y$ directions 
or $\vec e_n$ and $\vec e_t$ and present the corresponding results in the following. 

The expectations of different components of the tensor polarization is calculated using 
the differential cross section and the results given by Eqs.~(\ref{sllint})-(\ref{sttxyint}).
For example, if the hadron is in the eigenstate $|0;0,0\rangle$, 
we have $\lambda_h = 0$, $\vec S_\perp =\vec 0$, $\vec S_{LT} = 0$, 
$S_{TT}^{xx} = S_{TT}^{xy} = 0$, $S_{LL}=-1$.  
From Eqs.Eqs.~(\ref{csTpol1LL}-\ref{csTpol1TT}), we obtain immediately, at the leading twist,
\begin{equation}
{\cal P}^{(0)}(0;0,0)= 
\frac{\sum_q T_0^q (y) \Bigl[\hat D_1 (z,k'_\perp) - \hat D_{1LL}^q (z,k'_\perp) \Bigr]}
{3 \sum_q T_0^q (y) \hat D_1 (z,k'_\perp) },
\end{equation}
and from Eq.(\ref{sllint}), we obtain, 
\begin{equation}
S_{LL}^{(0)} (y,z,k'_\perp) = 
\frac{\sum_q T_0^q(y)\hat D_{1LL} (z,k'_\perp)}{2\sum_q T_0^q(y) \hat D_1(z,k'_\perp)} .
\end{equation}
Up to twist-3, we have,
\begin{widetext}
\begin{align}
S_{LL} (y,z,k'_\perp) &= 
S_{LL}^{(0)} (y,z,k'_\perp)\Bigl[1+\frac{M}{Q}\hat\Delta(y,z,k'_\perp)\Bigr]
- \frac{\sum_q 2 \Bigl[\tilde T_2^q (y) k'_x \hat D_{LL}^\perp (z,k'_\perp) - \tilde T_3^q (y) k'_y \Delta \hat D_{LL}^\perp  (z,k'_\perp) \Bigr] }{zQ \sum_q T_0^q (y) \hat D_1 (z,k'_\perp) }.
\end{align}
We see that $S_{LL}$ has both leading and twist 3 contributions and neither of 
the contributions depend on the polarization $P_q(y)$ of the fragmenting quark.
The leading twist contribution is determined by $\hat D_{1LL} (z,k'_\perp)$ 
while the twist 3 contributions depend on two components, 
$\hat D_{LL}^\perp (z,k'_\perp)$ and $\Delta\hat D_{LL}^\perp (z,k'_\perp)$.  

We can also change the variable $k'_\perp$ to $p_T$ and obtain,
\begin{align}
S_{LL}^{(0)}& (y,z,p_T) = 
\frac{\sum_q T_0^q(y)D_{1LL} (z,p_T)}{2\sum_q T_0^q(y) D_1(z,p_T)},\\
S_{LL} &(y,z,p_T) =  S_{LL}^{(0)} (y,z,p_T) \Bigl[1+\frac{M}{Q}\Delta(y,z,p_T)\Bigr]
- \frac{\sum_q 2 \Bigl[\tilde T_2^q (y) p_X D_{LL}^\perp (z,p_T) - \tilde T_3^q (y) p_Y \Delta D_{LL}^\perp  (z,p_T) \Bigr] }{ z^2Q \sum_q T_0^q (y) D_1 (z,p_T) }.
\end{align}

Other components can also be calculated in the same way from Eqs. (\ref{csTpol1LL}-\ref{csTpol1TT}). 
At the leading twist, in the collinear frame, we have,

\begin{align}
& S_{LT}^{x(0)} (y,z,k'_\perp) = - \frac{2\sum_q \Bigl[T_0^q (y) k'_x \hat D_{1LT}^\perp (z,k'_\perp) + T_1^q (y) k'_y \Delta \hat D_{1LT}^\perp (z,k'_\perp) \Bigr] }{3 M \sum_q T_0^q (y) \hat D_1 (z,k'_\perp)}, \\
& S_{LT}^{y(0)} (y,z,k'_\perp) = - \frac{2\sum_q \Bigl[T_0^q (y) k'_y \hat D_{1LT}^\perp (z,k'_\perp) - T_1^q (y) k'_x \Delta \hat D_{1LT}^\perp (z,k'_\perp) \Bigr] }{3 M \sum_q T_0^q (y) \hat D_1 (z,k'_\perp)}, \\
& S_{TT}^{xy(0)} (y,z,k'_\perp) = \frac{2\sum_q \Bigl[2T_0^q (y) k'_x k'_y \hat D_{1TT}^\perp (z,k'_\perp) - T_1^q (y) ( k'^2_x - k'^2_y) \Delta \hat D_{1TT}^\perp (z,k'_\perp) \Bigr] }{3 M^2 \sum_q T_0^q (y) \hat D_1 (z,k'_\perp)}, \\
& S_{TT}^{xx(0)} (y,z,k'_\perp) = \frac{2\sum_q \Bigl[T_0^q (y) ( k'^2_x - k'^2_y) \hat D_{1TT}^\perp (z,k'_\perp) + 2 T_1^q (y) k'_x k'_y \Delta \hat D_{1TT}^\perp (z,k'_\perp) \Bigr] }{3 M^2 \sum_q T_0^q (y) \hat D_1 (z,k'_\perp)}.
\end{align}
Up to twist-3, we have,
\begin{align}
&S_{LT}^x (y,z,k'_\perp) = S_{LT}^{x(0)} (y,z,k'_\perp) \Bigl[1+\frac{M}{Q} \hat \Delta (y,z,k'_\perp)\Bigr] \nonumber\\
&\qquad\qquad\qquad  + \frac{8\sum_q \Bigl[\tilde T_2^q (y) k'^2_x \hat D_{LT}^\perp (z, k'_\perp) - \tilde T_3^q (y) k'_x k'_y \Delta \hat D_{LT}^\perp (z, k'_\perp) - M^2 \tilde T_2^q (y) \hat D_{LT} (z,k'_\perp) \Bigr]}{3 zMQ \sum_q T_0^q (y) \hat D_1 (z,k'_\perp)}, \\
&S_{LT}^y (y,z,k'_\perp) = S_{LT}^{y(0)} (y,z,k'_\perp) \Bigl[1+\frac{M}{Q} \hat \Delta (y,z,k'_\perp)\Bigr]\nonumber\\
&\qquad\qquad\qquad + \frac{8\sum_q \Bigl[\tilde T_2^q (y) k'_x k'_y \hat D_{LT}^\perp (z, k'_\perp) - \tilde T_3^q (y) k'^2_y \Delta \hat D_{LT}^\perp (z, k'_\perp) + M^2 \tilde T_3^q (y) \Delta \hat D_{LT} (z,k'_\perp) \Bigr]}{3 zMQ \sum_q T_0^q (y) \hat D_1 (z,k'_\perp)}, \\
&S_{TT}^{xy} (y,z,k'_\perp) = S_{TT}^{xy(0)} (y,z,k'_\perp) \Bigl[1+\frac{M}{Q} \hat \Delta (y,z,k'_\perp)\Bigr] + \frac{8}{3zM^2Q \sum_q T_0^q (y) \hat D_1 (z,k'_\perp)} \nonumber\\
&\qquad\qquad\times  \sum_q \Bigl\{2k'_x k'_y \Bigl[- \tilde T_2^q (y) k'_x \hat D_{TT}^{\perp C} (z,k'_\perp) + \tilde T_3^q (y) k'_y \Delta \hat D_{TT}^{\perp C} (z,k'_\perp) \Bigr] 
+ M^2 \Bigl[\tilde T_2^q(y)k'_y  \hat D_{TT}^{\perp A} (z, k'_\perp) - \tilde T_3^q(y)k'_x  \Delta  \hat D_{TT}^{\perp A} (z, k'_\perp) \Bigr] \Bigr\},\\
&S_{TT}^{xx}  (y,z,k'_\perp) = S_{TT}^{xx(0)} (y,z,k'_\perp)\Bigl[1+\frac{M}{Q} \hat \Delta (y,z,k'_\perp)\Bigr] + \frac{8}{3zM^2Q \sum_q T_0^q (y) \hat D_1 (z,k'_\perp)} \nonumber\\
&\qquad\times \sum_q \Bigl\{ (k'^2_x - k'^2_y)\Bigl[- \tilde T_2^q (y) k'_x \hat D_{TT}^{\perp C} (z,k'_\perp) + \tilde T_3^q (y) k'_y \Delta \hat D_{TT}^{\perp C} (z,k'_\perp) \Bigr] + M^2 \Bigl[\tilde T_2^q(y) k'_x \hat D_{TT}^{\perp A} (z, k'_\perp) + \tilde T_3^q(y) k'_y \Delta  \hat D_{TT}^{\perp A} (z, k'_\perp) \Bigr] \Bigr\}.
\end{align}
In terms of $y$, $z$ and $p_T$, at the leading twist, we have, 
\begin{align}
& S_{LT}^{x(0)} (y,z,p_T) = - \frac{2\sum_q \Bigl[T_0^q (y) p_X D_{1LT}^\perp (z,p_T) + T_1^q (y) p_Y \Delta D_{1LT}^\perp (z,p_T) \Bigr] }{3 z M \sum_q T_0^q (y) D_1 (z,p_T)}, \\
& S_{LT}^{y(0)} (y,z,p_T) = - \frac{2\sum_q \Bigl[T_0^q (y) p_Y D_{1LT}^\perp (z,p_T) - T_1^q (y) p_X \Delta D_{1LT}^\perp (z,p_T) \Bigr] }{3 z M \sum_q T_0^q (y) D_1 (z,p_T)} , \\
& S_{TT}^{xy(0)} (y,z,p_T) = \frac{2\sum_q \Bigl[2T_0^q (y) p_X p_Y D_{1TT}^\perp (z,p_T) - T_1^q (y) ( p_X^2 - p_Y^2) \Delta D_{1TT}^\perp (z,p_T) \Bigr] }{3 z^2 M^2 \sum_q T_0^q (y) D_1 (z,p_T)}, \\
& S_{TT}^{xx(0)} (y,z,p_T) = \frac{2\sum_q \Bigl[T_0^q (y) ( p_X^2 - p_Y^2) D_{1TT}^\perp (z,p_T) + 2 T_1^q (y) p_X p_Y \Delta D_{1TT}^\perp (z,p_T) \Bigr] }{3 z^2 M^2 \sum_q T_0^q (y) D_1 (z,p_T)}.
\end{align}
Up to twist-3, we have
\begin{align}
S_{LT}^x (y,z,p_T) = & S_{LT}^{x(0)} (y,z,p_T) \Bigl[1+\frac{M}{Q} \Delta (y,z,p_T)\Bigr] - S_{LT}^{y(0)} (y,z,p_T) \frac{(2y-1)p_Y}{zQ\sqrt{y(1-y)}} +\nonumber\\
&+ \frac{8\sum_q \Bigl[\tilde T_2^q (y) p_X^2 D_{LT}^\perp (z, p_T) - \tilde T_3^q (y) p_X p_Y \Delta D_{LT}^\perp (z, p_T) - z^2 M^2 \tilde T_2^q (y) D_{LT} (z,p_T) \Bigr]}{3 z^3 MQ \sum_q T_0^q (y) D_1 (z,p_T)}, \\
S_{LT}^y (y,z,p_T) = & S_{LT}^{y(0)} (y,z,p_T) \Bigl[1+\frac{M}{Q} \Delta (y,z,p_T)\Bigr]+ S_{LT}^{x(0)} (y,z,p_T)  \frac{(2y-1)p_Y}{zQ\sqrt{y(1-y)}}+  \nonumber\\
& + \frac{8\sum_q \Bigl[\tilde T_2^q (y) p_X p_Y D_{LT}^\perp (z, p_T) - \tilde T_3^q (y) p_Y^2 \Delta D_{LT}^\perp (z, p_T) + z^2 M^2 \tilde T_3^q (y) \Delta D_{LT} (z,p_T) \Bigr]}{3 z^3MQ \sum_q T_0^q (y) D_1 (z,p_T)}, \\
S_{TT}^{xy} (y,z,p_T) = & S_{TT}^{xy(0)} (y,z,p_T) \Bigl[1+\frac{M}{Q} \Delta (y,z,p_T)\Bigr] + 2 S_{TT}^{xx(0)} (y,z,p_T)  \frac{(2y-1)p_Y}{zQ\sqrt{y(1-y)}}  + \frac{8 }{3z^4M^2Q \sum_q T_0^q (y) D_1 (z,p_T)} \nonumber\\
\times \sum_q \Bigl\{2p_Xp_Y&\Bigl[- \tilde T_2^q (y) p_X D_{TT}^{\perp C} (z,p_T) + 
\tilde T_3^q (y) p_Y \Delta D_{TT}^{\perp C} (z,p_T) \Bigr] 
+ z^2 M^2 \Bigl[\tilde T_2^q(y)p_Y D_{TT}^{\perp A} (z, p_T) - \tilde T_3^q(y)p_X \Delta  D_{TT}^{\perp A} (z, p_T) \Bigr] \Bigr\},\\
S_{TT}^{xx}  (y,z,p_T) = & S_{TT}^{xx(0)} (y,z,p_T) \Bigl[1+\frac{M}{Q} \Delta (y,z,p_T)\Bigr] - 2 S_{TT}^{xy(0)}  \frac{(2y-1)p_Y}{zQ\sqrt{y(1-y)}}  + \frac{8}{3z^4M^2Q \sum_q T_0^q (y) D_1 (z,p_T)}   \nonumber\\
\times \sum_q \Bigl\{ (p_X^2 - & p_Y^2)\Bigl[- \tilde T_2^q (y) p_X D_{TT}^{\perp C} (z,p_T) + \tilde T_3^q (y) p_Y \Delta D_{TT}^{\perp C} (z,p_T) \Bigr] +z^2 M^2 \Bigl[\tilde T_2^q(y) p_X D_{TT}^{\perp A} (z, p_T) + \tilde T_3^q(y) p_Y \Delta  D_{TT}^{\perp A} (z, p_T) \Bigr] \Bigr\}.
\end{align}
We see that for all the four independent components, $S_{LT}^x$, $S_{LT}^y$, $S_{TT}^{xx}$, and $S_{TT}^{xy}$,  
of the tensor polarization, each has a leading twist and a twist 3 contribution, 
and each contribution depends on several different components of the fragmentation function.
However, if we define,  
$\vec{S}_{LT} = S_{LT}^x \vec e_x + S_{LT}^y \vec e_y$, and 
$\tensor{S}_{TT} = S_{TT}^{xx} (\vec{e}_x \vec e_x - \vec e_y \vec e_y ) +  
S_{TT}^{xy} (\vec e_x \vec e_y + \vec e_y \vec e_x)$,
and we have 
$S_{LT}^n =\vec e_n\cdot \vec{S}_{LT}$, 
$S_{LT}^t = \vec e_t\cdot \vec{S}_{LT}$, 
$S_{TT}^{nn}=\vec e_n \cdot \tensor{S}_{TT} \cdot \vec e_n=-S_{TT}^{tt} = 
 -\vec e_t \cdot \tensor{S}_{TT} \cdot \vec e_t$, 
 and $S_{TT}^{nt}= S_{TT}^{tn}=\vec e_n\cdot \tensor{S}_{TT} \cdot \vec e_t
 = \vec e_t \cdot \tensor{S}_{TT} \cdot \vec e_n$. 
At the leading twist, they are given by, 
\begin{align}
& S_{LT}^{n(0)}(y, z, k'_\perp) = - \frac{2|\vec k'_\perp|}{3M} \frac{\sum_q P_q (y) T_0^q (y) \Delta \hat D_{1LT}^\perp (z,k'_\perp)}{\sum_q T_0^q (y) \hat D_1 (z, k'_\perp)}, \\
& S_{LT}^{t(0)}(y, z, k'_\perp) = - \frac{2|\vec k'_\perp|}{3M} \frac{\sum_q T_0^q (y) \hat D_{1LT}^\perp (z,k'_\perp)}{\sum_q T_0^q (y) \hat D_1 (z, k'_\perp)},\\
& S_{TT}^{nn(0)} (y, z, k'_\perp) = - \frac{2|\vec k'_\perp|^2}{3M^2} \frac{\sum_q T_0^q (y) \hat D_{1TT}^\perp (z, k'_\perp)}{\sum_q T_0^q (y) \hat D_1 (y,k'_\perp)},\\
& S_{TT}^{nt(0)}(y, z, k'_\perp) = \frac{2|\vec k'_\perp|^2}{3M^2} \frac{\sum_q P_q (y) T_0^q (y) \Delta \hat D_{1TT}^\perp (z, k'_\perp)}{\sum_q T_0^q (y) \hat D_1 (y,k'_\perp)}.
\end{align}
We see that the leading twist results in this case are indeed much simpler. 
Each component $S_{LT}^{n(0)}$, $S_{LT}^{t(0)}$, $S_{TT}^{nn(0)}$, or $S_{TT}^{nt(0)}$ is determined by 
one component of the fragmentation function and the unpolarized fragmentation function $\hat D_1$.
This shows that also in this case choosing $\vec e_n$ and $\vec e_t$ as the two independent 
transverse directions is more suitable for studying different components of the fragmentation function 
by measuring the polarization of hadrons.
We also see that $S_{LT}^{n(0)}$ and $S_{TT}^{nt(0)}$ are proportional to 
the polarization $P_q(y)$ of the fragmenting quark while the other two components $S_{LT}^{t(0)}$ and $S_{TT}^{nn(0)}$
are independent of $P_q(y)$. This implies that the leading twist contributions to the first two components vanish 
if the annihilation goes via electromagnetic interaction while the latter two survives. 
 
Up to twist 3, we have
\begin{align}
S_{LT}^n (y,z,k'_\perp) = & S_{LT}^{n(0)} (y,z,k'_\perp) \Bigl[1+\frac{M}{Q} \hat \Delta (y,z,k'_\perp)\Bigr]  - \frac{8M}{3zQ} \frac{\sum_q\Bigl[ \tilde T_2^q (y) k'_y \hat D_{LT} (z,k'_\perp) + \tilde T_3^q (y) k'_x \Delta \hat D_{LT} (z,k'_\perp) \Bigr]}{|\vec k'_\perp|\sum_q T_0^q (y) \hat D_1 (z, k'_\perp)},\\
S_{LT}^t (y,z,k'_\perp) = &S_{LT}^{t(0)} (y,z,k'_\perp) \Bigl[1+\frac{M}{Q} \hat \Delta (y,z,k'_\perp)\Bigr] 
 - \frac{8M}{3zQ} \frac{\sum_q \tilde T_2^q (y) k'_x \Bigl[\hat D_{LT} (z,k'_\perp) - \frac{|\vec k'_\perp|^2}{M^2} \hat D_{LT}^\perp (z,k'_\perp)\Bigr] }{|\vec k'_\perp|\sum_q T_0^q (y) \hat D_1 (z, k'_\perp)} + \nonumber\\
&  + \frac{8M}{3zQ} \frac{\sum_q \tilde T_3^q (y) k'_y \Bigl[ \Delta \hat D_{LT} (z,k'_\perp) - \frac{|\vec k'_\perp|^2}{M^2} \Delta \hat D_{LT}^\perp (z,k'_\perp)\Bigr] }{|\vec k'_\perp|\sum_q T_0^q (y) \hat D_1 (z, k'_\perp)},\\
S_{TT}^{nn} (y,z,k'_\perp) = &S_{TT}^{nn(0)} (y,z,k'_\perp) \Bigl[1+\frac{M}{Q} \hat \Delta (y,z,k'_\perp)\Bigr] 
 + \frac{8 |\vec k'_\perp|^2}{3zM^2Q} \frac{\sum_q \Bigl[\tilde T_2^q (y) k'_x \hat D_{TT}^{\perp C} (z,k'_\perp) - \tilde T_3^q (y) k'_y \Delta \hat D_{TT}^{\perp C} (z, k'_\perp)\Bigr]}{\sum_q T_0^q (y) \hat D_1 (z,k'_\perp)} + \nonumber\\
&  - \frac{8}{3zQ} \frac{\sum_q \Bigl[\tilde T_2^q (y) k'_x \hat D_{TT}^{\perp A} (z, k'_\perp) - \tilde T_3^q (y) k'_y \Delta \hat D_{TT}^{\perp A} (z, k'_\perp)\Bigr]}{\sum_q T_0^q (y) \hat D_1 (z,k'_\perp)},\\
S_{TT}^{nt} (y,z,k'_\perp)  = & S_{TT}^{nt(0)} (y,z,k'_\perp) \Bigl[1+\frac{M}{Q} \hat \Delta (y,z,k'_\perp)\Bigr]  
+ \frac{8}{3zQ} \frac{\sum_q \Bigl[\tilde T_2^q (y) k'_y \hat D_{TT}^{\perp A} (z, k'_\perp) + \tilde T_3^q (y) k'_x \Delta \hat D_{TT}^{\perp A} (z, k'_\perp)\Bigr]}{\sum_q T_0^q (y) \hat D_1 (z,k'_\perp)}.
\end{align}

In terms of $y$, $z$ and $p_T$, we have, at the leading twist,
\begin{align}
& S_{LT}^{n(0)} (y, z, p_T)= - \frac{2|\vec p_T|}{3zM} \frac{\sum_q P_q (y) T_0^q (y) \Delta   D_{1LT}^\perp (z, p_T )}{\sum_q T_0^q (y)   D_1 (z,  p_T )},\\
& S_{LT}^{t(0)}(y, z, p_T) = - \frac{2|\vec p_T|}{3zM} \frac{\sum_q T_0^q (y)   D_{1LT}^\perp (z, p_T )}{\sum_q T_0^q (y)   D_1 (z,  p_T )},\\
& S_{TT}^{nn(0)} (y, z, p_T)= - \frac{2|\vec p_T|^2}{3z^2M^2} \frac{\sum_q T_0^q (y)   D_{1TT}^\perp (z,  p_T )}{\sum_q T_0^q (y)   D_1 (y, p_T )},\\
& S_{TT}^{nt(0)} (y, z, p_T)= \frac{2|\vec p_T|^2}{3z^2M^2} \frac{\sum_q P_q (y) T_0^q (y) \Delta   D_{1TT}^\perp (z,  p_T )}{\sum_q T_0^q (y)   D_1 (y, p_T )},
\end{align}
and up to twist 3, 
\begin{align}
S_{LT}^n (y,z,p_T) =&  S_{LT}^{n(0)} (y,z,p_T) \Bigl[1+\frac{M}{Q}   \Delta (y,z,p_T)\Bigr] 
 - \frac{8M}{3zQ} \frac{\sum_q\Bigl[ \tilde T_2^q (y) p_Y   D_{LT} (z,p_T) + \tilde T_3^q (y) p_X \Delta   D_{LT} (z,p_T) \Bigr]}{|\vec p_T|\sum_q T_0^q (y)   D_1 (z, p_T)},\\
S_{LT}^t (y,z,p_T) = & S_{LT}^{t(0)} (y,z,p_T) \Bigl[1+\frac{M}{Q}   \Delta (y,z,p_T)\Bigr]
 - \frac{8M}{3zQ} \frac{\sum_q \tilde T_2^q (y) p_X \Bigl[  D_{LT} (z,p_T) - \frac{|\vec p_T|^2}{z^2M^2}   D_{LT}^\perp (z,p_T)\Bigr] }{|\vec p_T|\sum_q T_0^q (y)   D_1 (z, p_T)} + \nonumber\\
& + \frac{8M}{3zQ} \frac{\sum_q \tilde T_3^q (y) p_Y \Bigl[ \Delta   D_{LT} (z,p_T) - \frac{|\vec p_T|^2}{z^2M^2} \Delta   D_{LT}^\perp (z,p_T)\Bigr] }{|\vec p_T|\sum_q T_0^q (y)   D_1 (z, p_T)},\\
S_{TT}^{nn} (y,z,p_T)  =& S_{TT}^{nn(0)} (y,z,p_T) \Bigl[1+\frac{M}{Q}   \Delta (y,z,p_T)\Bigr] 
+ \frac{8 |\vec p_T|^2}{3z^4M^2Q} \frac{\sum_q \Bigl[\tilde T_2^q (y) p_X   D_{TT}^{\perp C} (z,p_T) - \tilde T_3^q (y) p_Y \Delta   D_{TT}^{\perp C} (z, p_T)\Bigr]}{\sum_q T_0^q (y)   D_1 (z,p_T)} + \nonumber\\
&  - \frac{8}{3z^2Q} \frac{\sum_q \Bigl[\tilde T_2^q (y) p_X   D_{TT}^{\perp A} (z, p_T) - \tilde T_3^q (y) p_Y \Delta   D_{TT}^{\perp A} (z, p_T)\Bigr]}{\sum_q T_0^q (y)   D_1 (z,p_T)},\\
S_{TT}^{nt} (y,z,p_T)  = &S_{TT}^{nt(0)} (y,z,p_T) \Bigl[1+\frac{M}{Q}   \Delta (y,z,p_T)\Bigr]  
+ \frac{8}{3z^2Q} \frac{\sum_q \Bigl[\tilde T_2^q (y) p_Y   D_{TT}^{\perp A} (z, p_T) + \tilde T_3^q (y) p_X \Delta   D_{TT}^{\perp A} (z, p_T)\Bigr]}{\sum_q T_0^q (y)   D_1 (z,p_T)}.
\end{align}

For the process via electromagnetic interaction, 
i.e., $e^+ + e^- \to \gamma^* \to h + \bar q + X$, at the leading twist level, we have,
\begin{align}
& S_{LL}^{(0,em)} (y,z,k'_\perp) = 
\frac{\sum_q e_q^2\hat D_{1LL} (z,k'_\perp)}{2\sum_q e_q^2 \hat D_1(z,k'_\perp)} ,\\
& S_{LT}^{x(0,em)} (y,z,k'_\perp) = - \frac{2k'_x}{3M} \frac{\sum_q e_q^2 \hat D_{1LT}^\perp (z,k'_\perp) }{\sum_q e_q^2 \hat D_1 (z,k'_\perp)}, \\
& S_{LT}^{y(0,em)} (y,z,k'_\perp) = - \frac{2k'_y}{3M}\frac{\sum_q e_q^2\hat D_{1LT}^\perp (z,k'_\perp)}{ \sum_q e_q^2 \hat D_1 (z,k'_\perp)}, \\
& S_{TT}^{xy(0,em)} (y,z,k'_\perp) = \frac{4k'_x k'_y}{3M^2} \frac{\sum_q e_q^2 \hat D_{1TT}^\perp (z,k'_\perp) }{\sum_q e_q^2 \hat D_1 (z,k'_\perp)}, \\
& S_{TT}^{xx(0,em)} (y,z,k'_\perp) = \frac{2(k'^2_x - k'^2_y)}{3M^2} \frac{\sum_q e_q^2 \hat D_{1TT}^\perp (z,k'_\perp) }{\sum_q e_q^2 \hat D_1 (z,k'_\perp)},\\
%
& S_{LT}^{n(0,em)}(y, z, k'_\perp) =0,\\ 
& S_{LT}^{t(0,em)}(y, z, k'_\perp) = - \frac{2|\vec k'_\perp|}{3M} \frac{\sum_q e_q^2 \hat D_{1LT}^\perp (z,k'_\perp)}{\sum_q e_q^2 \hat D_1 (z, k'_\perp)},\\
&S_{TT}^{nn(0,em)} (y, z, k'_\perp) = - \frac{2|\vec k'_\perp|^2}{3M^2} \frac{\sum_q e_q^2 \hat D_{1TT}^\perp (z, k'_\perp)}{\sum_q e_q^2 \hat D_1 (y,k'_\perp)},\\
&S_{TT}^{nt(0,em)}(y, z, k'_\perp) =0.
\end{align}

Up to twist-3, we have,
\begin{align}
&S_{LL}^{(em)} (y,z,k'_\perp) = S_{LL}^{(0,em)} (y,z,k'_\perp)\Bigl[1+\frac{M}{Q}\hat\Delta^{(em)} (y,z,k'_\perp)\Bigr] 
- \frac{2k'_x}{zQ} \frac{\tilde B (y)}{A(y)} \frac{\sum_q e_q^2 \hat D_{LL}^\perp (z,k'_\perp)  }{\sum_q e_q^2 \hat D_1 (z,k'_\perp) },\\
&S_{LT}^{x (em)} (y,z,k'_\perp) = S_{LT}^{x(0,em)} (y,z,k'_\perp) \Bigl[1+\frac{M}{Q} \hat \Delta (y,z,k'_\perp)\Bigr] 
+ \frac{8M}{3 zQ}\frac{\tilde B (y)}{A(y)}
\frac{\sum_q e_q^2 \Bigl[ k'^2_x \hat D_{LT}^\perp (z, k'_\perp) - M^2 \hat D_{LT} (z,k'_\perp) \Bigr]}{M^2\sum_q e_q^2 \hat D_1 (z,k'_\perp)}, \\
&S_{LT}^{y (em)} (y,z,k'_\perp) = S_{LT}^{y(0,em)} (y,z,k'_\perp) \Bigl[1+\frac{M}{Q} \hat \Delta (y,z,k'_\perp)\Bigr]
+ \frac{8 k'_x k'_y }{3 zMQ}\frac{\tilde B (y)}{A(y)}
\frac{\sum_q e_q^2 \hat D_{LT}^\perp (z, k'_\perp)}{\sum_q e_q^2 \hat D_1 (z,k'_\perp)}, \\
&S_{TT}^{xy (em)}  (y,z,k'_\perp) = S_{TT}^{xy(0,em)} (y,z,k'_\perp) \Bigl[1+\frac{M}{Q} \hat \Delta (y,z,k'_\perp)\Bigr] 
- \frac{8k'_y}{3zQ}\frac{\tilde B (y)}{A(y)}
\frac{\sum_q e_q^2 \Bigl[2k'^2_x \hat D_{TT}^{\perp C} (z,k'_\perp) 
- M^2 \hat D_{TT}^{\perp A} (z, k'_\perp) \Bigr]}{M^2\sum_q e_q^2 \hat D_1 (z,k'_\perp)}  ,\\
&S_{TT}^{xx (em)} (y,z,k'_\perp) = S_{TT}^{xx(0,em)} (y,z,k'_\perp) \Bigl[1+\frac{M}{Q} \hat \Delta (y,z,k'_\perp)\Bigr] + \frac{8k'_x}{3zQ}\frac{\tilde B (y)}{A(y)}
\frac{\sum_q e_q^2 \Bigl[  M^2 \hat D_{TT}^{\perp A} (z, k'_\perp) -(k'^2_x - k'^2_y) \hat D_{TT}^{\perp C} (z,k'_\perp)\Bigr]}{M^2\sum_q e_q^2 \hat D_1 (z,k'_\perp)},\\
&S_{LT}^{n (em)} (y,z,k'_\perp) = - \frac{8M}{3zQ} \frac{\tilde B (y)}{A(y)} \frac{\sum_q e_q^2 k'_y \hat D_{LT} (z,k'_\perp)}{|\vec k'_\perp|\sum_q e_q^2 \hat D_1 (z, k'_\perp)},\\
&S_{LT}^{t (em)} (y,z,k'_\perp) = S_{LT}^{t(0,em)} (y,z,k'_\perp) \Bigl[1+\frac{M}{Q} \hat \Delta^{(em)} (y,z,k'_\perp)\Bigr]  
- \frac{8M}{3zQ} \frac{\tilde B (y)}{A(y)} \frac{\sum_q e_q^2 k'_x \Bigl[\hat D_{LT} (z,k'_\perp) - \frac{|\vec k'_\perp|^2}{M^2} \hat D_{LT}^\perp (z,k'_\perp)\Bigr] }{|\vec k'_\perp|\sum_q e_q^2 \hat D_1 (z, k'_\perp)} ,\\
&S_{TT}^{nn(em)} (y,z,k'_\perp)  =S_{TT}^{nn(0,em)} (y,z,k'_\perp) \Bigl[1+\frac{M}{Q} \hat \Delta^{(em)} (y,z,k'_\perp)\Bigr] 
 - \frac{8 k'_x}{3zQ} \frac{\tilde B (y)}{A(y)}  \frac{\sum_q e_q^2 \Bigl[\hat D_{TT}^{\perp A} (z, k'_\perp) - \frac{|\vec k'_\perp|^2}{M^2} \hat D_{TT}^{\perp C} (z, k'_\perp)\Bigr]}{\sum_q e_q^2 \hat D_1 (z,k'_\perp)},\\
&S_{TT}^{nt(em)} (y,z,k'_\perp)  = \frac{8 k'_y }{3zQ} \frac{\tilde B(y)}{A(y)} \frac{\sum_q e_q^2 \hat D_{TT}^{\perp A} (z, k'_\perp)}{\sum_q e_q^2 \hat D_1 (z,k'_\perp)}.
\end{align}

In terms of $y$, $z$ and $p_T$, they are given by,
\begin{align}
& S_{LL}^{(0,em)} (y,z, p_T ) = 
\frac{\sum_q e_q^2D_{1LL} (z, p_T )}{2\sum_q e_q^2 D_1(z, p_T )} ,\\
& S_{LT}^{x(0,em)} (y,z, p_T ) = - \frac{2p_X}{3zM} \frac{\sum_q e_q^2 D_{1LT}^\perp (z, p_T ) }{\sum_q e_q^2 D_1 (z, p_T )}, \\
& S_{LT}^{y(0,em)} (y,z, p_T ) = - \frac{2p_Y}{3zM}\frac{\sum_q e_q^2D_{1LT}^\perp (z, p_T )}{ \sum_q e_q^2 D_1 (z, p_T )}, \\
& S_{TT}^{xy(0,em)} (y,z, p_T ) = \frac{4 p_X p_Y}{3z^2M^2} \frac{\sum_q e_q^2 D_{1TT}^\perp (z, p_T ) }{\sum_q e_q^2 D_1 (z, p_T )}, \\
& S_{TT}^{xx(0,em)} (y,z, p_T ) = \frac{2(p^2_X - p^2_Y)}{3z^2M^2} \frac{\sum_q e_q^2 D_{1TT}^\perp (z, p_T ) }{\sum_q e_q^2 D_1 (z, p_T )},\\
& S_{LT}^{n(0,em)}(y, z,  p_T ) =  S_{TT}^{nt(0,em)}(y, z,  p_T ) = 0, \\
& S_{LT}^{t(0,em)}(y, z,  p_T ) = - \frac{2|\vec  p_T |}{3zM} \frac{\sum_q e_q^2   D_{1LT}^\perp (z, p_T )}{\sum_q e_q^2   D_1 (z,  p_T )},\\
&S_{TT}^{nn(0,em)} (y, z,  p_T ) = - \frac{2|\vec  p_T |^2}{3z^2M^2} \frac{\sum_q e_q^2   D_{1TT}^\perp (z,  p_T )}{\sum_q e_q^2   D_1 (y, p_T )}.
\end{align}

Up to twist-3, we have,
\begin{align}
S&_{LL}^{em}(y,z, p_T ) = S_{LL}^{(0,em)} (y,z, p_T )\Bigl[1+\frac{M}{Q}\Delta^{(em)} (y,z, p_T )\Bigr]  
- \frac{2p_X}{z^2Q} \frac{\tilde B (y)}{A(y)} 
\frac{\sum_q e_q^2 \hat D_{LL}^\perp (z, p_T )}{\sum_q e_q^2 \hat D_1 (z, p_T ) },\\
S&_{LT}^{x (em)}(y,z, p_T ) = S_{LT}^{x(0,em)} (y,z, p_T ) \Bigl[1+\frac{M}{Q}   \Delta^{(em)} (y,z, p_T )\Bigr] 
 + \frac{8M}{3z^3Q}\frac{\tilde B (y)}{A(y)}
\frac{\sum_q e_q^2 \Bigl[p^2_X \hat D_{LT}^\perp (z,p_T)-z^2 M^2\hat D_{LT} (z, p_T ) \Bigr]}{M^2\sum_q e_q^2 \hat D_1 (z, p_T )}, \\
S&_{LT}^{y (em)}(y,z, p_T ) = S_{LT}^{y(0,em)} (y,z, p_T ) \Bigl[1+\frac{M}{Q}   \Delta^{(em)} (y,z, p_T )\Bigr] 
+ \frac{8p_X p_Y }{3z^3MQ}\frac{\tilde B (y)}{A(y)}
\frac{\sum_q e_q^2 \hat D_{LT}^\perp (z,  p_T )}{\sum_q e_q^2 \hat D_1 (z, p_T )}, \\
S&_{TT}^{xx (em)}(y,z, p_T ) = S_{TT}^{xx(0,em)} (y,z, p_T ) \Bigl[1+\frac{M}{Q}   \Delta^{(em)} (y,z, p_T )\Bigr] 
+ \frac{8p_X}{3z^4Q}\frac{\tilde B (y)}{A(y)}
\frac{\sum_q e_q^2 \Bigl[ z^2 M^2 \hat D_{TT}^{\perp A} (z,  p_T ) -(p^2_X - p^2_Y) \hat D_{TT}^{\perp C} (z, p_T )\Bigr]}{M^2\sum_q e_q^2 \hat D_1 (z, p_T )},\\
S&_{TT}^{xy (em)}(y,z, p_T ) = S_{TT}^{xy(0,em)} (y,z, p_T ) \Bigl[1+\frac{M}{Q}   \Delta^{(em)} (y,z, p_T )\Bigr]
- \frac{8p_Y}{3z^4Q}\frac{\tilde B (y)}{A(y)}
\frac{\sum_q e_q^2 \Bigl[2p^2_X \hat D_{TT}^{\perp C} (z, p_T ) 
- z^2 M^2 \hat D_{TT}^{\perp A} (z,  p_T ) \Bigr]}{M^2\sum_q e_q^2 \hat D_1 (z, p_T )}  ,\\
S&_{LT}^{n (em)}(y,z, p_T ) = - \frac{8M}{3zQ} \frac{\tilde B (y)}{A(y)} \frac{\sum_q e_q^2  p_Y    D_{LT} (z, p_T )}{|\vec  p_T |\sum_q e_q^2   D_1 (z,  p_T )},\\
S&_{LT}^{t (em)}(y,z, p_T ) = S_{LT}^{t(0,em)} (y,z, p_T ) \Bigl[1+\frac{M}{Q} \Delta^{(em)} (y,z, p_T )\Bigr] 
 - \frac{8M}{3zQ} \frac{\tilde B (y)}{A(y)} 
\frac{\sum_q e_q^2  p_X  \Bigl[ D_{LT} (z,p_T ) +\frac{ p_T ^2}{z^2M^2} D_{LT}^\perp (z, p_T )\Bigr] }{|\vec  p_T |\sum_q e_q^2 D_1(z,p_T )} ,\\
S&_{TT}^{nn(em)}(y,z, p_T ) =S_{TT}^{nn(0,em)} (y,z, p_T ) \Bigl[1+\frac{M}{Q}   \Delta^{(em)} (y,z, p_T )\Bigr] 
- \frac{8  p_X }{3z^2Q} \frac{\tilde B (y)}{A(y)}  \frac{\sum_q e_q^2 \Bigl[  D_{TT}^{\perp A} (z,  p_T ) + \frac{p_T^2}{z^2M^2}   D_{TT}^{\perp C} (z,  p_T )\Bigr]}{\sum_q e_q^2   D_1 (z, p_T )},\\
S&_{TT}^{nt(em)}(y,z, p_T ) = \frac{8  p_Y  }{3z^2Q} \frac{\tilde B(y)}{A(y)} \frac{\sum_q e_q^2   D_{TT}^{\perp A} (z,  p_T )}{\sum_q e_q^2   D_1 (z, p_T )}.
\end{align}
\end{widetext}

We see that in $e^++e^-\to\gamma^*\to V+\bar q+X$, at the leading twist, we 
have three non-zero components, $S_{LL}^{(0,em)}$, $S_{LT}^{t(0,em)}$ and $S_{TT}^{nn(0,em)}$,  
proportional to the three components $D_{1LL}$, $D_{1LT}^\perp$, and $D_{1TT}^\perp$ respectively, 
while the other two components $S_{LT}^{n(0,em)}$ and $S_{TT}^{nt(0,em)}$ are zero. 
On the other hand, at twist 3, all the five components, 
$S_{LL}^{(em)}$, $S_{LT}^{n(em)}$, $S_{LT}^{t(em)}$, $S_{TT}^{nn(em)}$ and $S_{TT}^{tn(em)}$ 
receive contributions from different components of the fragmentation function. 
Clearly, precise measurements of them provide deep understanding of the TMD fragmentation function.

\subsection{Confronting with experiments}
In experiments, not only the azimuthal asymmetries but also different components of the polarization 
can be measured in a conceptually easy way. 
Hyperon polarization can be measured by studying the angular distribution of the decay products 
of its spin self analysing parity violating decay. 
All the five independent components $S_{LL}$, $S_{LT}^{n}$, $S_{LT}^{t}$, $S_{TT}^{nn}$ and $S_{TT}^{nt}$, 
of the tensor polarization of vector meson can also be measured via the 
angular distribution in its strong decay into two pseudoscalar mesons~\cite{Bacchetta:2000jk}. 

In experiments with $e^+e^-$ annihilation at high energies, 
measurements in this direction have been carried out at LEP by ALEPH and OPAL collaborations 
for $\Lambda$ hyperon production~\cite{Buskulic:1996vb, Ackerstaff:1997nh,ALEPH:2005ab}.
There are also measurements on the spin alignment $\rho_{00}=(1-2S_{LL})/3$ 
for vector mesons such as $K^*$, $\rho$ 
and so on~\cite{Ackerstaff:1997kj, Ackerstaff:1997kd, Abreu:1997wd, ALEPH:2005ab}. 
Results for $z$ dependences have been obtained in both cases. 
These data are definitely still far from enough to limit the precise forms 
of the fragmentation functions involved, 
they can however be used to extract some hints for the forms of the corresponding components. 

These data have attracted much attention theoretically 
and many phenomenological model studies 
have been carried out in last years~\cite{Gustafson:1992iq,Boros:1998kc,Liu:2000fi,Liu:2001yt,Liang:2002ub,Xu:2002hz,
Ma:1998pd,Ma:1999gj,Ma:1999wp,Ma:1999hi,Ma:2000uu,Ma:2000cg,Chi:2013hka,
Ellis:2002zv,Xu:2001hz,Xu:2003fq}.
Among them, a series of calculations have been carried out~\cite{Gustafson:1992iq,Boros:1998kc,Liu:2000fi,Liu:2001yt,Liang:2002ub,Xu:2002hz} 
using a formalism where hyperon polarization is calculated according to the origin of its production 
and event generator JETSET~\cite{Sjostrand:1985ys} based on Lund model~\cite{Andersson:1983ia,Andersson:1998tv} 
is used to trace back the origin(s).  
In these calculations, not only the polarizations of the directly produced hyperons but also those from the 
decays are taken into account. 
Another class of studies are carried out by relating the fragmentation function to the parton distribution 
function using the reciprocity relation 
such as Gribov relation\cite{Ma:1998pd,Ma:1999gj,Ma:1999wp,Ma:1999hi,Ma:2000uu,Ma:2000cg}.

In the theoretical framework presented in this paper, for directly produced hyperons, 
the longitudinal polarization is given by Eq.~(\ref{Plh0}) at the leading twist and by Eq.~(\ref{Plh}) 
if twist 3 contributions are considered. 
The calculations presented in~\cite{Gustafson:1992iq,Boros:1998kc,Liu:2000fi,Liu:2001yt,Liang:2002ub,Xu:2002hz} 
are equivalent to take 
\begin{equation}
\Delta D_{1L}^{q\to H_j}(z,p_T)=t_F^{q\to H_j}f^{q\to H_j}(z)f_\perp(p_T), \label{D1Lpara}
\end{equation}
where $f^{q\to H_j}(z)$ is the number density of the first rank $H_j$ in the fragmentation of 
the quark $q$ in recursive cascade model such as Lund model~\cite{Andersson:1983ia,Andersson:1998tv}; 
$t_F^{q\to H_j}=\Delta Q^{H_j}/n_q^{H_j}$ is a spin transfer constant, 
$\Delta Q^{H_j}$ is the average value of the contribution of spin of quark of flavor $q$ 
to the spin of $H_j$ and $n_q^{H_j}$ is the number of valence quark of flavor $q$ in $H_j$; 
$f_\perp(p_T)$ denotes the transverse momentum dependence and is taken as factorized. 
It is interesting to see that, with the simple SU(6) quark model results for $\Delta Q^{H_j}$ 
and Lund model\cite{Andersson:1983ia,Andersson:1998tv} 
input for $f^{q\to H_j}(z)$, the simple parameterization given by Eq.~(\ref{D1Lpara}) 
can already provide a rather good fit to the 
data available~\cite{Buskulic:1996vb, Ackerstaff:1997nh,ALEPH:2005ab}. 
Similar ansatz was taken for vector meson spin alignment, i.e., 
\begin{equation}
D_{1LL}^{q\to V}(z,p_T)\propto f^{q\to V}(z)f_\perp(p_T), \label{D1LLpara}
\end{equation}
and reasonable fit was obtained to the data available. 
We see that these data~\cite{Buskulic:1996vb, Ackerstaff:1997nh,Ackerstaff:1997kj,
Ackerstaff:1997kd, Abreu:1997wd, ALEPH:2005ab} indeed provide some clue to parameterize 
the $z$-dependence of these components of the fragmentation function.

For the TMD session, even less data is available yet in particular 
when azimuthal asymmetries and polarizations are concerned. 
For the unpolarized fragmentation function $D_1(z,p_T)$, 
there are few data available on the $p_T$ spectra of the produced hadrons 
in $e^+e^-$ annihilation~\cite{Barate:1996fi}
and they can provide useful information on the $p_T$ dependence of $D_1(z,p_T)$.  
Recently there are measurements carried out in semi-inclusive deep-inelastic 
lepton-nucleon scatterings~\cite{Arneodo:1986cf,Ashman:1991cj,Mkrtchyan:2007sr,Osipenko:2008aa,
Airapetian:2009jy,Airapetian:2012ki,Asaturyan:2011mq} 
and analysis~\cite{Anselmino:2005nn,Schweitzer:2010tt,Signori:2014kda,Anselmino:2013lza}
have been made to extract information on the $p_T$ dependence of $D_1(z,p_T)$.
It is usually taken as factorized from $z$ dependence and a flavor independent 
Gaussian ansatz is taken for the $p_T$ dependence.
To the accuracy of current studies, such Gaussian ansatz seems to provides us 
a reasonable description~\cite{Anselmino:2005nn,Schweitzer:2010tt,Anselmino:2013lza} 
of the data available~\cite{Arneodo:1986cf,Ashman:1991cj,Mkrtchyan:2007sr,Osipenko:2008aa,
Airapetian:2009jy,Airapetian:2012ki,Asaturyan:2011mq}.
However, it has already been pointed out that a careful check seems to suggest that 
the falvor dependence of the transverse momentum dependence and 
even violation of its fatctorization from 
the longitudinal momentum dependence do exist~\cite{Signori:2014kda}.
For spin dependent components of the fragmentation function discussed above, 
no data on transverse momentum dependence is available yet. 

Recent measurements have been carried out on azimuthal asymmetries 
for two hadron production by BELLE, BARBAR and BES III Collaborations\cite{Vossen:2011fk,TheBABAR:2013yha,BES3}. 
However, no data is available yet on the azimuthal asymmetries for 
single hadron with respect to the lepton plane. 
Such measurements are in principle easy to carry out in the corresponding reactions. 
Analysis can be made easily for spinless hadrons but for hadrons with spins they should be made 
consistently with polarization measurements.

\section{Summary and outlook} 

In summary, we show that the collinear expansion can be extended to 
the semi-inclusive hadron production process $e^++e^-\to h+\bar q+X$. 
We show that a theoretical framework can be obtained in this way 
where leading and higher twist contributions can be calculated systematically. 
We carry out the calculations for hadrons with spin-0, spin-$1/2$ and spin-$1$ up to twist 3 
and present the results for the hadronic tensor, the differential cross section, 
the azimuthal asymmetries and the polarizations in terms of the different components 
of the gauge invariant transverse momentum dependent fragmentation function. 

For hadrons with spin zero, the differential cross section show 
a left-right and an up-down azimuthal asymmetry in reaction via $Z$-boson exchange 
but only the up-down asymmetry survives in reaction via virtual photon exchange. 
Such azimuthal asymmetries are also influenced by the polarizations when hadrons
with spins are studied.

For spin-$1/2$ hadrons, the polarization is described by the polarization vector $S$.
The hadrons produced show longitudinal and transverse polarizations at leading twist 
and also twist 3 addenda to them.  
The longitudinal polarization $P_{Lh}$ is usually defined with respect to helicity. 
The leading twist contribution to $P_{Lh}$ exist only in $e^++e^-\to Z^0\to h+\bar q+X$ 
since quark produced in $Z$-decay is longitudinally polarized. 
This leading twist contribution is proportional to the quark polarization $P_q(y)$ 
and is determined by $\Delta D_{1L}$ that describes the spin transfer in fragmentation. 
The results show that the transverse polarization is conveniently described by 
$P_{hn}$ along the normal direction of the production plane and 
$P_{ht}$ along the transverse direction in the production plane. 
We have a leading twist contribution to $P_{hn}^{(0)}$ that is independent of $P_q(y)$ 
and is determined by $D_{1T}^\perp$, a counterpart of Sivers function in fragmentation function. 
The leading twist contribution to $P_{ht}^{(0)}$ is proportional to $P_q(y)$ thus vanishes in 
reaction via electromagnetic interaction, i.e., $e^++e^-\to\gamma^*\to h+\bar q+X$,
where $P_q(y)=0$ and parity violating terms disappear.  
At leading twist, the only nonzero component of the polarization $e^++e^-\to\gamma^*\to h+\bar q+X$ 
is the transverse polarization $P_{hn}^{(0)}$ along the normal direction of the production plane. 
However, different contributions at twist 3 to all the three components exist.

The results for spin-1 hadrons show even abundant features: 
in addition to the spin alignment given by $S_{LL}$, 
we obtain results also for all the other four independent components 
$S_{LT}^{n}$, $S_{LT}^{t}$, $S_{TT}^{nn}$ and $S_{TT}^{tn}$. 
In $e^++e^-\to Z^0\to V+\bar q+X$, all the five components have leading twist 
contributions and each component is determined by one component 
of the fragmentation function and the unpolarized fragmentation function. 
Among them the leading twist contributions to $S_{LL}$, $S_{LT}^{t}$, and $S_{TT}^{nn}$  
are independent of the polarization $P_q(y)$ of the fragmenting quark 
while those to the other two components, $S_{LT}^{n}$and $S_{TT}^{tn}$, are proportional to $P_q(y)$.
As a result, in $e^++e^-\to \gamma^*\to V+\bar q+X$, 
the leading twist contributions to $S_{LL}$, $S_{LT}^{t}$, and $S_{TT}^{nn}$ survive, 
while those to $S_{LT}^{n}$and $S_{TT}^{tn}$ vanish.
In all the different cases, twist 3 contributions exist. 

Measurements of longitudinal Lambda hyperon polarization and 
vector meson spin alignments have been carried out 
at LEP by ALEPH and OPAL Collaborations\cite{Buskulic:1996vb, Ackerstaff:1997nh,Ackerstaff:1997kj,
Ackerstaff:1997kd, Abreu:1997wd, ALEPH:2005ab}, 
and $z$ dependences have been obtained. 
These data provides good hints for $z$-dependence of the corresponding components 
of the spin dependent session of the fragmentation function. 
However, there is no data available yet for the transverse momentum dependence 
of the spin dependent components of the fragmentation function discussed above.
Such measurements are important in studying different components of the fragmentation function 
and can provide useful information on the properties of QCD. 
They can be carried out in the existing $e^+e^-$ colliders such as BELLE and BEPC, 
and can certainly also be carried out in future $e^+e^-$ colliders~\cite{ZhangZX2012} at high energies 
discussed in the community.

\section*{Acknowledgements}
This work was supported in part by the National Natural Science Foundation of China
(project 11035003 and 11375104),  and 
the Major State Basic Research Development Program in China (No. 2014CB845406).


\end{document}